\documentclass[12pt]{article}
\usepackage{amssymb}
\usepackage{epsfig}
\usepackage{bbm}

\usepackage{graphicx}

\def\L{\mathcal L}
\def\e{\varepsilon}
\def\simlt{\stackrel{<}{{}_\sim}}

\newcommand{\wt}{\widetilde}

\begin{document}

\def\a{\alpha}
\def\b{\beta}
\def\c{\chi}
\def\d{\delta}
\def\e{\epsilon}
\def\f{\phi}
\def\g{\gamma}
\def\h{\eta}
\def\i{\iota}
\def\j{\psi}
\def\k{\kappa}
\def\l{\lambda}
\def\m{\mu}
\def\n{\nu}
\def\o{\omega}
\def\p{\pi}
\def\q{\theta}
\def\r{\rho}
\def\s{\sigma}
\def\t{\tau}
\def\u{\upsilon}
\def\x{\xi}
\def\z{\zeta}
\def\D{\Delta}
\def\F{\Phi}
\def\G{\Gamma}
\def\J{\Psi}
\def\L{\Lambda}
\def\O{\Omega}
\def\P{\Pi}
\def\Q{\Theta}
\def\S{\Sigma}
\def\U{\Upsilon}
\def\X{\Xi}

%Varletters
\def\ve{\varepsilon}
\def\vf{\varphi}
\def\vr{\varrho}
\def\vs{\varsigma}
\def\vq{\vartheta}

\def\dg{\dagger}                                     % hermitian conjugate
\def\ddg{\ddagger}                                   % double dagger
\def\wt#1{\widetilde{#1}}                    % big tilde
\def\mt{\widetilde{m}_1}
\def\mti{\widetilde{m}_i}
\def\rt{\widetilde{r}_1}
\def\mtt{\widetilde{m}_2}
\def\mttt{\widetilde{m}_3}
\def\rtt{\widetilde{r}_2}
\def\mb{\overline{m}}
\def\VEV#1{\left\langle #1\right\rangle}        % < >
\def\be{\begin{equation}}
\def\ee{\end{equation}}
\def\ds{\displaystyle}
\def\ra{\rightarrow}

\def\bea{\begin{eqnarray}}
\def\eea{\end{eqnarray}}
\def\NO{\nonumber}
\def\Bar#1{\overline{#1}}

% Journal abbreviations (preprints)

\def\pl#1#2#3{Phys.~Lett.~{\bf B {#1}} ({#2}) #3}
\def\np#1#2#3{Nucl.~Phys.~{\bf B {#1}} ({#2}) #3}
\def\prl#1#2#3{Phys.~Rev.~Lett.~{\bf #1} ({#2}) #3}
\def\pr#1#2#3{Phys.~Rev.~{\bf D {#1}} ({#2}) #3}
\def\zp#1#2#3{Z.~Phys.~{\bf C {#1}} ({#2}) #3}
\def\cqg#1#2#3{Class.~and Quantum Grav.~{\bf {#1}} ({#2}) #3}
\def\cmp#1#2#3{Commun.~Math.~Phys.~{\bf {#1}} ({#2}) #3}
\def\jmp#1#2#3{J.~Math.~Phys.~{\bf {#1}} ({#2}) #3}
\def\ap#1#2#3{Ann.~of Phys.~{\bf {#1}} ({#2}) #3}
\def\prep#1#2#3{Phys.~Rep.~{\bf {#1}C} ({#2}) #3}
\def\ptp#1#2#3{Progr.~Theor.~Phys.~{\bf {#1}} ({#2}) #3}
\def\ijmp#1#2#3{Int.~J.~Mod.~Phys.~{\bf A {#1}} ({#2}) #3}
\def\mpl#1#2#3{Mod.~Phys.~Lett.~{\bf A {#1}} ({#2}) #3}
\def\nc#1#2#3{Nuovo Cim.~{\bf {#1}} ({#2}) #3}
\def\ibid#1#2#3{{\it ibid.}~{\bf {#1}} ({#2}) #3}

\title
{\vspace*{-12mm}
{\normalsize \mbox{ }\hfill
\begin{minipage}{5cm}
CERN-PH-TH/2010-292
\end{minipage}}\\
\bf  Testing $SO(10)$-inspired leptogenesis with low energy neutrino experiments
}
\author{
{\Large Pasquale Di Bari$^{a,b}$ and Antonio Riotto$^{c,d}$}
\\
$^a$
{\it\small School of Physics and Astronomy},
{\it\small University of Southampton,}
{\it\small  Southampton, SO17 1BJ, U.K.}
 \\
$^b$
{\it\small Department of Physics and Astronomy},
{\it\small University of Sussex,}
{\it\small  Brighton, BN1 9QH, U.K.} \\
$^c$
{\it\small  INFN, Sezione di Padova},
{\it\small  Dipartimento di Fisica Galileo Galilei} \\
{\it\small  Via Marzolo 8, I-35131 Padua, Italy}\\
$^d$
{\it\small CERN, PH-TH Division,  CH-1211, Geneva 23, Switzerland}
}

\maketitle \thispagestyle{empty}

\vspace{-10mm}
%\centerline{\date{\today}}

\begin{abstract}
\noindent
We extend the results of a previous analysis  of ours showing
that, when both heavy and light flavour effects are taken into account,
successful minimal (type I + thermal) leptogenesis with $SO(10)$-inspired relations is possible.
Barring fine tuned choices of the parameters, these relations enforce a hierarchical RH neutrino
mass spectrum that results into a final asymmetry dominantly produced by the
next-to-lightest RH neutrino decays ($N_2$ dominated leptogenesis).
%The neutrino Dirac mass
%matrix is assumed to be diagonal  in the same basis
%where the charged  lepton mass matrix is diagonal.
We present the constraints on the whole set of low energy
neutrino parameters. Allowing a small misalignment between the Dirac basis and
the charged lepton basis as in the quark sector, the allowed regions enlarge
and the lower bound on the reheating temperature gets relaxed to values
as low as $\sim 10^{10}\,{\rm GeV}$. It is confirmed that for normal ordering (NO)
there are two allowed ranges of values for the lightest neutrino mass:
$m_1\simeq (1-5)\times 10^{-3}\,{\rm eV}$ and  $m_1\simeq (0.03-0.1)\,{\rm eV}$.
For $m_1\lesssim 0.01\,{\rm eV}$
the allowed region in the plane $\theta_{13}$-$\theta_{23}$
is approximately given by $\theta_{23} \lesssim 49^{\circ}+ 0.65\,(\theta_{13}-5^{\circ})$,
while the neutrinoless double beta decay effective neutrino mass falls in the
range $m_{ee}=(1-3)\times 10^{-3}\,{\rm eV}$ for
$\theta_{13}=(6^{\circ}-11.5^{\circ})$. For $m_1 \gtrsim 0.01\,{\rm eV}$,
one has quite sharply $m_{ee}\simeq m_1$ and an upper
bound $\theta_{23}\lesssim 46^{\circ}$.
These constraints will be tested by low energy
neutrino experiments during next years.
We also find that inverted ordering (IO), though quite strongly constrained,
is not completely ruled out. In particular, we find approximately
$\theta_{23}\simeq 43^{\circ}+12^{\circ}\,\log(0.2\,{\rm eV}/m_1)$,
that will be fully tested by future experiments.
\end{abstract}

\newpage

%%%%%%%%%%%%%%%%%%%%%%%%%%%%%%%%%%%%%%%%%%%%%%%%%%
\section{Introduction}
%%%%%%%%%%%%%%%%%%%%%%%%%%%%%%%%%%%%%%%%%%%%%%%%%%

With the discovery of neutrino masses and mixing in neutrino
oscillation experiments, leptogenesis \cite{fy,review} has become
the most attractive model of baryogenesis to explain the
observed matter-antimatter asymmetry of the Universe.
This can be expressed for example in terms of the baryon-to-photon
number ratio and is very well measured by CMB observations \cite{WMAP7} to be
\be\label{etaBobs}
\eta_B^{\rm CMB} = (6.2 \pm 0.15)\times 10^{-10} \, .
\ee
Leptogenesis originates from  the see-saw mechanism \cite{seesaw}
that is based on a simple extension of the Standard Model
where right-handed (RH) neutrinos  with a Majorana mass matrix $M$ and Yukawa
couplings $h$ to leptons and Higgs are added.
Within $SO(10)$ models, three RH neutrinos $N_{i}$ ($i=1,2,3$) are nicely predicted
and for this reason they are traditionally regarded as
the most appealing theoretical framework to embed the seesaw mechanism.

However,  within the simplest set of assumptions
inspired by $SO(10)$ models \cite{branco},
barring strong fine-tuned degeneracies in the RH neutrino mass spectrum
and using the experimental information from neutrino oscillation experiments,
the traditional $N_1$-dominated leptogenesis scenario predicts an asymmetry
that falls many orders of magnitudes below the observed one  \cite{orlof,afs}.
This is because, within  $N_1$-dominated leptogenesis, where
the spectrum of RH  neutrinos is hierarchical and the asymmetry is produced
from the decays of the lightest ones, successful leptogenesis
implies a stringent lower bound on their mass \cite{di},
$M_1 > {\cal O}(10^9)\,{\rm GeV}$.
On the other hand, $SO(10)$ grand-unified theories  typically yield, in their simplest version
and for the measured values of the neutrino mixing parameters,
a hierarchical spectrum with the RH neutrino masses proportional
to the squares of the up-quark masses, leading to
$M_1={\cal O}(10^5)\,{\rm GeV}$ and therefore
to a final asymmetry much below the observed one.

However, it has been shown \cite{SO10} that, when the production from
the next-to-lightest RH neutrinos \cite{geometry} and lepton flavour effects \cite{flavour}
are simultaneously taken into account \cite{vives}, the final asymmetry
can be generated by the decays of the next-to-lightest RH neutrinos and
allowed regions in the low energy neutrino parameter space open up.

In this paper we proceed with the analysis of \cite{SO10} and
present the resulting constraints on all low energy neutrino
parameters. The paper is organized as follows.
In Section 2 we discuss the current experimental status on low
energy neutrino parameters, we set up the notation and describe the general procedure
to calculate the the asymmetry and find the constraints.
In Section 3
we  first consider  the case already studied in \cite{SO10}, when the
Dirac basis and the charged lepton basis coincide and then, in Section 4,
we allow  for a misalignment between the two bases not larger than that one described
by the CKM matrix in the quark sector. Finally, in Section 5 we
present a global scan in the space of parameters where all
possible cases between the case of no misalignment and the case of a
misalignment at the level of the CKM matrix are taken into account.
We also discuss two scenarios, one at small $m_1$ and one at large $m_1$,
and show  how, within $SO(10)$-inspired models, minimal leptogenesis could be
tested in future low energy neutrino experiments.

Notice that our discussion is made within a non-supersymmetric framework.
Recently a study of $SO(10)$-inspired models within a supersymmetric framework
has also  enlightened interesting potential connections with lepton flavour violating decays
and Dark Matter \cite{blanchet}. An analysis of leptogenesis within left-right symmetric
models, where a type II seesaw contribution to the neutrino mass matrix is also present,
has been performed in \cite{abada}. Within these models, the minimal type I scenario
considered here represents a particular case recovered under specific conditions.

%%%%%%%%%%%%%%%%%%%%%%%%%%%%%%%%%%%%%%%%%%%%%%%%%%%%
\section{Experimental information and general setup}
%%%%%%%%%%%%%%%%%%%%%%%%%%%%%%%%%%%%%%%%%%%%%%%%%%%%

After spontaneous symmetry breaking,
a Dirac mass term $m_D=h\, v$, is generated
by the vacuum expectation value (VEV)  $v=174$ GeV of the Higgs boson.
In the see-saw limit, $M\gg m_D$, the spectrum of neutrino mass eigenstates
splits in two sets: three very heavy neutrinos, $N_1,N_2$ and $N_3$
respectively with masses $M_1\leq M_2 \leq M_3$ almost coinciding with
the eigenvalues of $M$, and three light neutrinos with masses $m_1\leq m_2\leq m_3$,
the eigenvalues of the light neutrino mass matrix
given by the see-saw formula \cite{seesaw},
\be
m_{\nu}= - m_D\,{1\over D_M}\,m_D^T \, ,
\ee
that we wrote in a basis where the Majorana mass matrix is diagonal
defining $D_M\equiv {\rm diag}(M_1,M_2,M_3)$.
The symmetric light neutrino mass matrix $m_{\nu}$ is
diagonalized by a unitary matrix $U$,
\begin{equation}
U^{\dagger}\,m_{\nu}\,U^{\star}=-D_m \,
\end{equation}
with $D_m\equiv {\rm diag}(m_1,m_2,m_3)$, that,
in the basis where the charged lepton mass matrix is diagonal,
can be identified with the lepton mixing matrix.

Neutrino oscillation experiments measure two
neutrino mass-squared differences.
For NO one has
$m^{\,2}_3-m_2^{\,2}=\Delta m^2_{\rm atm}$ and
$m^{\,2}_2-m_1^{\,2}=\Delta m^2_{\rm sol}$. The two heavier neutrino masses
can therefore be expressed in terms of the lightest neutrino mass $m_1$ as
\be\label{m2m3nor}
m_2 = \sqrt{m_1^2 + m_{\rm sol}^2} \, , \;\;\; \mbox{\rm and} \;\;\;
m_3 =  \sqrt{m_1^2 + m_{\rm atm}^2} \, ,
\ee
where we defined $m_{\rm atm} \equiv
\sqrt{\Delta m^2_{\rm atm}+\Delta m^2_{\rm sol}}=
(0.050\pm 0.001)\,{\rm eV}$ and
$m_{\rm sol}\equiv \sqrt{\D m^2_{\rm sol}}
=(0.00875\pm 0.00012)\,{\rm eV}$ \cite{oscillations}.
Recently, a conservative upper bound on the sum of neutrino masses,
$\sum_i\,m_i\leq 0.58\,{\rm eV}\; (95\%\, {\rm CL})$,
has been obtained by the WMAP collaboration \cite{WMAP7} combining WMAP 7 years data
plus  baryon acoustic oscillations observations and the latest HST measurement of $H_0$.
Considering that it falls in the quasi-degenerate regime, it straightforwardly translates into
\be\label{upperbound}
m_1 < 0.19\,{\rm eV} \;\; (95\%\, {\rm CL}) \, .
\ee

We will adopt the following parametrization  for the matrix $U$
in terms of the mixing angles, the Dirac phase $\delta$ and
 the Majorana phases $\rho$ and $\sigma$ \cite{PDG}
\begin{equation}\label{Umatrix}
U=\left( \begin{array}{ccc}
c_{12}\,c_{13} & s_{12}\,c_{13} & s_{13}\,e^{-{\rm i}\,\d} \\
-s_{12}\,c_{23}-c_{12}\,s_{23}\,s_{13}\,e^{{\rm i}\,\d} &
c_{12}\,c_{23}-s_{12}\,s_{23}\,s_{13}\,e^{{\rm i}\,\d} & s_{23}\,c_{13} \\
s_{12}\,s_{23}-c_{12}\,c_{23}\,s_{13}\,e^{{\rm i}\,\d}
& -c_{12}\,s_{23}-s_{12}\,c_{23}\,s_{13}\,e^{{\rm i}\,\d}  &
c_{23}\,c_{13}
\end{array}\right)
\cdot {\rm diag}\left(e^{i\,\rho}, 1, e^{i\,\sigma}
\right)\,
\end{equation}
and the following $2\,\sigma$ ranges for the three mixing angles \cite{oscillations}
\be\label{twosigma}
\theta_{12}= (31.3^\circ-36.3^\circ)  \, , \;\;\;
\theta_{23}= (38.5^\circ-52.5^\circ) \, , \;\;\;
\theta_{13}= (0^\circ-11.5^\circ) \, .
\ee
In the case of IO the expression of $m_2$ in terms of $m_1$ becomes
\be\label{m2m3inv}
m_2 = \sqrt{m_1^2 +m_{\rm atm}^2- m_{\rm sol}^2} \, ,
\ee
while the expression for $m_3$ does not change.
With the adopted convention for the light neutrino masses, $m_1<m_2<m_3$,
the case of IO corresponds to relabel the column of the leptonic mixing matrix
performing a column cyclic permutation, explicitly
\begin{equation}\label{Umatrix}
U=\left( \begin{array}{ccc}
s_{13}\,e^{-{\rm i}\,\d} & c_{12}\,c_{13} & s_{12}\,c_{13}  \\
s_{23}\,c_{13} & -s_{12}\,c_{23}-c_{12}\,s_{23}\,s_{13}\,e^{{\rm i}\,\d} &
c_{12}\,c_{23}-s_{12}\,s_{23}\,s_{13}\,e^{{\rm i}\,\d} \\
c_{23}\,c_{13} & s_{12}\,s_{23}-c_{12}\,c_{23}\,s_{13}\,e^{{\rm i}\,\d}
& -c_{12}\,s_{23}-s_{12}\,c_{23}\,s_{13}\,e^{{\rm i}\,\d}
\end{array}\right)
\cdot {\rm diag}\left(e^{i\,\sigma}, e^{i\,\rho}, 1  \right)\, .
\end{equation}
The predicted baryon-to-photon ratio $\eta_B$ is related to the
value of the final $(B-L)$ asymmetry $N^{\rm f}_{B-L}$ by  \cite{review}
\be\label{etaB}
\eta_B \simeq 0.96\times 10^{-2} N_{B-L}^{\rm f} \, ,
\ee
where $N_{B-L}$ is the $B-L$ number in a co-moving volume
that contains on average one RH neutrino $N_i$
in thermal ultra-relativistic equilibrium abundance ($T\gg M_i$).

The Dirac mass matrix can be diagonalized by a bi-unitary transformation
\be
m_D=V_L^{\dagger}\,D_{m_D}\,U_R \, ,
\ee
where $D_{m_D}={\rm diag}(\l_1,\l_2,\l_3)$.
The matrix $U_R$ can be obtained from $V_L$, $U$ and $m_i$,
considering that it provides a Takagi factorization \cite{takagi} of \cite{di2,branco}
\be
M^{-1} \equiv D^{-1}_{m_D}\,V_L\,U\,D_m\,U^T\,V_L^T\,D^{-1}_{m_D} \, ,
\ee
explicitly
\be\label{Takagi}
M^{-1} = U_R\,D_M^{-1}\,U_R^T \, .
\ee
For non degenerate $M_i$, the matrix $U_R$ can be determined
noticing that it diagonalizes $M^{-1}\,(M^{-1})^{\, \dagger}$, i.e.
\be
M^{-1}\,(M^{-1})^{\, \dagger} = U_R\,D_M^{-2}\,U_R^{\dagger} \, .
\ee
This relation determines $U_R$ unless a diagonal unitary transformation, since
any $\widetilde{U}_R=U_R\,D^{-1}_{\phi}$ is also a solution. However, given a $\widetilde{U}_R$,
 one can fix $D_{\phi}$ from the eq.~(\ref{Takagi}),
\be\label{Dphi}
D_{\phi} = \sqrt{D_M\,\widetilde{U}_R^{\dagger}\,M^{-1}\,\widetilde{U}^{\star}_R}
\ee
and in doing so $U_R$ is unambiguously determined.
Inspired by $SO(10)$ relations, we can parameterize the eigenvalues of
$m_D$ in terms of the up quark masses as
\be\label{SO(10)}
\l_{1}= \alpha_1\,m_u , \;\; \l_{2}= \alpha_2\, m_c ,\;\; \l_{3}= \alpha_3\,m_t \, .
\ee
Within $SO(10)$ models one can expect $\alpha_i={\cal O}(1)$ and we will refer to this case. The reader
is invited to read Ref. \cite{SO10} for a more comprehensive discussion about these $SO(10)$-inspired relations.
Notice however that our results will be valid for a much broader range of values, since, quite importantly,
it turns out that they are independent
of $\a_1$ and $\a_3$ provided $M_3\gg M_2$ and $M_1\lesssim 10^9\,{\rm GeV}$.
With the parametrization eq.~(\ref{SO(10)}) and barring
very special choices of parameters where
the RH neutrino masses can become degenerate \cite{afs}
\footnote{As in \cite{SO10}, we consider only solutions where
$M_3/M_2$ and $M_2/M_1 > 10$. This is clearly a conservative condition,
since the asymmetry gets enhanced when $M_2\simeq M_3$ or $M_2\simeq M_1$. However, in this
way, we only neglect very special points in the parameter space yielding
$M_3/M_2$ and $M_2/M_1 < 10$. We will comment again later on this point.},
the RH neutrino mass spectrum is hierarchical and
of the form (for generic expressions in terms of the low energy parameters,
see Ref. \cite{afs})
\be
\label{alpha}
M_1\,:\,M_2\,:\,M_3=(\alpha_1\,m_u)^2\,:\,(\alpha_2\,m_c)^2\,:\,(\alpha_3\;m_t)^2\, .
\ee
As we said, the values of $\alpha_1$ and $\alpha_3$
are actually irrelevant for the determination of the final asymmetry
(unless $\alpha_1$ is unrealistically large to push $M_1$ from $\sim 10^5\,{\rm GeV}$
above the lower bound $\sim  10^{9}$ GeV to achieve successful $N_1$ leptogenesis).
On the other hand, the value of $\alpha_2$ is  relevant
to set the scale of the mass $M_2\simeq 2(\alpha_2\,m_c)^2/m_3$ (valid for $\theta_{13}\simeq 0$)
of the next-to-lightest RH neutrino mass, but it does not alter
other quantities crucial for thermal leptogenesis,
such as the amount of wash-out from the lightest RH neutrinos.

Defining the flavoured $C\!P$ asymmetries as
\be
\ve_{2\a}\equiv -{\G_{2\alpha}-\overline{\G}_{2\alpha}
\over \G_{2}+\overline{\G}_{2}} \, ,
\ee
these can be calculated using \cite{crv}
\be\label{eps2a}
\ve_{2\a} \simeq
\frac{3}{16 \p (h^{\dag}h)_{22}}  \left\{ {\rm Im}\left[h_{\a 2}^{\star}
h_{\a 3}(h^{\dag}h)_{2 3}\right] \frac{\x(x_3/x_2)}{\sqrt{x_3/x_2}}+
\frac{2}{3(x_3/x_2-1)}{\rm Im}
\left[h_{\a 2}^{\star}h_{\a 3}(h^{\dag}h)_{3 2}\right]\right\}\, ,
\ee
where
\be
\xi(x)=\frac{2}{3}x\left[(1+x)\ln\left(\frac{1+x}{x}\right)-\frac{2-x}{1-x}\right]\,
\ee
and $\G_{2\alpha}$ is the decay rate of the RH neutrino $N_2$ into the flavor $\alpha$ with couplings
given by the Yukawa's matrix $h$.
We will assume an initial vanishing $N_2$-abundance instead of
an initial thermal abundance as in \cite{SO10}.
In this way, a comparison of the results in the two analyses
gives a useful information about the dependence of the final asymmetry on the initial $N_2$ abundance
when successful leptogenesis is imposed.

Let us now define the flavored decay parameters as
\be
K_{i \alpha} = \frac{\Gamma_{i\alpha}+\overline{\G}_{i\alpha}}{H(T=M_i)}=
                      \frac{\left|(m_D)_{\alpha i}\right|^2}{m_\star\, M_i}\, ,
\ee
where $H$ is the Hubble rate,
\be
m_\star=\frac{16\,\pi^{5/2}\sqrt{g_*}}{3\sqrt{5}}\frac{v^2}{M_{\rm Pl}}
\simeq 1.08\times 10^{-3}\,{\rm eV} \, ,
\ee
$g_*$ is the number of the effective relativistic degrees of freedom and $M_{\rm Pl}$ is the Planck mass.
The total decay parameters  are then just
simply given by $K_i = \sum_{\alpha}\,K_{i \alpha}$. It is also
convenient to introduce the quantities $P^0_{2\alpha}=K_{2\alpha}/K_2$.

From the decay parameters one can then calculate the
efficiency factors that are the second needed ingredient, together with the
$C\!P$ asymmetries, for the calculation of the final asymmetry. These can be well approximated
by the following analytical expression \cite{flavorlep}
\footnote{It is in quite a good agreement with the numerical results shown in \cite{akr}.
The maximum difference is  $\sim 30\%$ at the peak for $K_{2\a}\sim 1$.
For $K_{2\a}\gg 1$, the difference is below $10\%$.}
\be
\k(K_2,K_{2\a})
=\k_{-}^{\rm f}(K_2,K_{2\a})+
 \k_{+}^{\rm f}(K_2,K_{2\a}) \, ,
\ee
where the negative and the positive contributions are respectively
approximately given  by
\be\label{k-}
\k_{-}^{\rm f}(K_2,K_{2\a})\simeq
-{2\over P_{2\a}^{0}}\ e^{-{3\,\pi\,K_{2\a} \over 8}}
\left(e^{{P_{2\a}^{0}\over 2}\,N_{N_2}(z_{\rm eq})} - 1 \right) \, ,
\ee
\begin{equation}\label{nka}
N_{N_2}(z_2^{\rm eq}) \simeq \overline{N}(K)\equiv
{N(K_2)\over\left(1 + \sqrt{N(K_2)}\right)^2} \, ,
\end{equation}
and
\be\label{k+}
\k_{+}^{\rm f}(K_2,K_{2\a})\simeq
{2\over z_B(K_{2\a})\,K_{2\a}}
\left(1-e^{-{K_{2\a}\,z_B(K_{2\a})\,N_{N_2}(z_{\rm eq})\over 2}}\right) \, ,
\ee
where
\be
z_{B}(K_{2\a}) \simeq 2+4\,K_{2\a}^{0.13}\,e^{-{2.5\over K_{2\a}}}={\cal O}(1\div 10) \, .
\ee
The $SO(10)$-inspired conditions $\a_i={\cal O}(1)$, yield a RH neutrino mass spectrum
with $M_1\ll 10^9\,{\rm GeV}\lesssim M_2 \lesssim 10^{12}\,{\rm GeV}\ll M_3$,
though, as we already noticed, this spectrum is obtained for a broader range of $\a_i$ values.
In this situation, the asymmetry is dominantly produced from
$N_2$ decays at $T\sim M_2$  in a two flavour regime, i.e. when final lepton states
can be described as an incoherent mixture of a tauon component and of
coherent superposition of a an electron and a muon component. Therefore, at the freeze-out
of the $N_2$ wash-out processes, the produced asymmetry can be calculated as
the sum of two contributions,
\be\label{NBmLTM2}
N_{B-L}^{T\sim M_2} \simeq
\ve_{2\tau}\,\kappa(K_2,K_{2\tau})+ \ve_{2e+\mu}\,\kappa(K_2,K_{2e+\mu}) \, ,
\ee
where $\ve_{2e+\mu}$ stands for $\ve_{2e+\mu}=\ve_{2e}+\ve_{2\mu}$
and  $K_{2e+\mu}=K_{2e}+K_{2\m}$.

More precisely, notice that each flavour contribution to the asymmetry  is produced in
an interval of temperatures  between $M_2/[z_B(K_{2\a})-2]$ and
$M_2/[z_B(K_{2\a})+2]$, with $\a=\tau, e+\m$.

At $T\lesssim 10^9\,{\rm GeV}$ the coherence of the $e+\m$ quantum states breaks down
and a three flavour regime holds, with the lepton quantum states given by an incoherent
mixture of $e$, $\m$ and $\t$ flavours. The asymmetry has then to be calculated at the
$N_1$ wash-out stage as a sum of three flavoured contributions.

The assumption of an initial vanishing $N_2$-abundance allows to neglect the
phantom terms in the muon and in the electron components \cite{flcoupling}
so that the final asymmetry can be calculated using the expression
\be
N_{B-L}^{\rm f} \simeq
{P^0_{2e}\over P^0_{2e+\m}}\,\ve_{2e+\m}\,\kappa(K_{2 e+\mu})\, e^{-{3\pi\over 8}\,K_{1 e}}+
{P^0_{2\mu}\over P^0_{2e+\m}}\,\ve_{2e+\m}\,\kappa(K_{2 e+\mu})\, e^{-{3\pi\over 8}\,K_{1 \mu}}+
\ve_{2 \tau}\,\kappa(K_{2 \tau})\,e^{-{3\pi\over 8}\,K_{1 \tau}} \, .
\ee

Notice that successful leptogenesis relies on points in the parameter space
where one out of the three $K_{1\a}\lesssim 1$. From this point of view
the constraints on low  energy neutrino experiments that we will obtain should
be quite stable against effects
that could enhance the asymmetry such as a resonant enhancement for special
points where $(M_3-M_2)/M_2 \ll 1$. Such effects are however
still able to relax the lower bound on $M_2$ and on the $T_{\rm RH}$, since the
$K_{1\a}$'s do not depend on $M_2$.

%%%%%%%%%%%%%%%%%%%%%%%%%%
\section{The case $V_L=I$}
%%%%%%%%%%%%%%%%%%%%%%%%%%

We start from the case $V_L=I$ that has been studied already in \cite{SO10}
deriving constraints in the plane $m_1-\theta_{13}$ for NO.
Here we show constraints on all low
energy neutrino parameters, including the case of IO.

\subsection{Normal ordering}

Let us first discuss the case of NO.
In  Fig.~1 we plotted the final asymmetry $\eta_B$ for the same three
sets of values of the involved parameters as in the Fig. 4 of Ref. \cite{SO10}, where these
three choices were corresponding to three different kinds of solutions for successful leptogenesis.
\begin{figure}
\begin{center}
\psfig{file=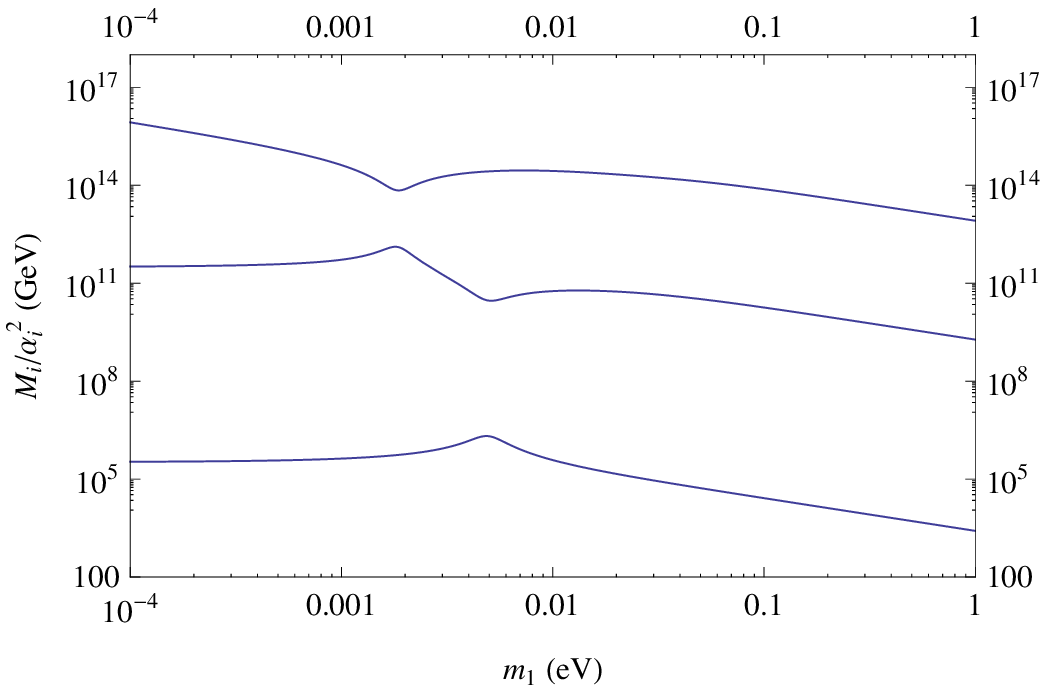,height=38mm,width=45mm}
\hspace{3mm}
\psfig{file=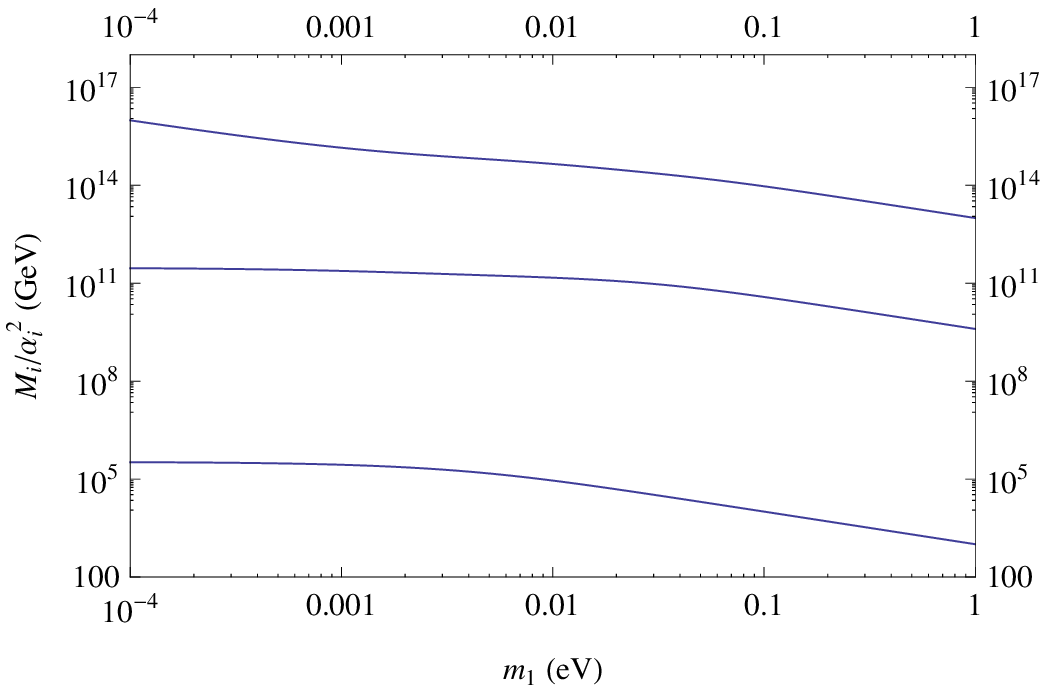,height=38mm,width=45mm}
\hspace{3mm}
\psfig{file=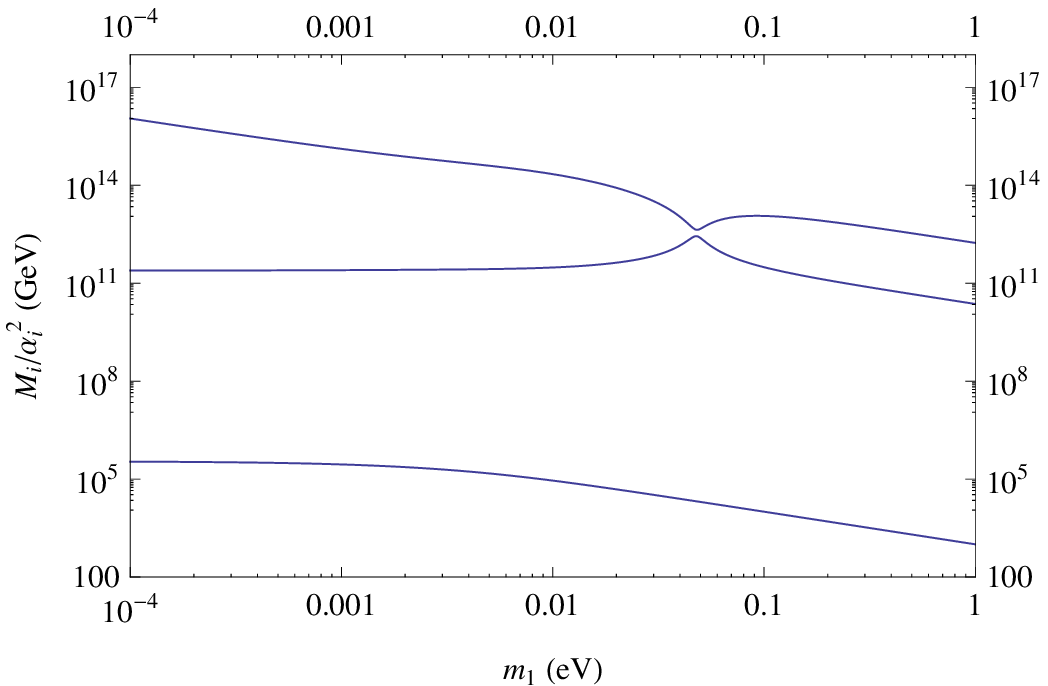,height=38mm,width=45mm} \\
\psfig{file=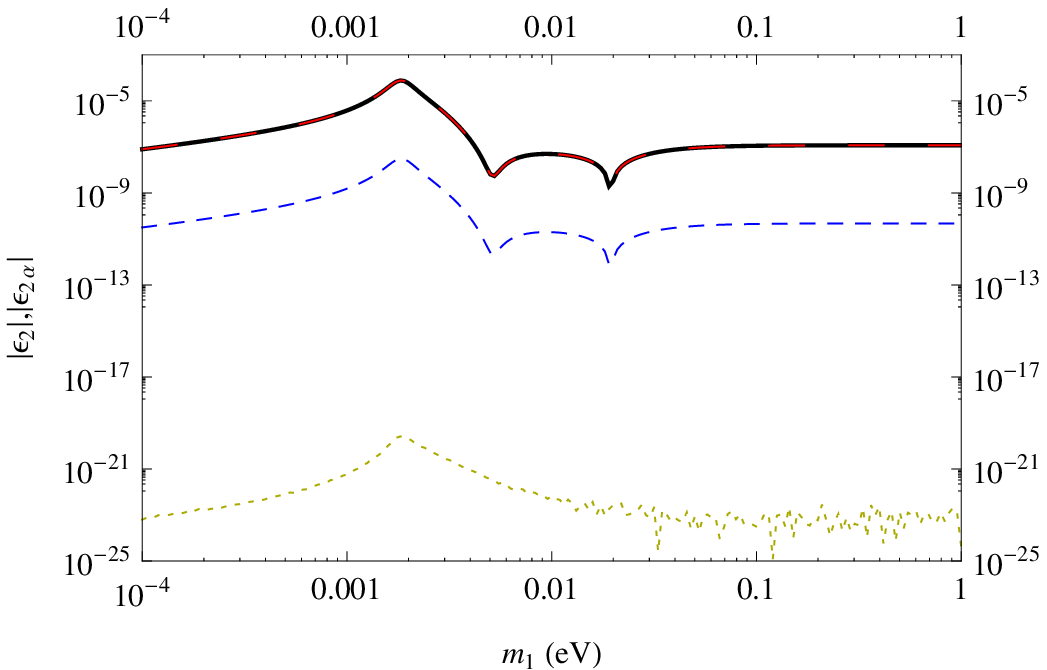,height=38mm,width=45mm}
\hspace{3mm}
\psfig{file=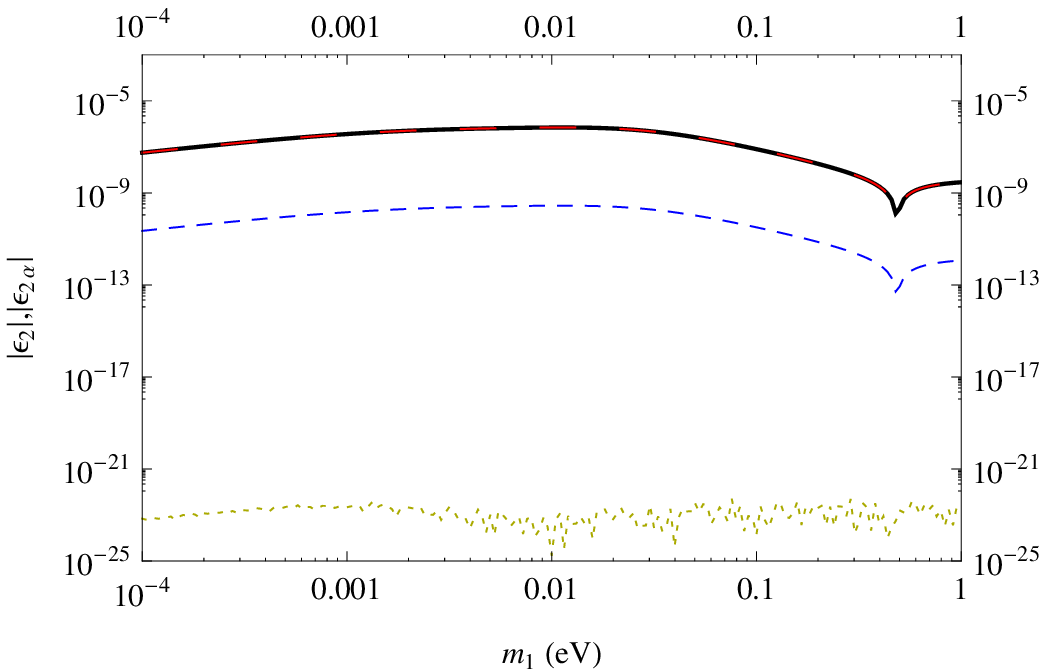,height=38mm,width=45mm}
\hspace{3mm}
\psfig{file=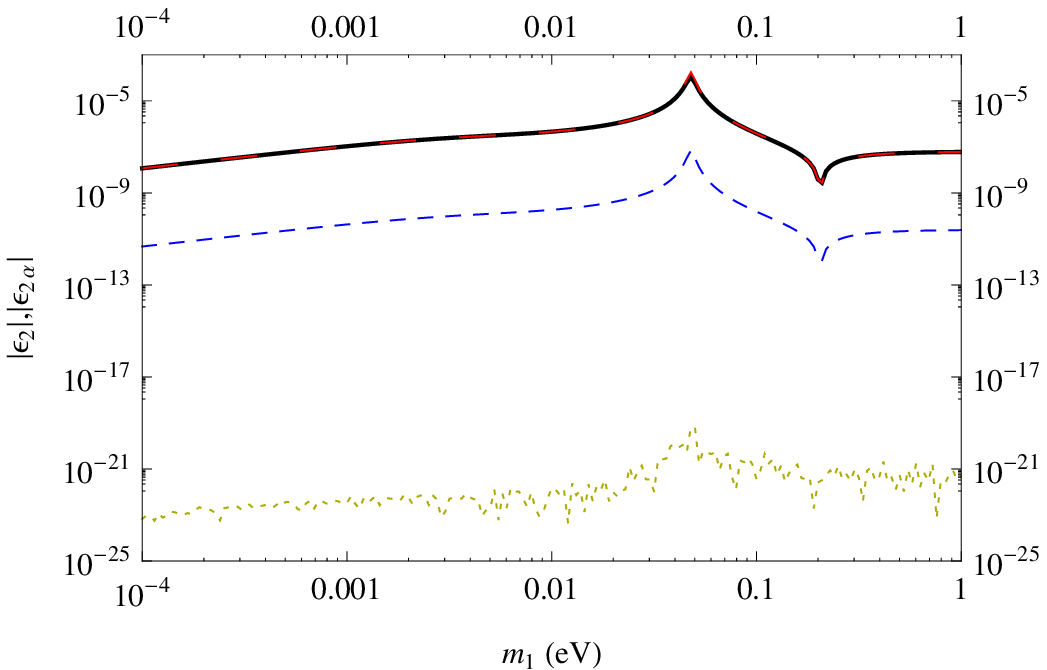,height=38mm,width=45mm} \\
\psfig{file=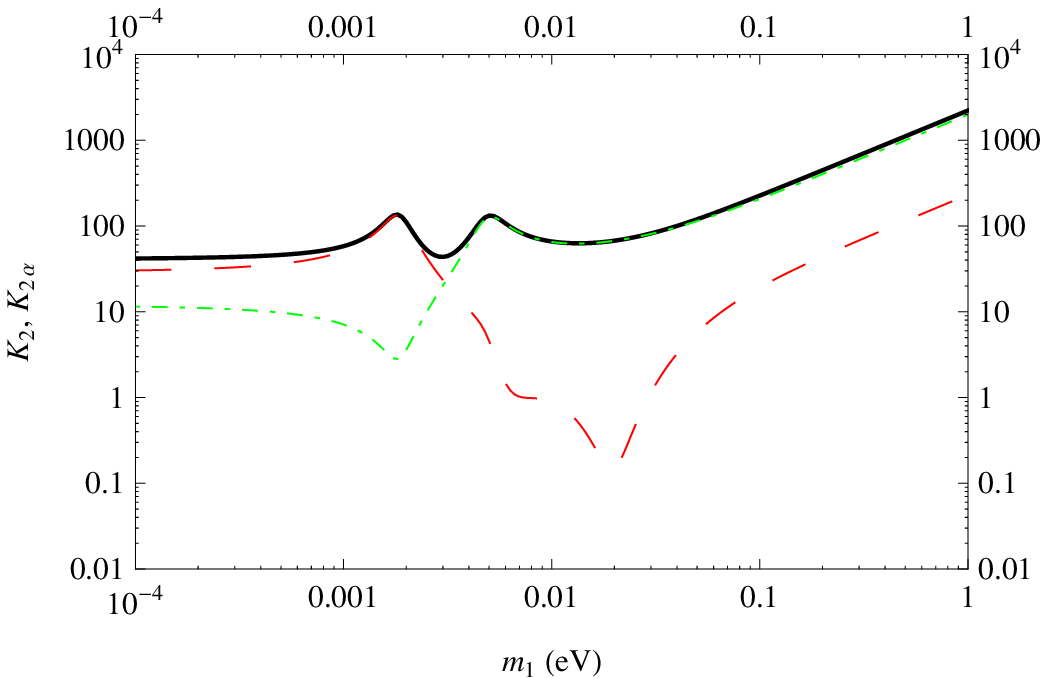,height=38mm,width=45mm}
\hspace{3mm}
\psfig{file=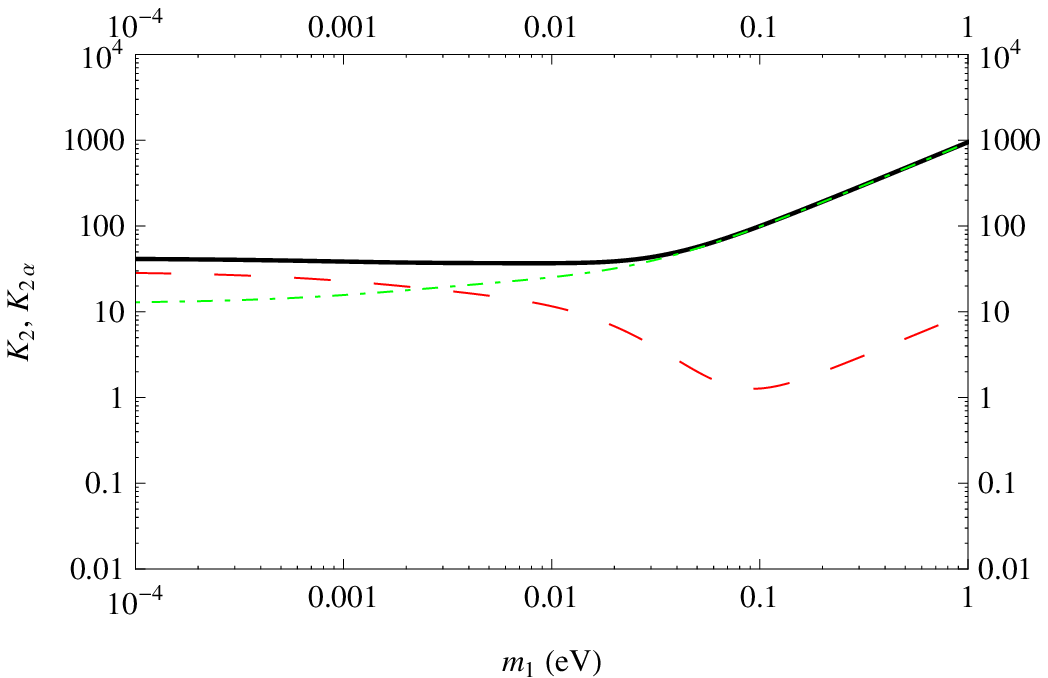,height=38mm,width=45mm}
\hspace{3mm}
\psfig{file=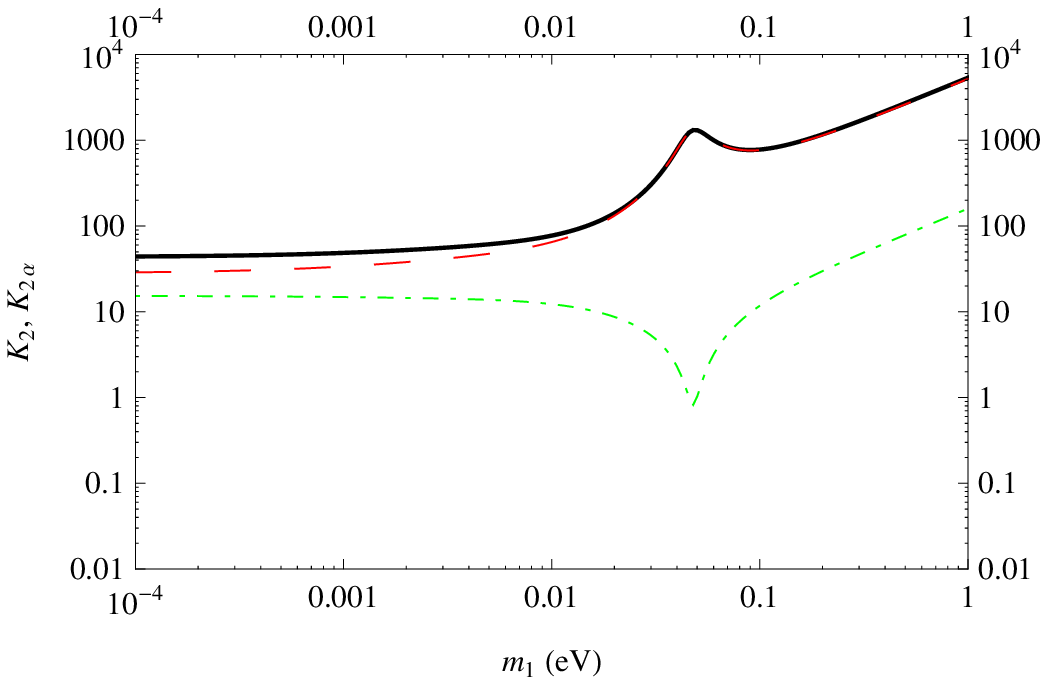,height=38mm,width=45mm} \\
\psfig{file=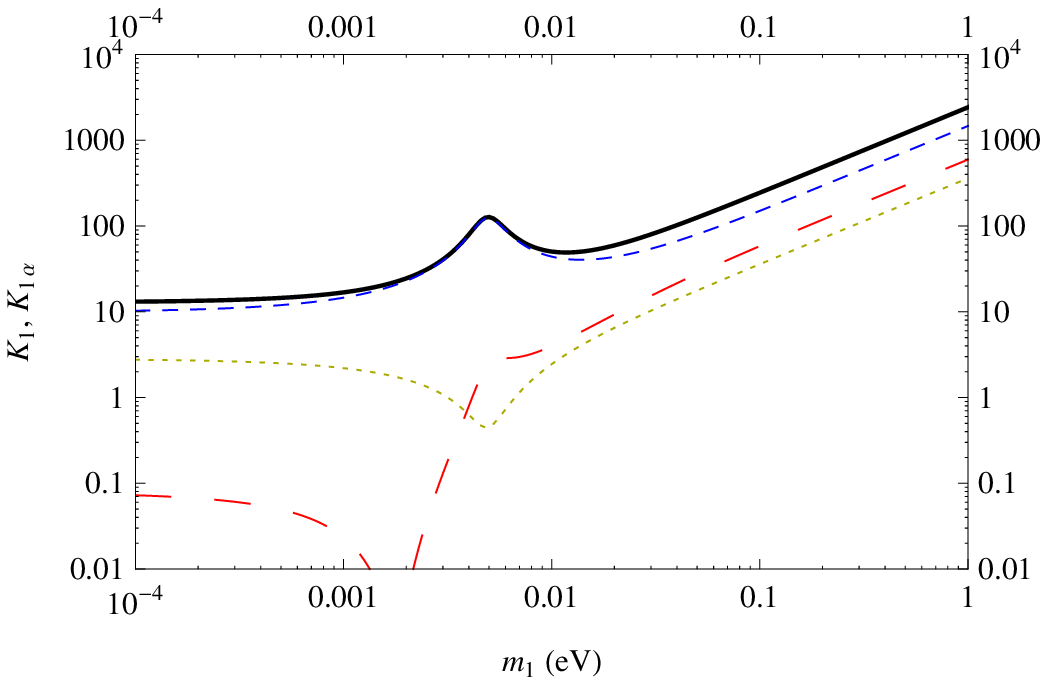,height=38mm,width=45mm}
\hspace{3mm}
\psfig{file=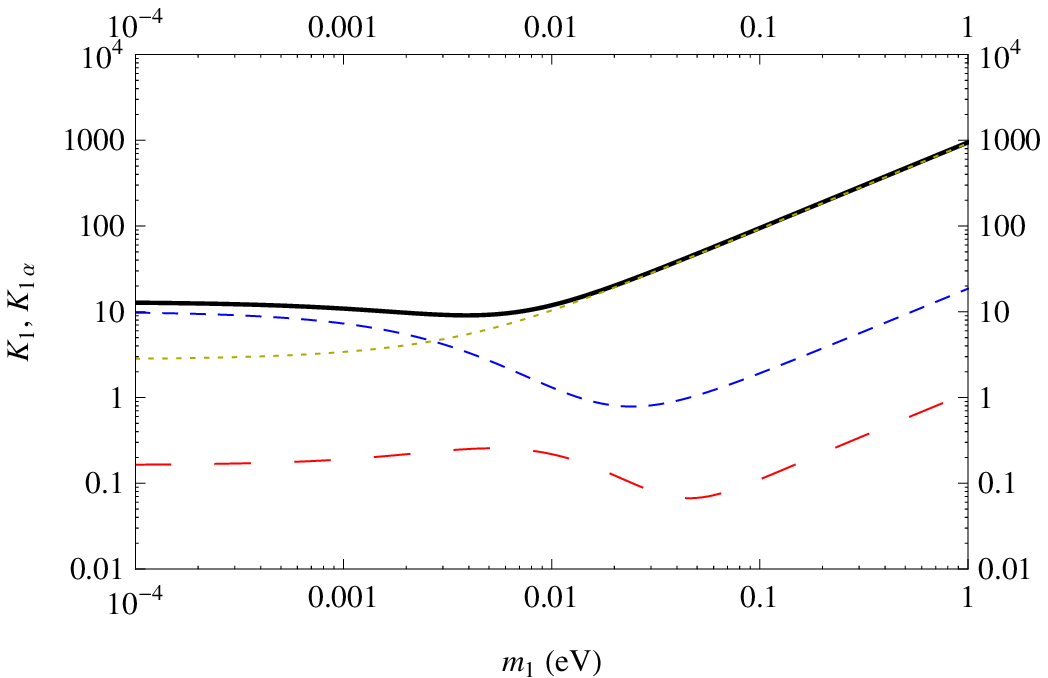,height=38mm,width=45mm}
\hspace{3mm}
\psfig{file=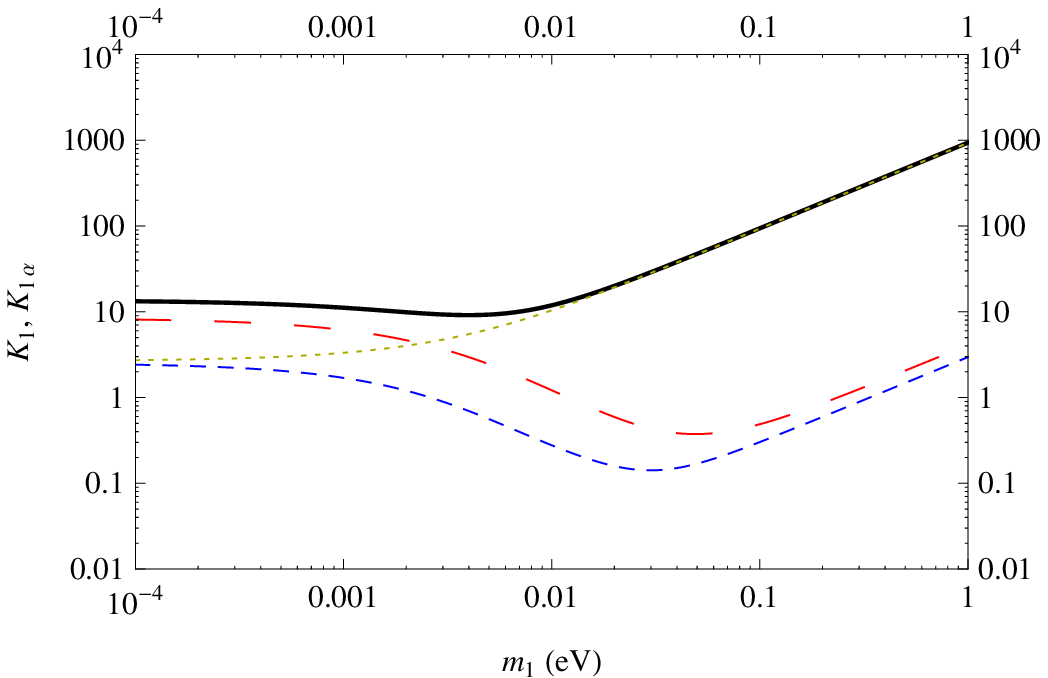,height=38mm,width=45mm} \\
\psfig{file=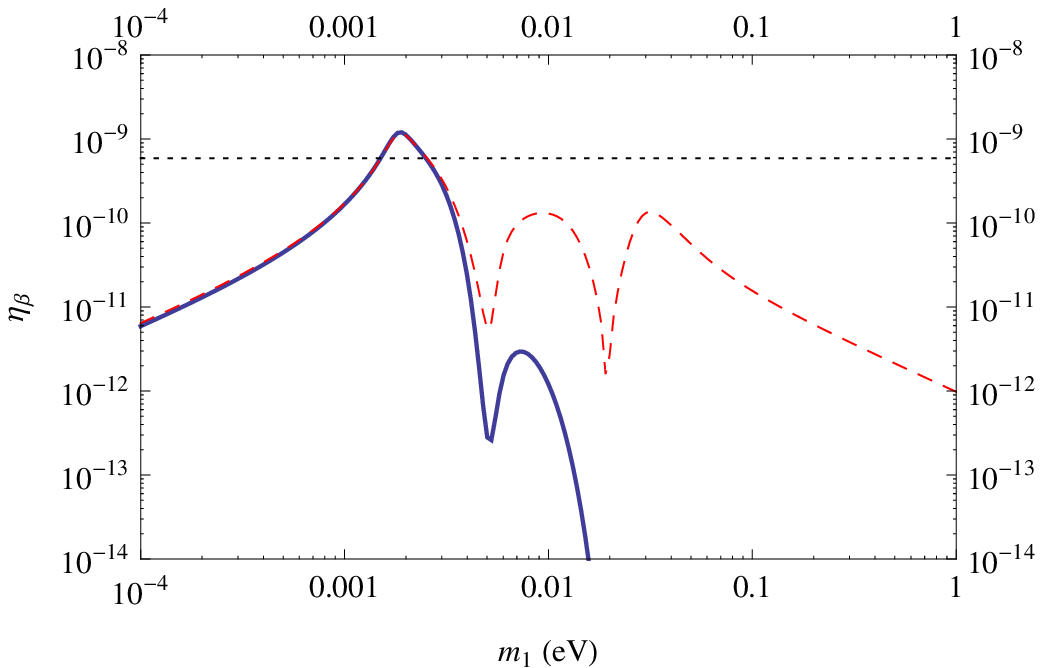,height=38mm,width=45mm}
\hspace{3mm}
\psfig{file=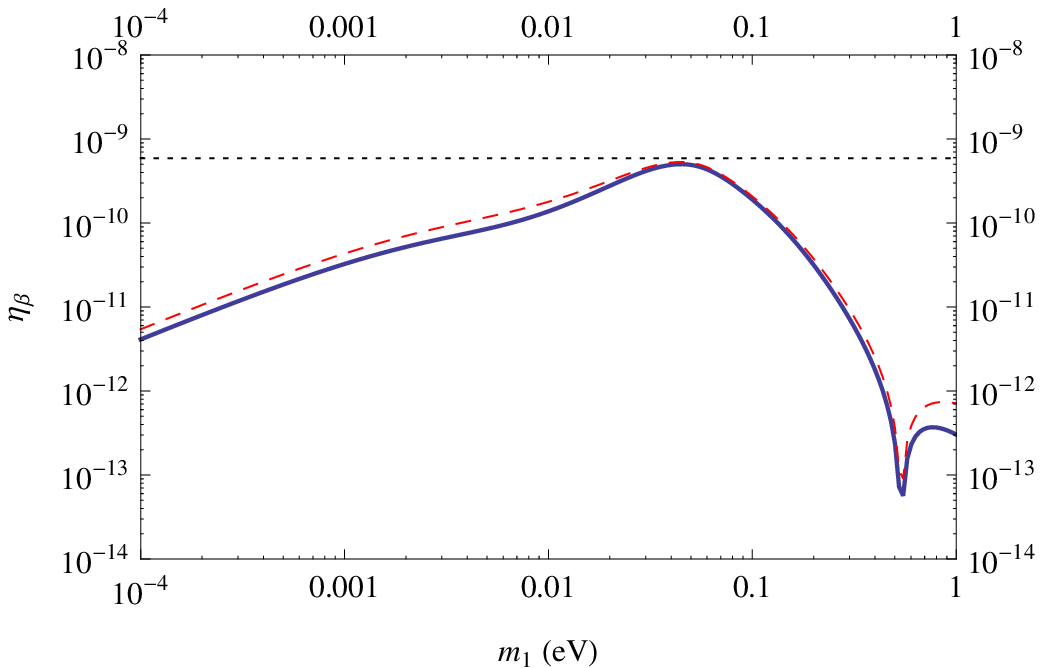,height=38mm,width=45mm}
\hspace{3mm}
\psfig{file=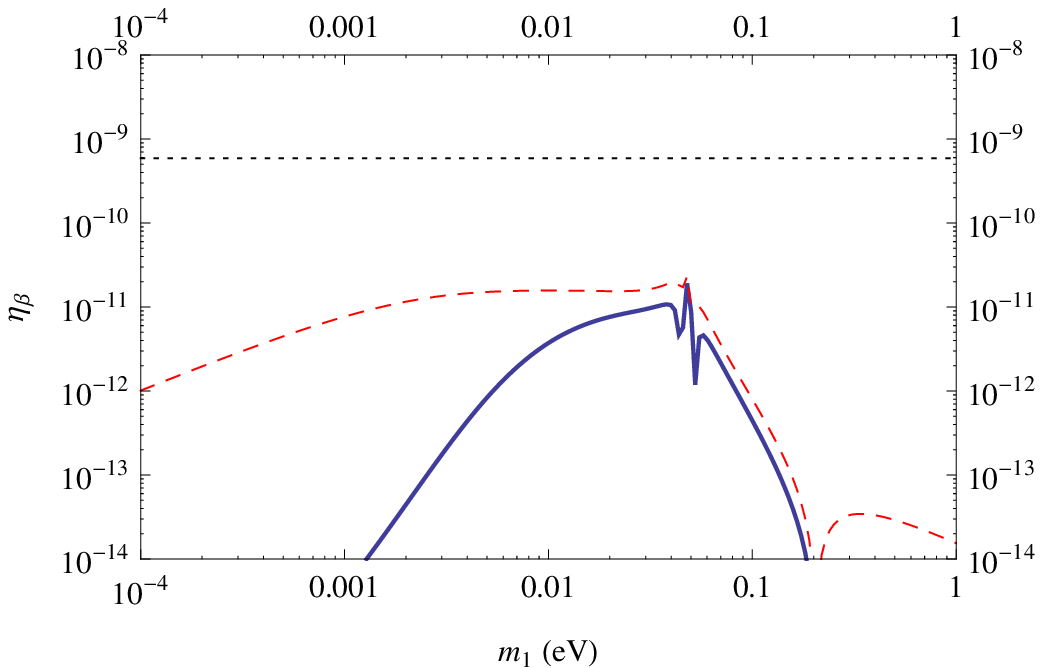,height=38mm,width=45mm}
\end{center}
\caption{Case $V_L=I$, NO. Plots of the relevant quantities for three
choices of the involved parameters as in Fig.~4 of Ref. \cite{SO10}:
$\theta_{13}=5^{\circ},\theta_{23}=40^{\circ}$ $\theta_{12}=33.5^{\circ}$
in all three cases. The values of the phases are different in the three panels (radiants):
$\delta=\sigma=0, \rho=1.5$ (left); $\delta=5.86,
\rho=\sigma=3$ (center); $\delta=\pi/3,\rho=0.02,\sigma=\pi/2$ (right).
The long-dashed red lines correspond to $\a=\t$,
the dashed blue lines to $\a=\m$ and the short-dashed dark yellow lines to $\a=e$.
In the bottom panels the horizontal dotted line is the
$2\sigma$ lowest value $\eta_B^{CMB}=5.9\times 10^{-10}$ (cf. (\ref{etaBobs})),
the solid line is $\eta_B^{\rm f}$ while the dashed line is  $\eta_B^{T\sim M_2}$.}
\label{fig:NBmL}
\end{figure}
This time the third solution (right panel),
is suppressed and successful leptogenesis is not attained.
In \cite{SO10}, this  was the only solution
corresponding to a final asymmetry dominantly in the muon flavour instead than in the tauon
flavour (as for the first two).
The suppression that we find now is explained partly because we are adopting a correct determination of the phases
in the $U_R$ matrix (cf. eq.~(\ref{Dphi})) and partly because we are now assuming
an initial vanishing $N_2$-abundance instead than an initial thermal one. We will see
however that, allowing for  $V_L\neq I$, this kind of solution will again yield successful
leptogenesis in some allowed regions of the parameter space,
characterized in particular by large values of $m_1\sim 0.1\,{\rm eV}$.

The solution in the central panel is also partly suppressed and successful leptogenesis
is not attained. However, in a parameter scan, we find that this kind of solution can still give
successful leptogenesis for slightly different values of the parameters than those
indicated in the figure caption. In this case
the difference with respect to the results in \cite{SO10} is explained just in terms
of the different assumption on the initial abundance. This
dependence on the initial conditions is due to the fact
that $K_{2\tau}\sim 1$, i.e. the solution falls in the weak wash-out regime at the production.

Finally, the first solution (left panel) is fully unchanged. It therefore
exhibits a full independence of the initial conditions and this is in agreement
with the fact that it respects all the necessary conditions
for the independence on the initial conditions found in \cite{problem}.
Notice that these conditions also enforce
an efficient wash-out of a possible pre-existing asymmetry.

A scan in the space of parameters confirms that these
three solutions obtained for special sets of values
are actually representative of the three general kinds of
solutions that come out and,
therefore, the drawn conclusions apply in general.

In the panels of Figure 2 we show the
results of such a scan that highlight the allowed regions
in the parameter space projected on different two-parameter planes.
\begin{figure}
\begin{center}
\psfig{file=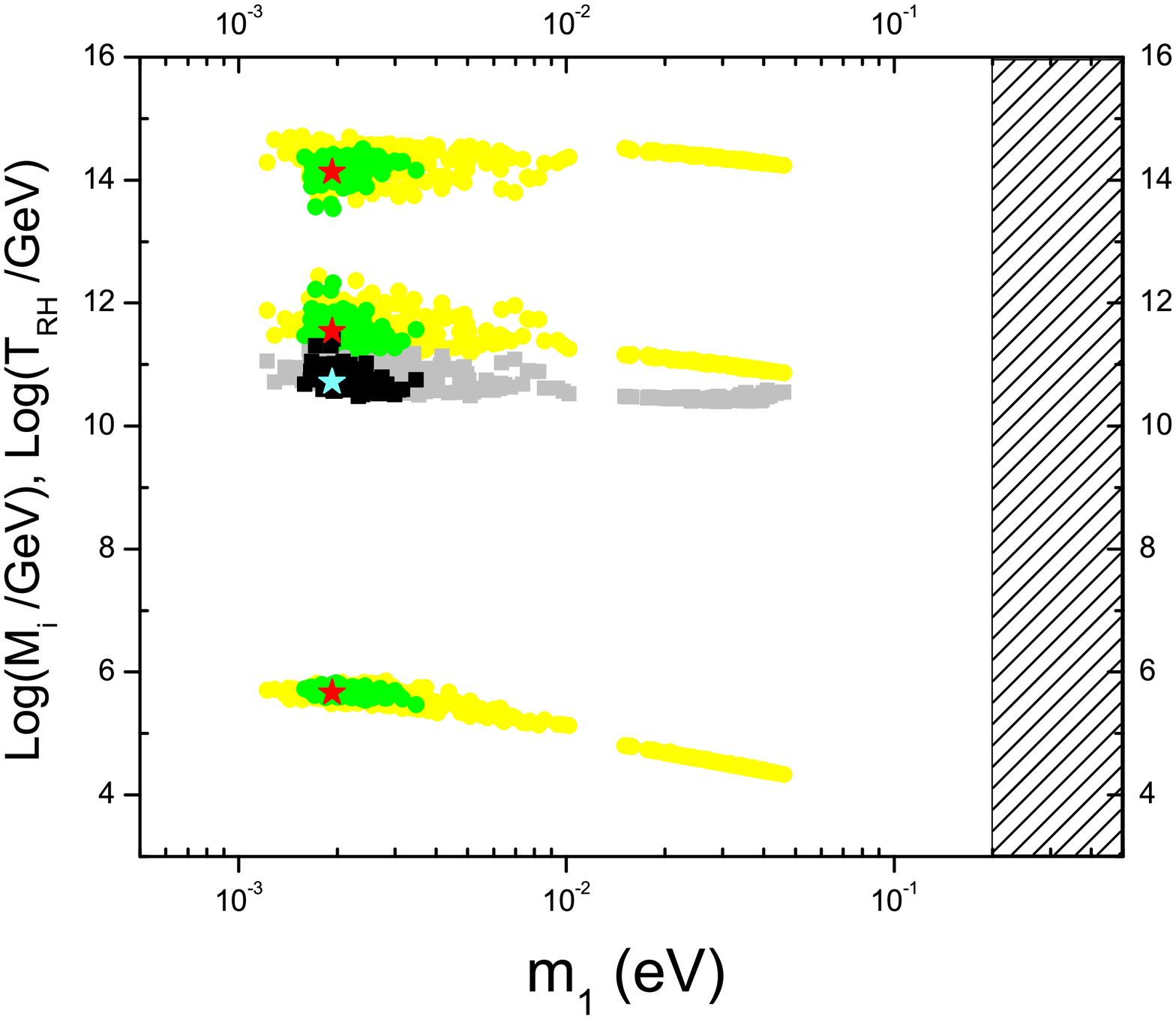,height=48mm,width=54mm}
\hspace{-4mm}
\psfig{file=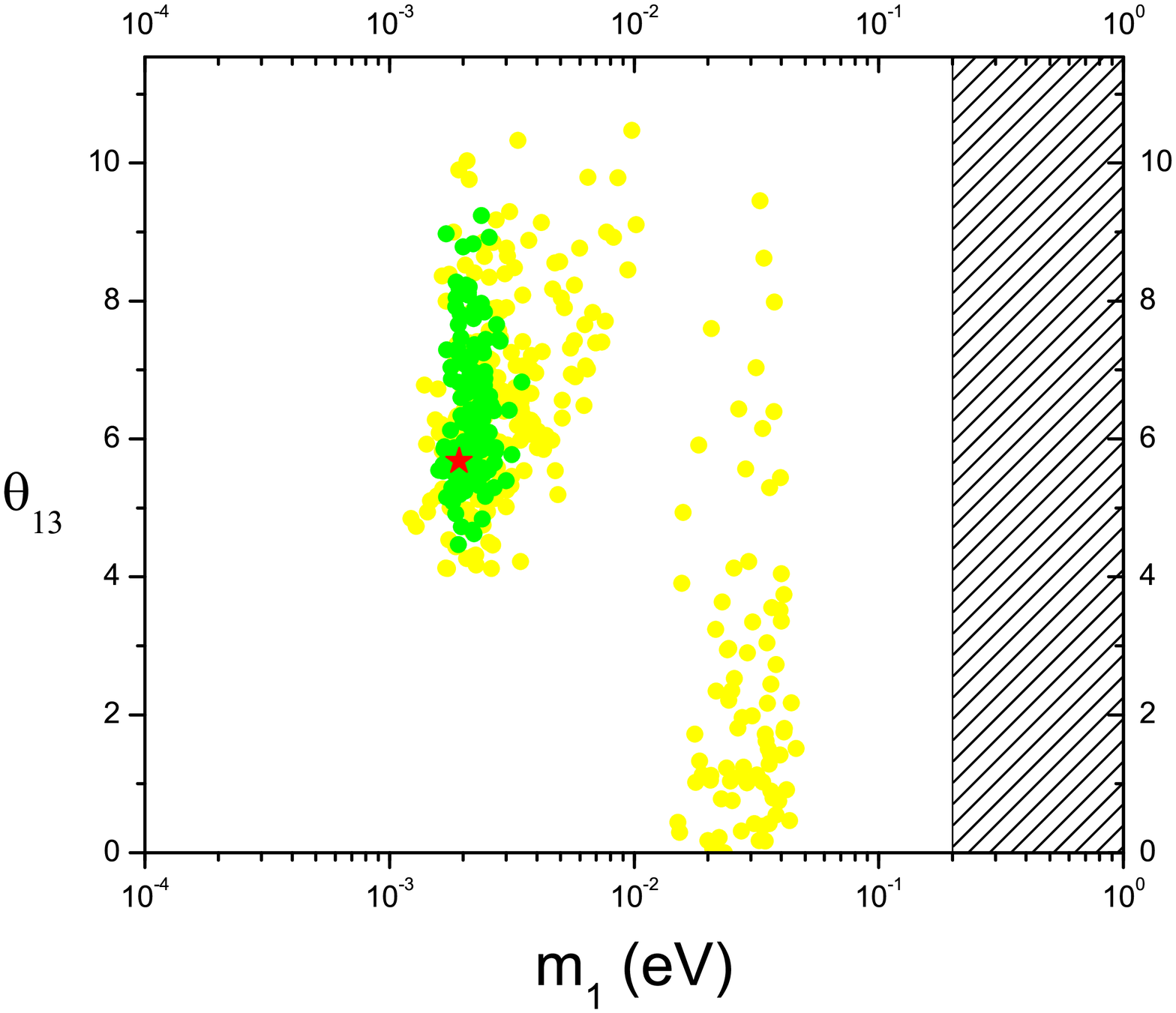,height=48mm,width=54mm}
\hspace{-4mm}
\psfig{file=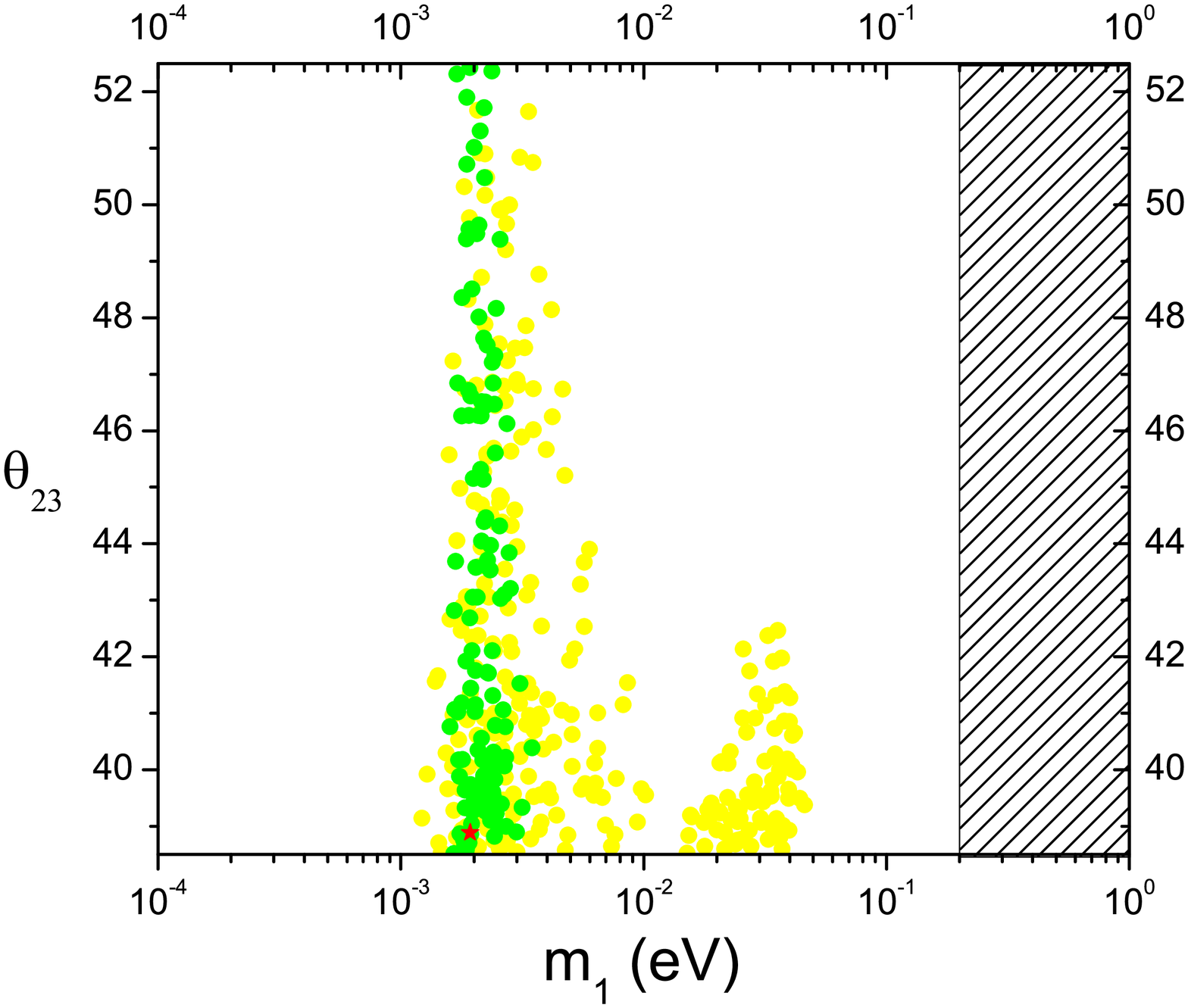,height=48mm,width=54mm} \\
\psfig{file=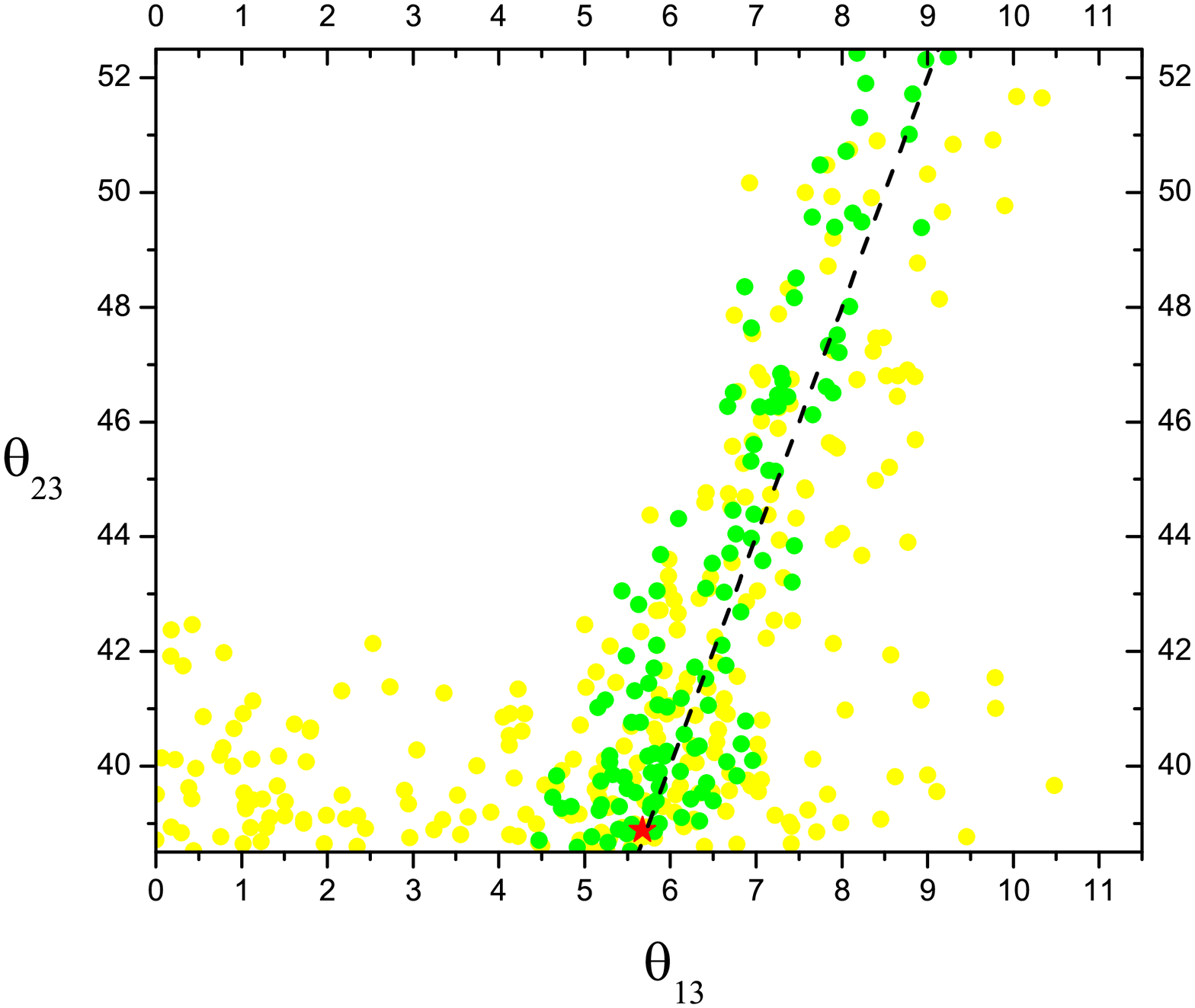,height=48mm,width=54mm}
\hspace{-4mm}
\psfig{file=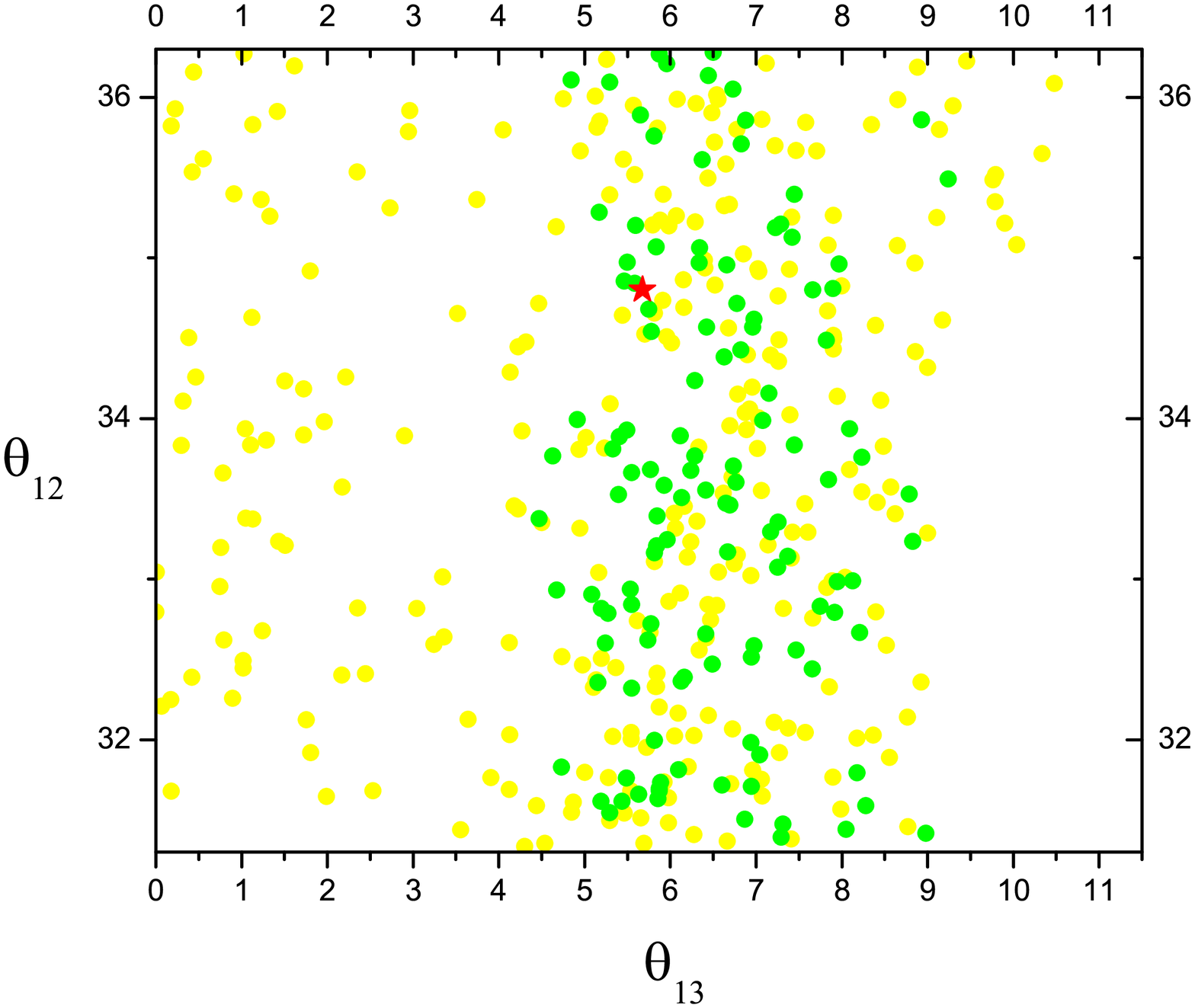,height=48mm,width=54mm}
\hspace{-4mm}
\psfig{file=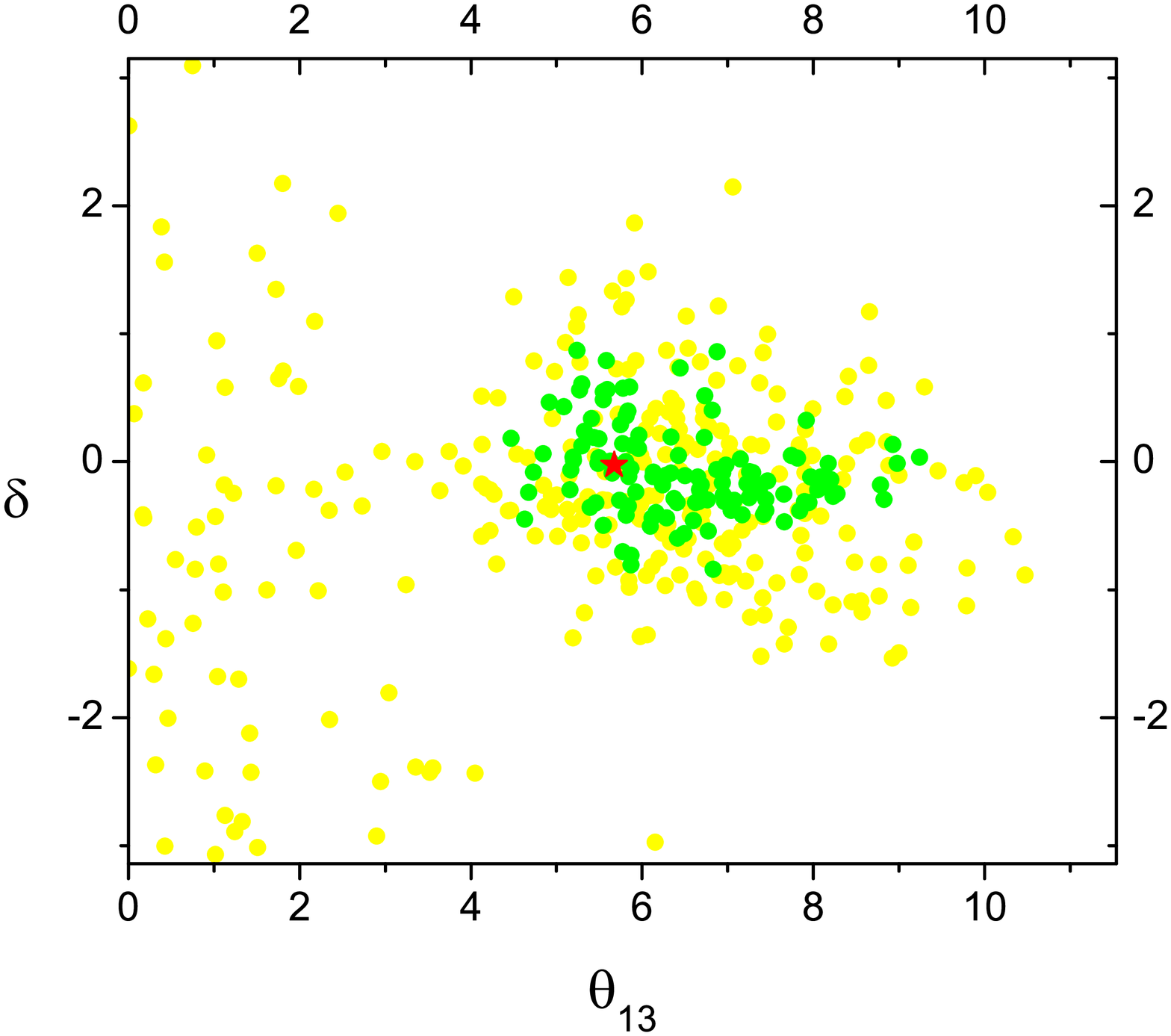,height=48mm,width=54mm} \\
\psfig{file=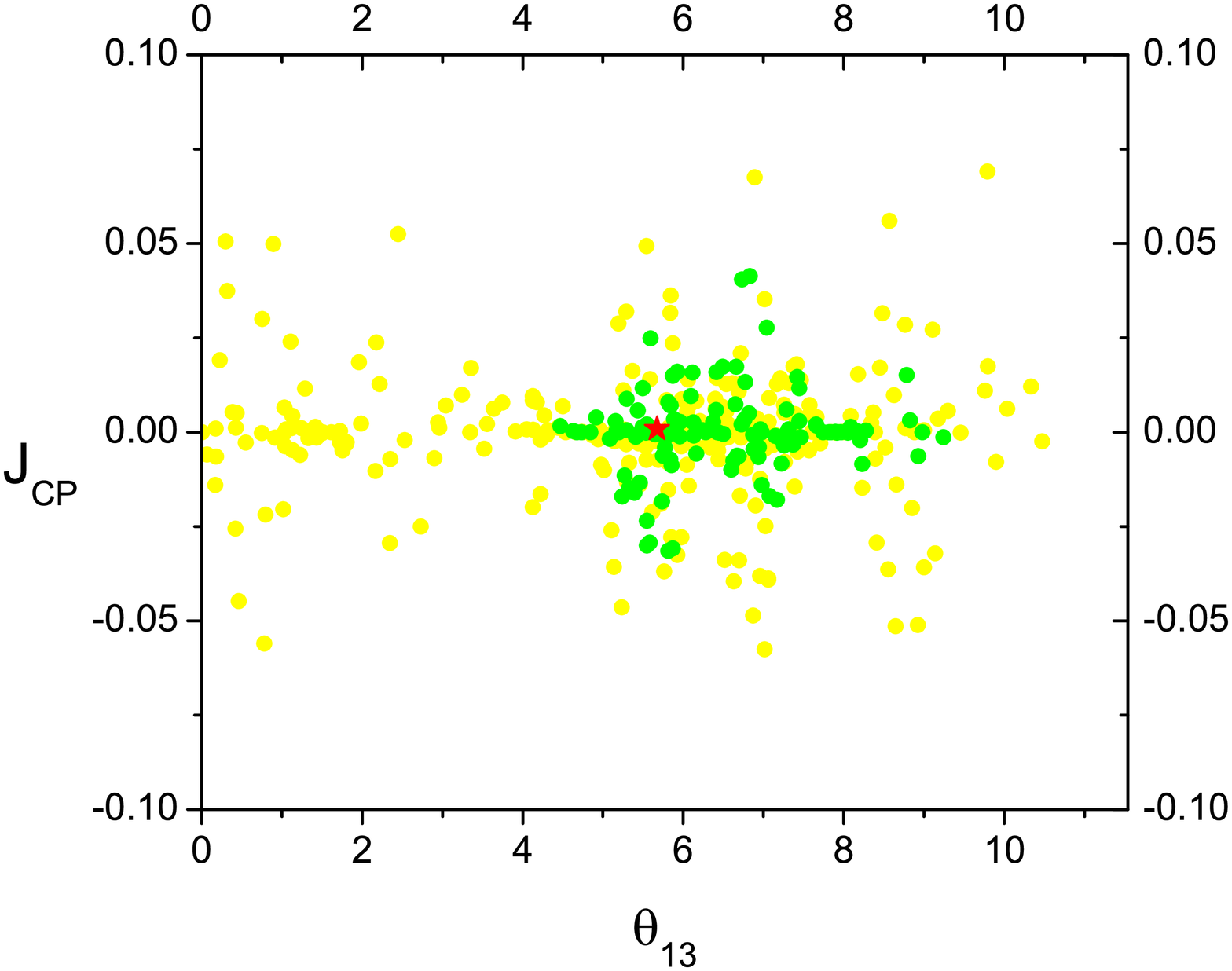,height=48mm,width=54mm}
\hspace{-4mm}
\psfig{file=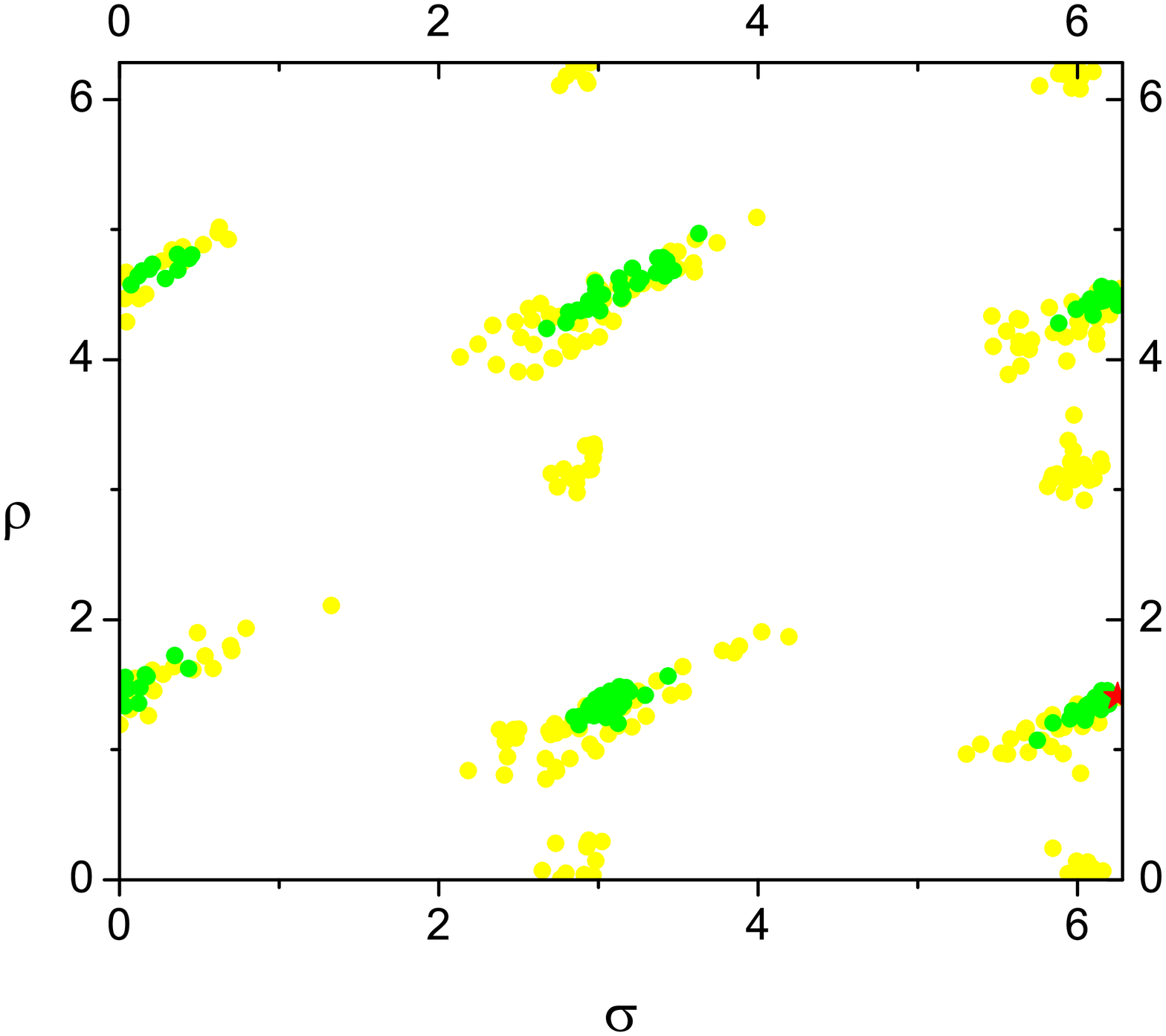,height=48mm,width=54mm}
\hspace{-4mm}
\psfig{file=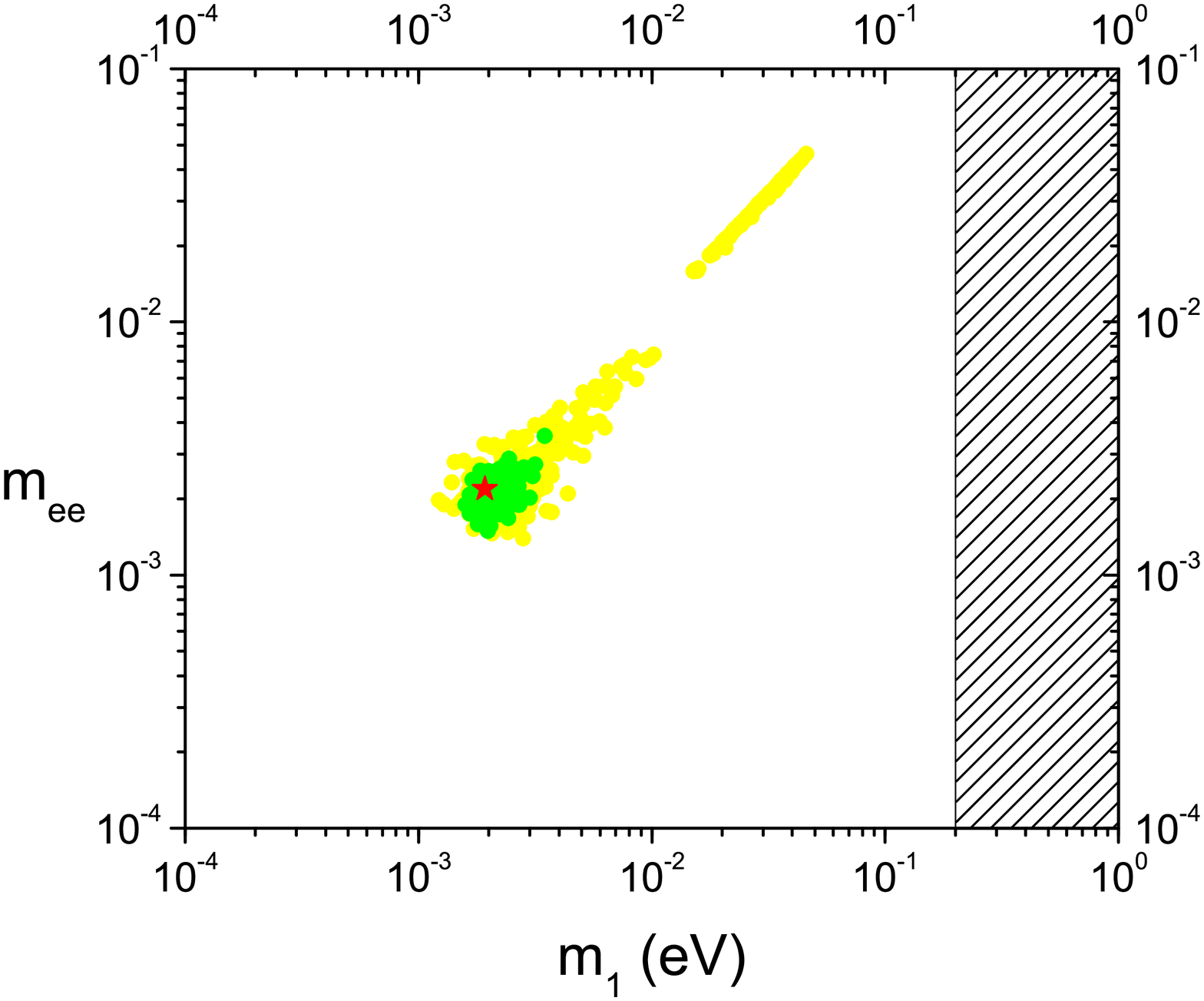,height=48mm,width=54mm}
\end{center}
\caption{Case $V_L=I$, NO. Scatter plot of points in the parameter space that satisfy
the condition $\eta_B>5.9\times 10^{-9}$
for three values of the crucial parameter $\a_2$: $\a_2=5$ (yellow circles); $\a_2=4$ (green circles);
$\a_2=3.4$ (red stars). In the top left panel the
lower bound on $T_{RH}$ (cf. eq.~(\ref{TRHmin})) is also indicated for the same values of $\alpha_2$
but with different symbols: $\a_2=5$ (grey squares), $\a_2=4$ (black squares),
$\a_2=3.4$ (blue star). The three mixing angles are in degrees, the
three phases in radiants. The dashed line in the left central
panel is the eq.~(\ref{linear}).}
\label{constrNO}
\end{figure}
The scatter plots have been obtained scanning the three mixing angles $\theta_{12},\theta_{23}$ and $\theta_{13}$
over the $2\sigma$ ranges eqs.(\ref{twosigma}), the three phases $\delta, \rho, \sigma$ over the ranges $[0,2\pi]$
and the absolute neutrino mass scale for $m_1 < 1 \,{\rm eV}$. These ranges also coincide with those shown in the plots,
except for $m_1$ where the plots are for $m_1>10^{-4}\,{\rm eV}$ simply because no allowed solutions
have been found for lower values.
The shown results have been obtained in two steps. A first scan of ${\cal O}(10^{6})$ points
has yielded a first determination of the allowed regions. With a second scan of additional ${\cal O}(5\times 10^{6})$ points,
restricted to the excluded regions, we have then more robustly and sharply determined the contours of the
allowed regions. Notice that these regions have no statistical significance and the random values of the
parameters have been generated uniformly.

 In the top left panel the three RH neutrino masses are plotted versus $m_1$.
 We have also plotted the lower bound on the reheating temperature
calculated as
\be\label{TRHmin}
T_{RH}^{\rm min}\simeq {M_2\over z_B(K_{2\t})-2} \, .
\ee
This calculation relies on the fact that in the case $V_L=I$ the solutions, as we commented,
always  fall in a tauon $N_2$-dominated scenario. It can be seen that the lowest
bound is given by $T_{RH}^{\rm min}\simeq 2\times 10^{10}\,{\rm GeV}$ that
in a supersymmetric version, if unchanged,  would be marginally reconcilable with the upper bound
from the gravitino problem \cite{gravitino}. This is another  reason to extend
our investigation to cases with $V_L\neq I$ in next sections.

In the top central panel we have then plotted the allowed region in the $m_1-\theta_{13}$
plane that can be compared with an analogous figure in \cite{SO10}. Here, however,
we show only those points that respect the condition $\eta_B>5.9\times 10^{-10}$ but
for 2 different values of $\a_2=4,5$ (in \cite{SO10} we were only showing points for $\a_2=5$).
The (red) star represents a point found for a minimum value $\a_2=3.4$.
This point basically roughly indicates where the maximum of the asymmetry
occurs in the parameter space for a fixed value of $\a_2$. We will continue
to use this convention (yellow circles for $\a_2=5$, green squares for $\a_2=4$
and red stars for minimum found $\a_2$ value) throughout the next figures.
The structure of the allowed region in the $m_1-\theta_{13}$ plane can be understood as follows.
Since $\ve_{2\t}\propto (M_2/M_3)$ and $M_3\propto m^{-1}_1$,
we immediately deduce that a large lepton asymmetry in the tau flavor may be
produced only for sufficiently large values of $m_1$. This is rather easy to understand.
If $m_1$ tends to zero, we go into the so-called decoupling limit, $M_2/M_3\simeq 0$.
As the $C\!P$ asymmetry needs (at least) two heavy states to be generated at the one-loop level,
and disregarding  the contribution from the $N_1$, $\ve_{2\t}$ must vanish.
The wash-out parameter $K_{2\tau}$ is ${\cal O}(25) $ \cite{SO10}
and therefore the final baryon asymmetry may be estimated to be
\be
\eta_B\simeq 5\times 10^{-3}\,\ve_{2\tau}
\simeq 5\, \left(\frac{\alpha^2_2 \,
m_1}{m_3}\right)\cdot 10^{-10}\, ,
\ee
which requires
\be
m_1\gtrsim \,\left(\frac{5}{\alpha_2}\right)^2\,10^{-3}\,{\rm eV}\, ,
\ee
for NO. This estimate holds if the wash-out from the interaction with $N_1$ is negligible,
{\it i.e.} $K_{1\tau}\simlt 1$. Of course, the smaller is $m_1$, the smaller $K_{1\tau}$ needs to be.
For $m_1={\cal O}\left(10^{-3}\right)$ eV, the only possibility is that $K_{1\tau}$ is significantly below unity. Extending the analysis of Ref. \cite{SO10}, one finds
\be
\label{pp}
s_{13}\cos\left(\delta-2\sigma\right)> \frac{m_2{\rm tan}\theta_{23}}{3\sqrt{2} m_3}\simeq 0.04\, .
\ee
To get the feeling of the figures involved, we may set
$\delta\simeq 2\sigma$ and  find that the wash-out
mediated by the $N_1$'s vanishes
for an experimentally allowed value of the mixing between
the first and the third generation of LH neutrinos,
$\theta_{13}> 2.3^\circ$ in agreement with our numerical results. If $m_1$ is larger than ${\cal O}\left(10^{-3}\right)$ eV, then $K_{1\tau}={\cal O}(1)$ is allowed and $\theta_{13}$ can be taken to be vanishing.
Notice also that the lower bound eq.~(\ref{pp}) on $\theta_{13}$ increases with ${\rm tan}\theta_{23}$. This nicely reproduces the linear dependence emerging from the
numerical results in the {\em left column middle panel} for the plane $\theta_{13}-\theta_{23}$
and that is  described, roughly for $\a_2=5$ and more accurately for $\a_2=4$, by
\be\label{linear}
\theta_{23}\simeq 44^{\circ}+4\,(\theta_{13}-7^{\circ}) \, ,
\ee
represented with a dashed line in the panel.
In the {\em top right panel} we show the allowed region in the plane $m_1-\theta_{23}$.

The $C\!P$ non conserving terms in neutrino oscillation probabilities can be
expressed in terms of the {\em Jarlskog invariant} $J_{CP}$ given by \cite{giunti}
\bea
J_{CP} & = & {\rm Im}[U_{\m 3}\,U_{e 2}\,U_{\m 2}^{\star}\,U_{e3}^{\star}] \\
& = & c_{12}\,s_{12}\,c_{23}\,s_{23}\,c^2_{13}\,s_{13}\,\sin\d  ,
\eea
such that
\be
P_{\nu_\a\rightarrow \nu_\b}-P_{\bar{\nu}_\a\rightarrow \bar{\nu}_\b}=
4\,J_{CP}\,\sum_{k>j}\,s_{\a\b;kj}\,\sin\left({\Delta m^2_{kj}\,L\over 2E}\right) \, ,
\ee
where $s_{\a\b;kj}=\pm 1$. In the bottom left panel we show the allowed points
in the plane $J_{CP}-\theta_{13}$. It can be noticed that a non zero value of $J_{CP}$
is not crucial. %Therefore, $C\!P$ violation can also stem from the Majorana phases.
Looking at the bottom-central panel, it is interesting to notice that the allowed regions for the
Majorana phases are centered approximately around  $\sigma=n\,\pi$ and $\rho=(n+1/2)\pi$.

These play a role in the determination of the
{\em effective Majorana mass} of $\nu_e$ in $\b\b 0\n$ decays
that is given by
\bea
m_{ee} & = & \left|\sum_i\,m_i\,U_{ei}^2 \right| \\
         & = & \left| m_1\,c_{12}^2\,c_{13}^2\,e^{2\,i\,\rho}+m_2\,s_{12}^2\,c_{13}^2 + m_3\,s^2_{13}\,e^{2\,i\,(\sigma-\delta)} \right|\, .
\eea
In the bottom-right panel one can see how there is a precise relation between
$m_{ee}$ and $m_1$, given approximately by
$m_{ee}\simeq m_1$. It can be also noticed
that there is quite a strict lower bound
$m_{ee}\gtrsim 1.5\,\times \,10^{-3}\,{\rm eV}$.
Lowest values $m_{ee}\gtrsim 2\times 10^{-3}\,{\rm eV}$
are the most favoured ones in this case. Though current planned experiments will not be
able to test the full allowed range, it is still interesting that they
will test it partially, tightening  the constraints on the other parameters as well.

\begin{figure}
\begin{center}
\psfig{file=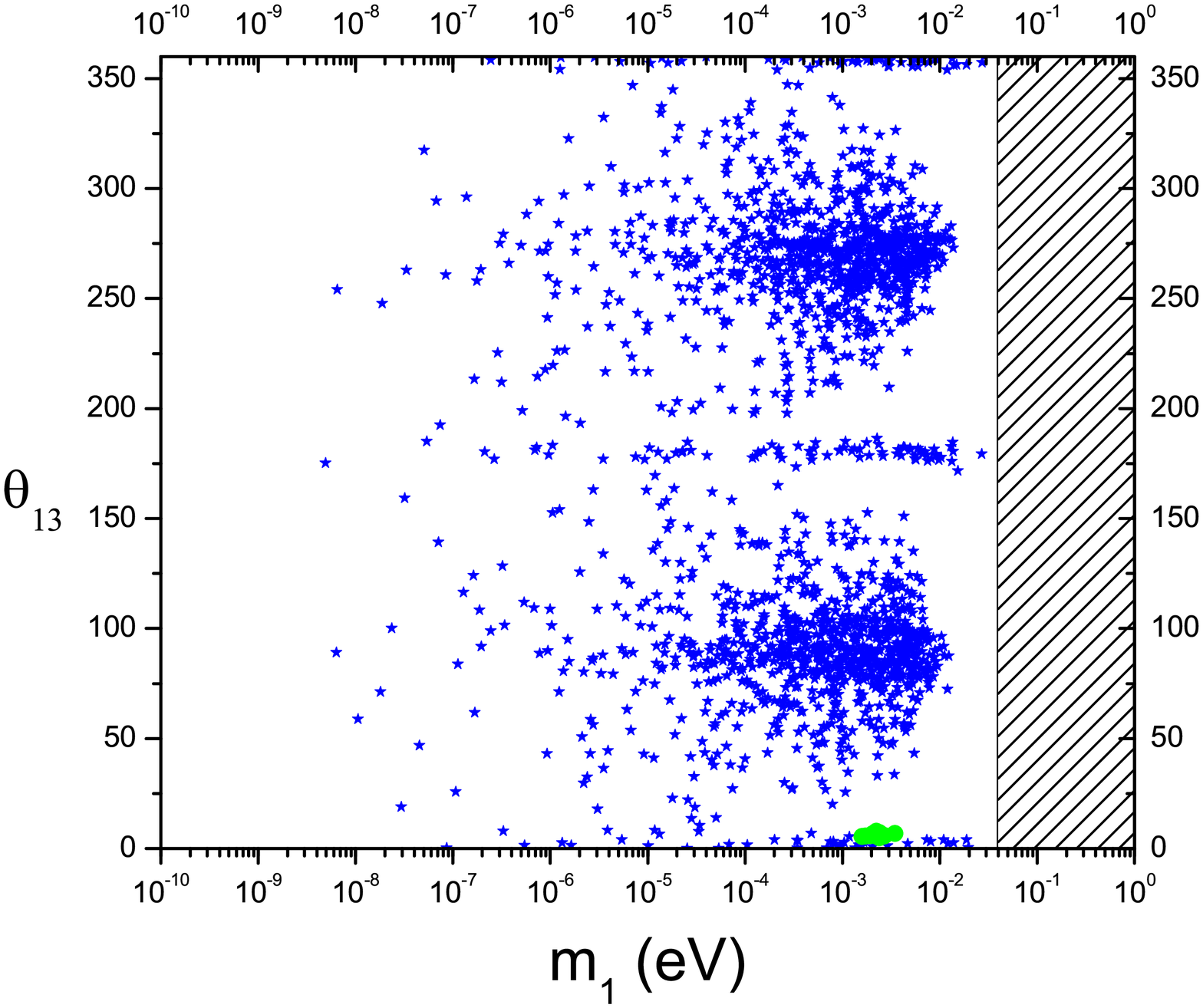,height=48mm,width=54mm}
\hspace{-4mm}
\psfig{file=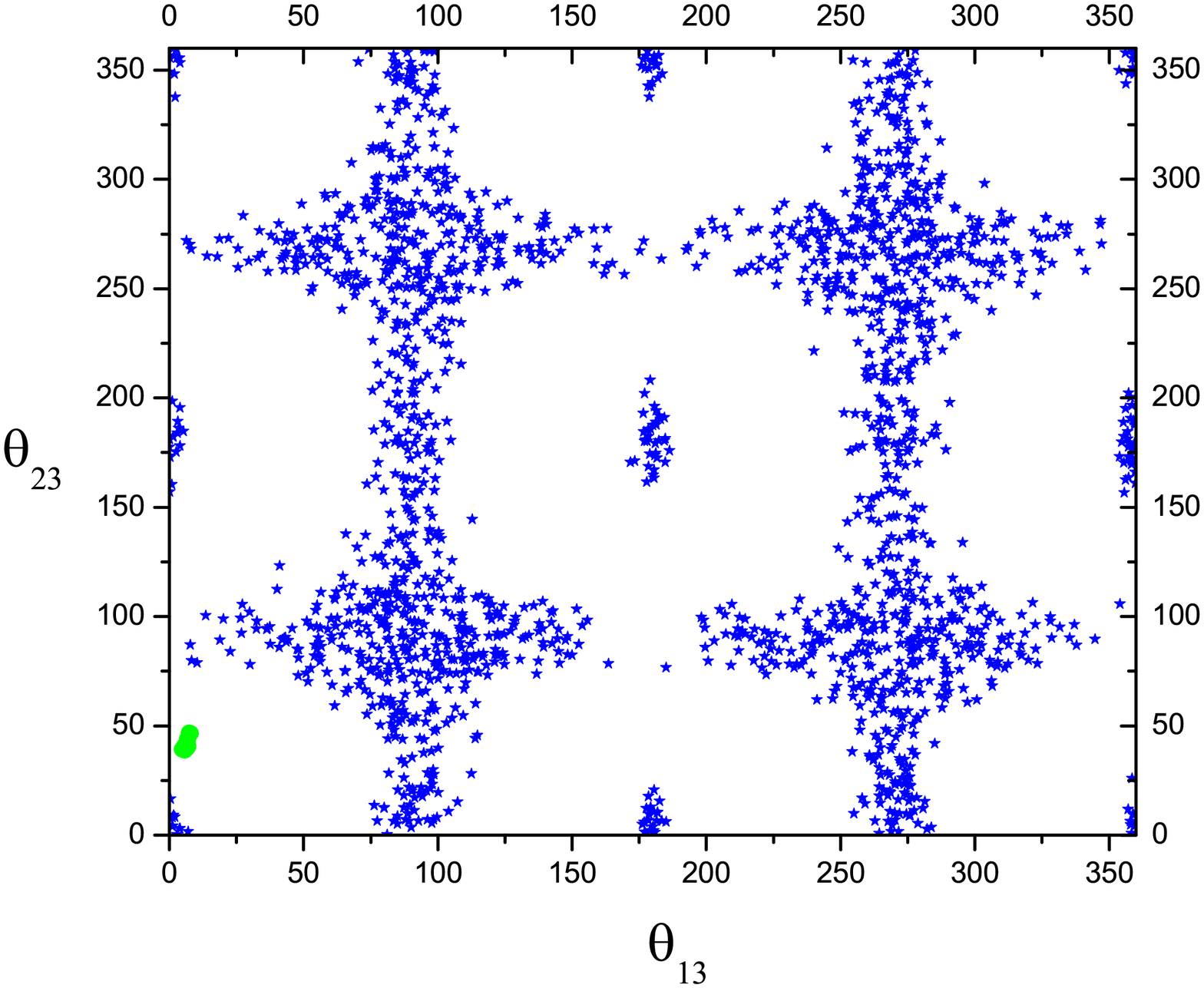,height=48mm,width=54mm}
\hspace{-4mm}
\psfig{file=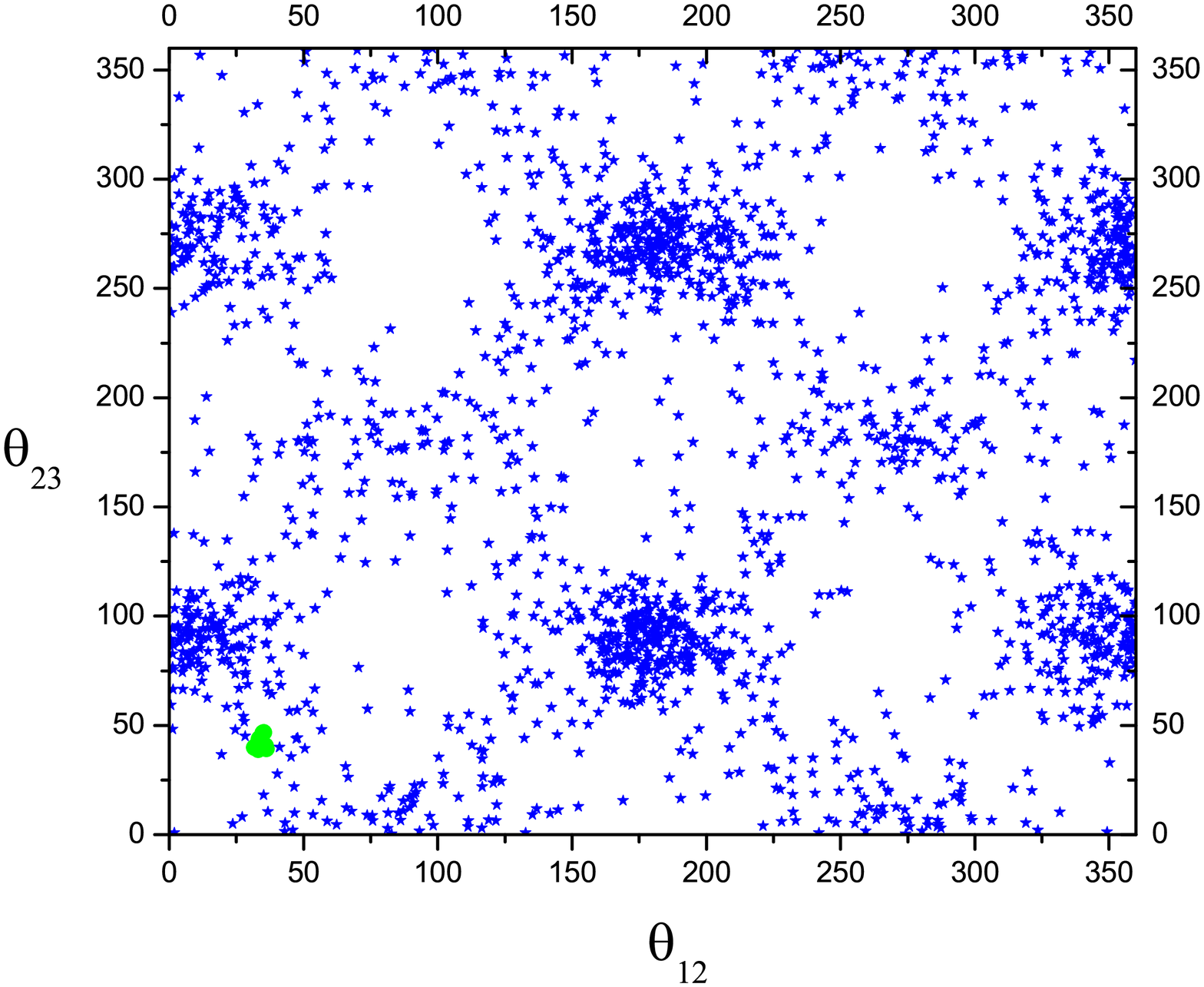,height=48mm,width=54mm}
\end{center}
\caption{Case $V_L=I$, NO, $\a_2=4$. Constraints on the mixing angles
obtained without imposing the current
experimental information from neutrino oscillation experiments (blue points) compared to
those previously obtained (green points). Notice that the regions exhibit a $\pi$ periodicity
and they are specular around $\pi/2$ so that all mixing angles
can be limited to the physical range $[0,\pi/2]$.
This can be proven to hold on very general grounds
\cite{giunti} and therefore this plot can
be regarded as a consistency check as well.}
\label{thijarb}
\end{figure}

We have also made an interesting exercise. We determined the constraints without making
use of any experimental information on the mixing angles and letting them just simply
variate between $0^{\circ}$ and $360^{\circ}$. The results are shown in Fig.~\ref{thijarb}.
First, notice that the lower bound on $m_1$ relaxes of
a few orders of magnitude  (see left panel).
Then notice  quite interestingly that small values $\theta_{13}\lesssim 10^{\circ}$
are well allowed for $m_1\gtrsim 10^{-3}\,{\rm eV}$ but values $30^{\circ} \gtrsim \theta_{13}\gtrsim 10^{\circ}$
would have been very marginally consistent. Therefore, the current bound
$\theta_{13}\lesssim 10^{\circ}$ seems to match quite well with successful $SO(10)$-inspired leptogenesis

On the other hand,
values $\theta_{23}\lesssim 30^{\circ}$ would have
been more optimal for $\theta_{13}\lesssim 10^{\circ}$ than
the current experimental large atmospheric values
(see the central panel in the figure).
However, they are still allowed thanks to the observed range
of values of the solar neutrino mixing angle (see the right panel).
For the solar neutrino mixing angle there is no real favourite range of values
for $\theta_{13}\lesssim 10^{\circ}$.

\subsection{Inverted ordering}

Let us now discuss the results for IO.
It has been shown \cite{albright} that in grand unified models with conventional type I seesaw mechanism
one can always find, for any NO model satisfying the low energy neutrino
experimental constraints, a corresponding IO model.
Therefore, though they exhibit some unattractive features that quite strongly disfavour them
(e.g. instability under radiative corrections), IO models within grand unified
theories are not unequivocally excluded.
It is therefore legitimate to check whether the requirement
of successful leptogenesis can somehow  provide some completely
independent information.

\begin{figure}
\begin{center}
\psfig{file=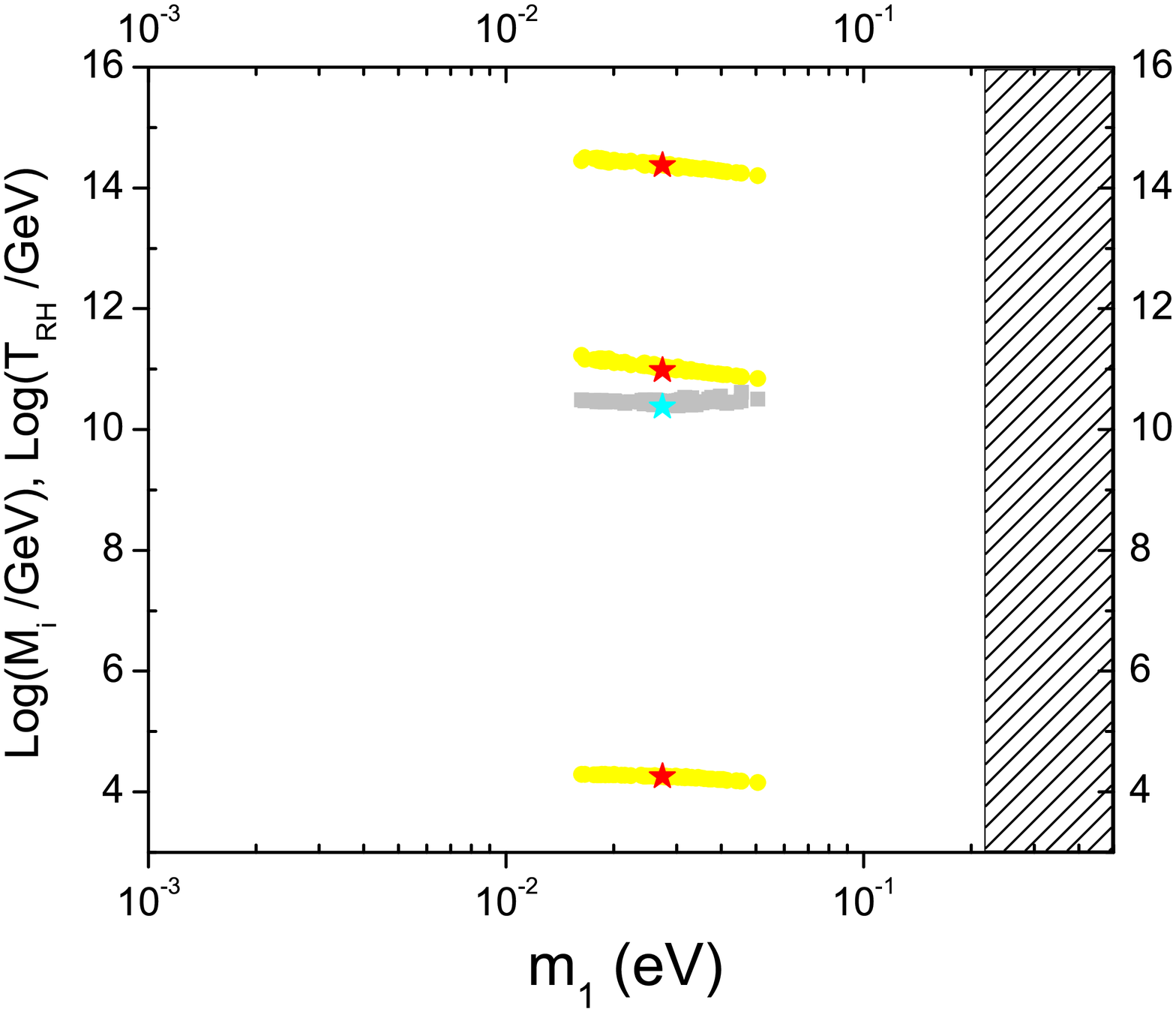,height=48mm,width=54mm}
\hspace{-4mm}
\psfig{file=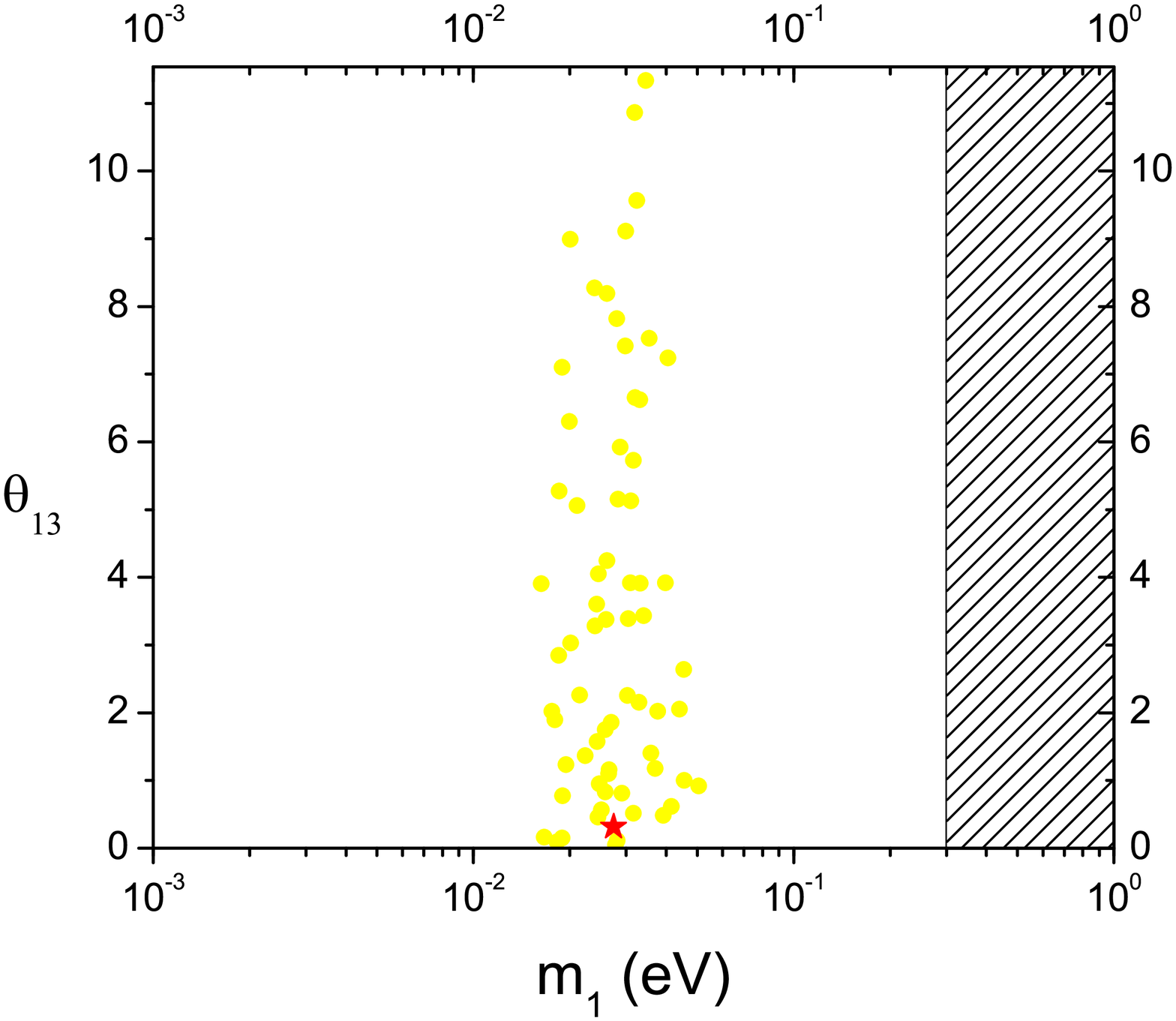,height=48mm,width=54mm}
\hspace{-4mm}
\psfig{file=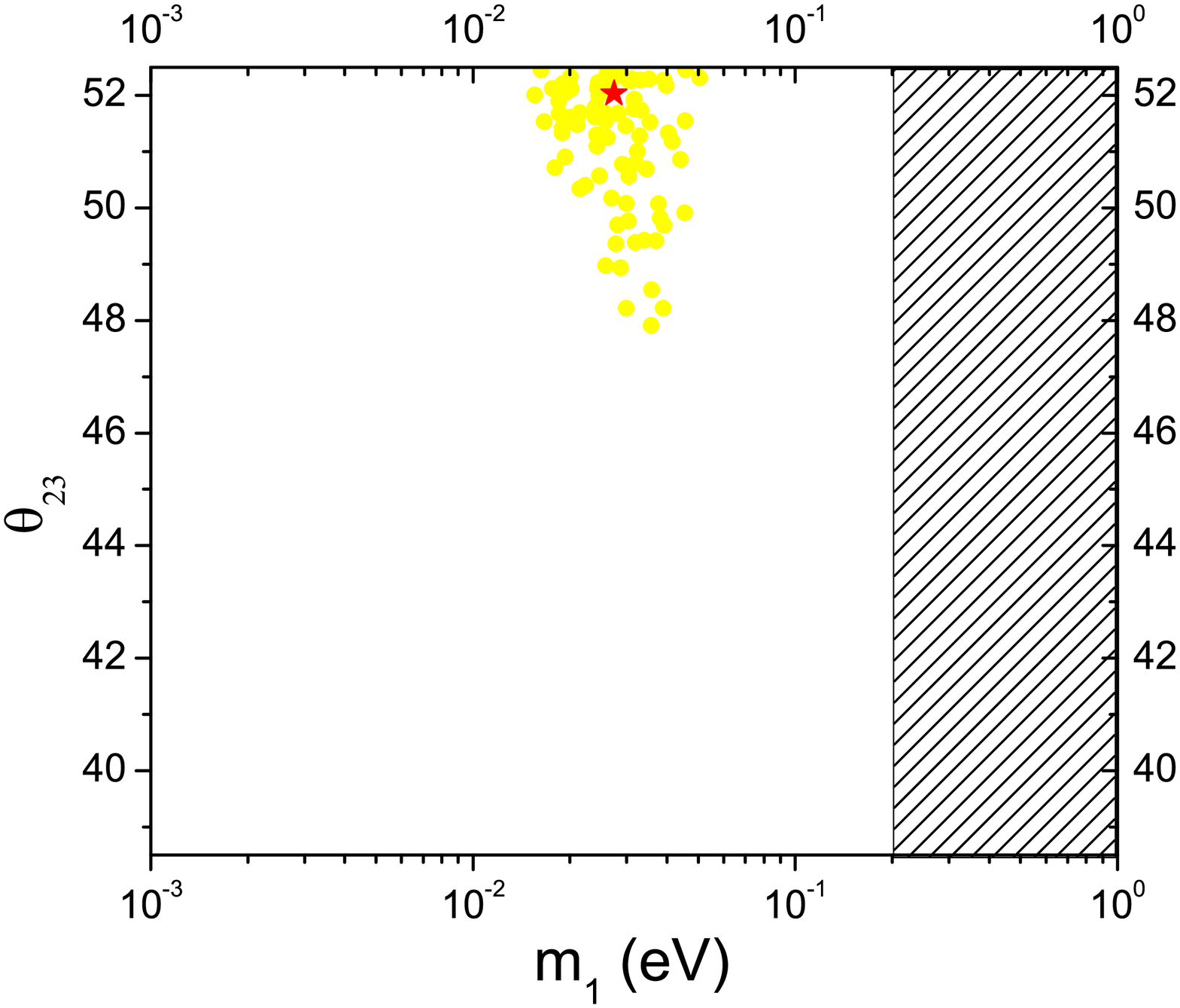,height=48mm,width=54mm} \\
\psfig{file=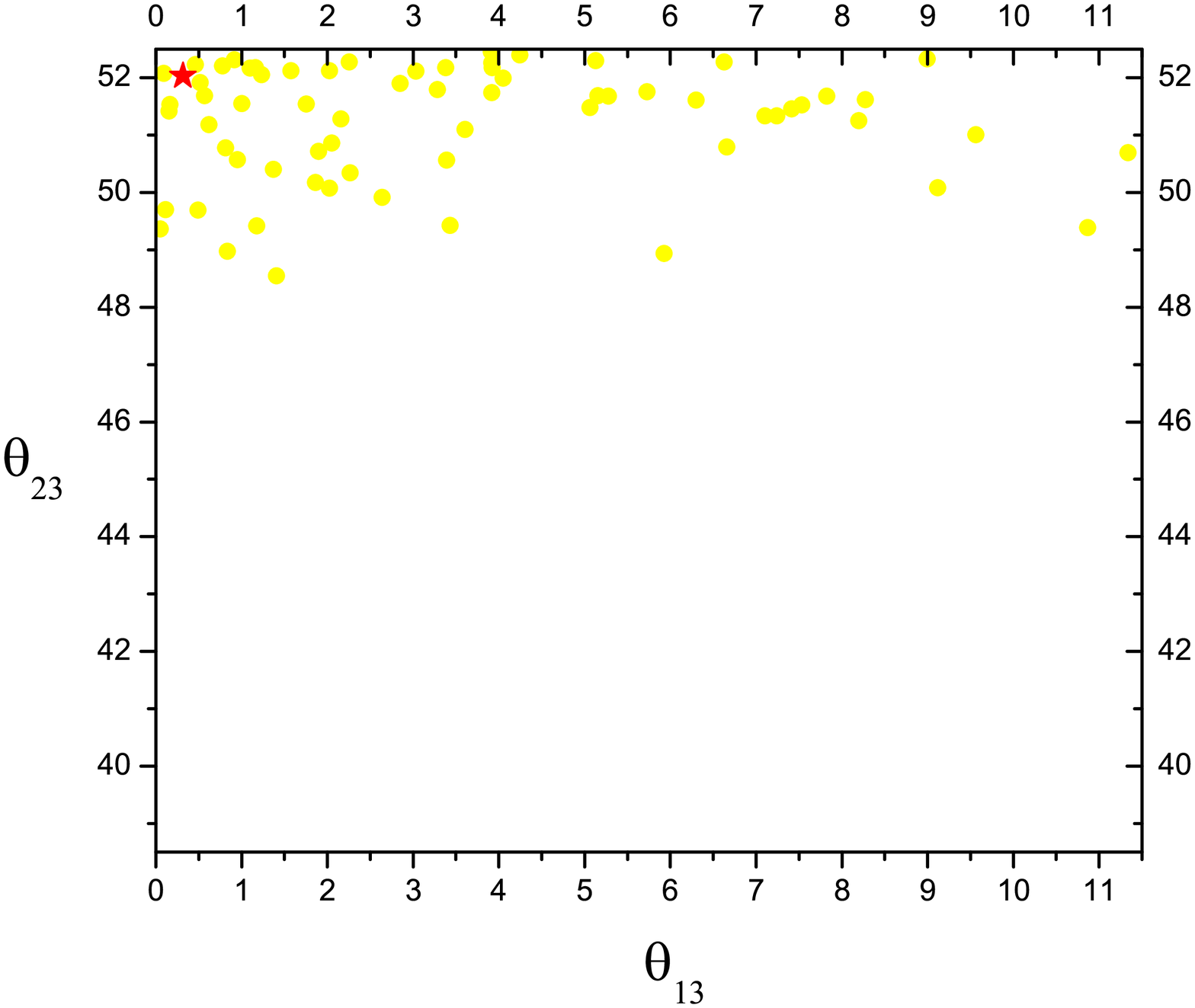,height=48mm,width=54mm}
\hspace{-4mm}
\psfig{file=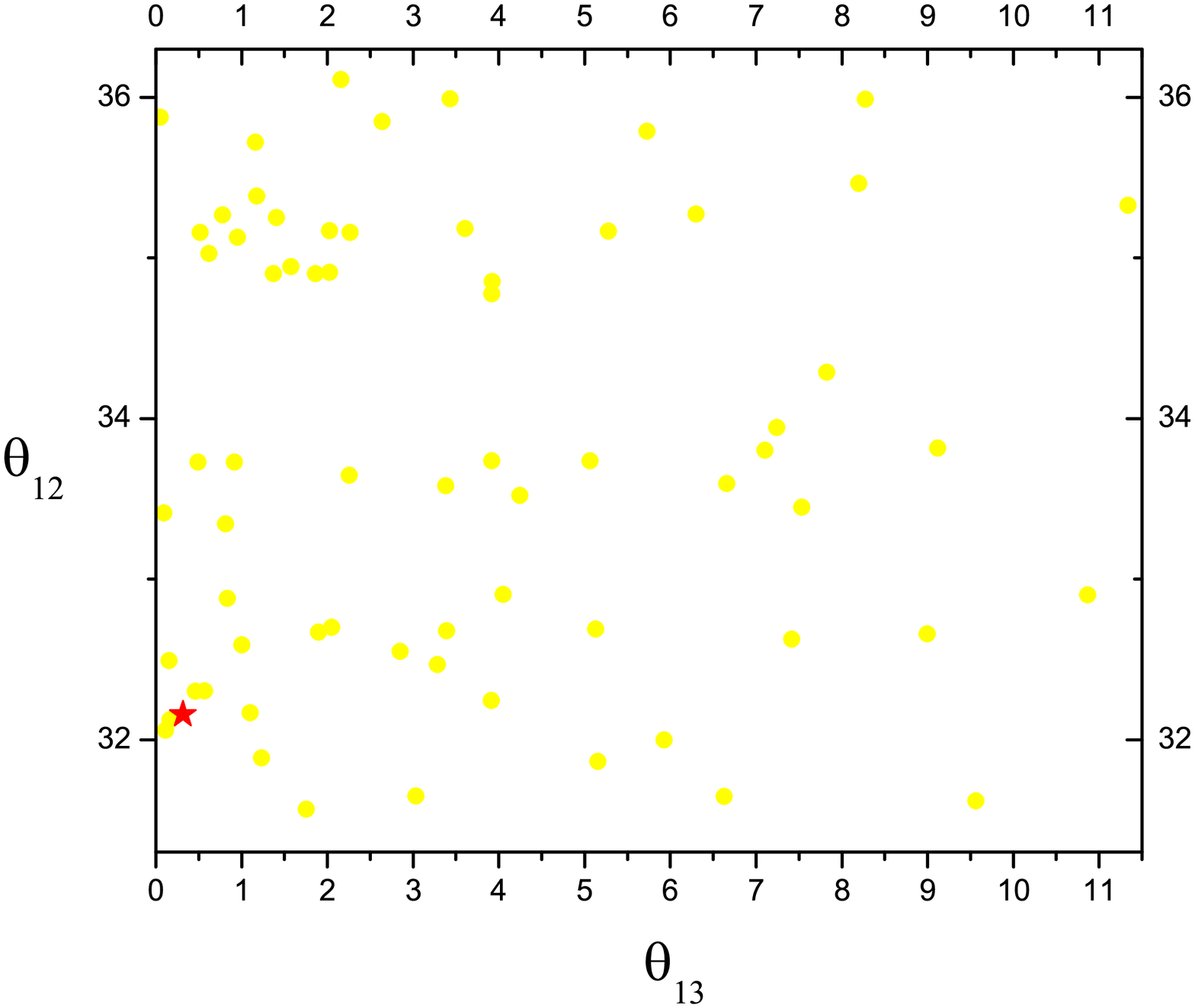,height=48mm,width=54mm}
\hspace{-4mm}
\psfig{file=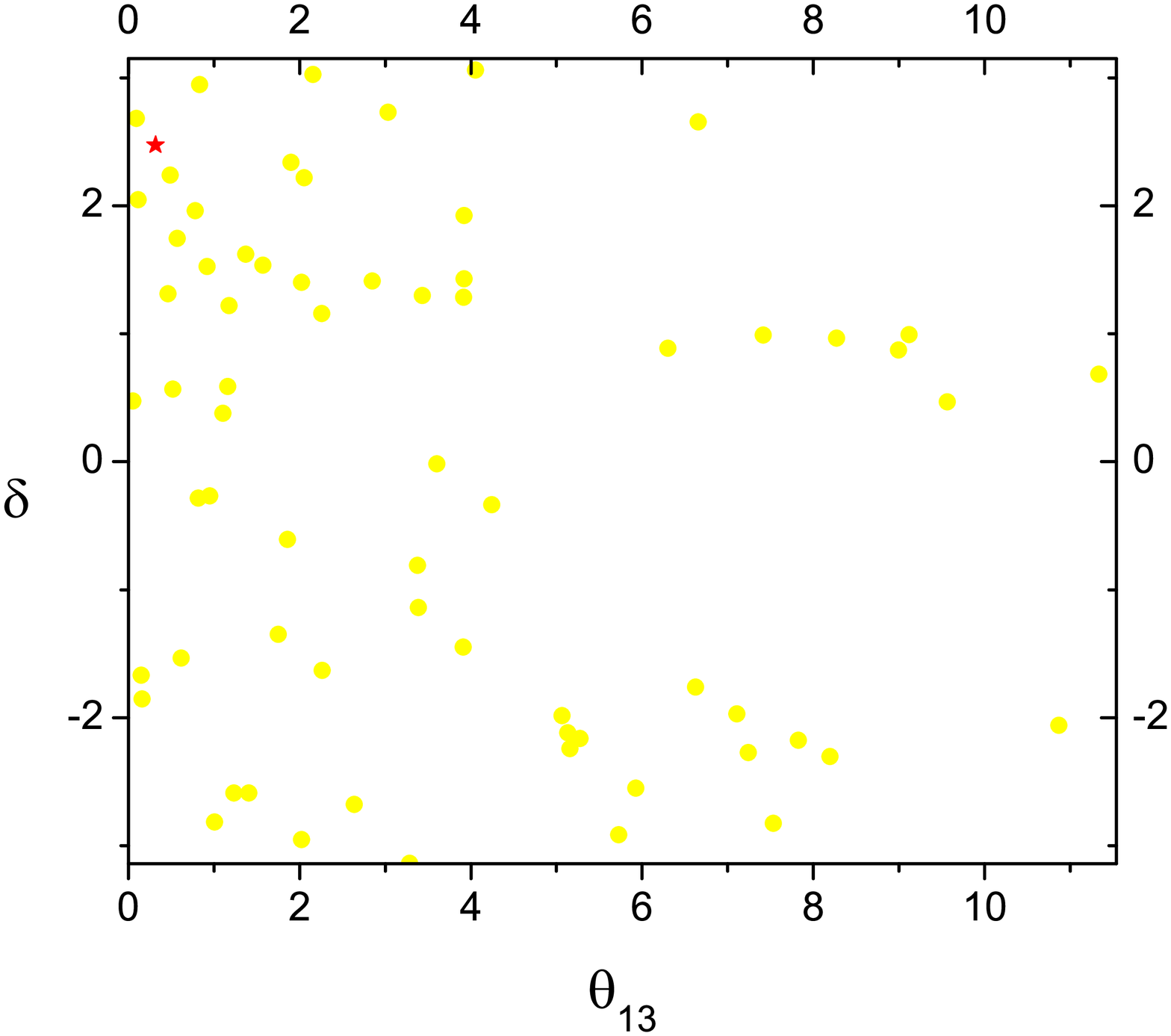,height=48mm,width=54mm} \\
\psfig{file=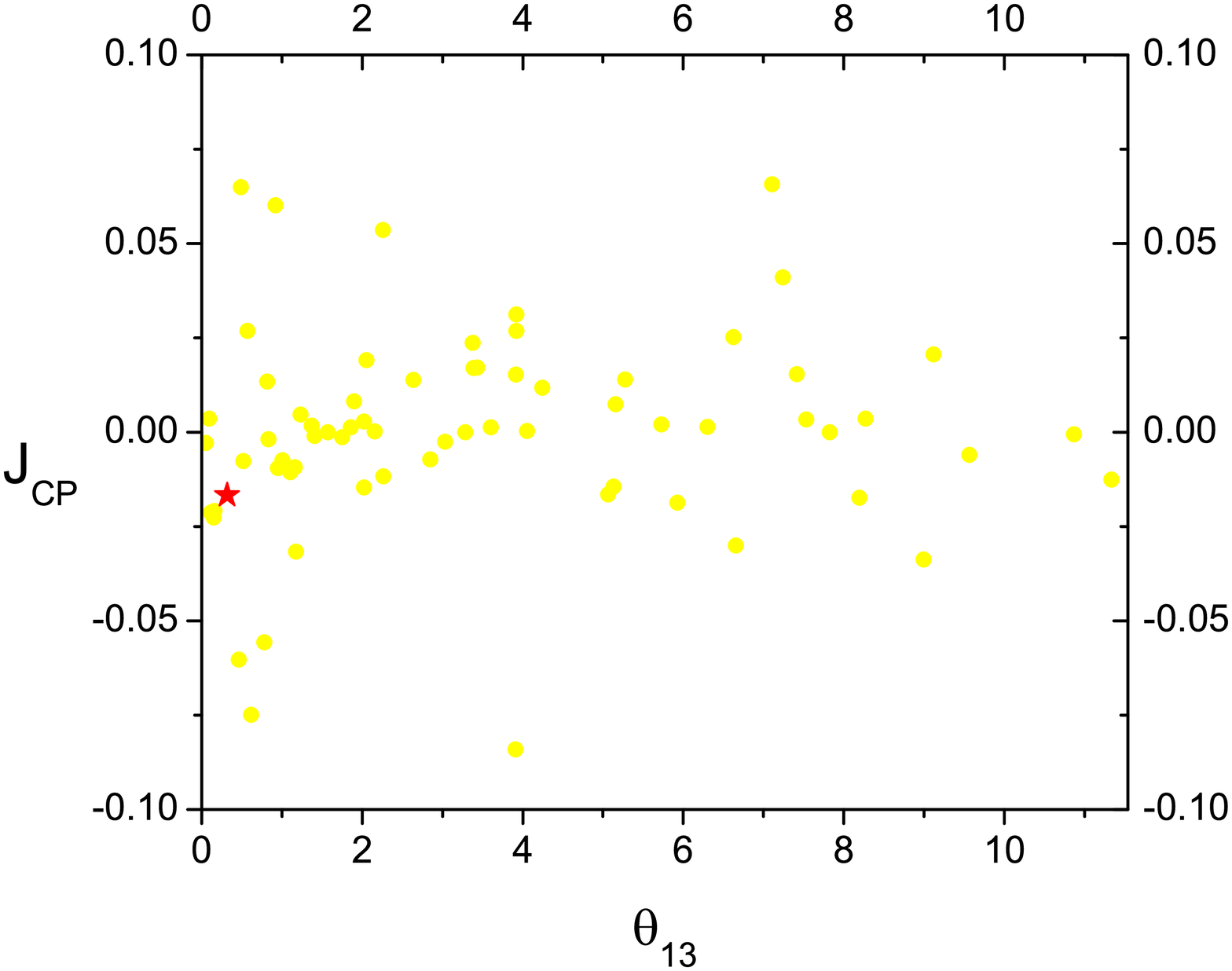,height=48mm,width=54mm}
\hspace{-4mm}
\psfig{file=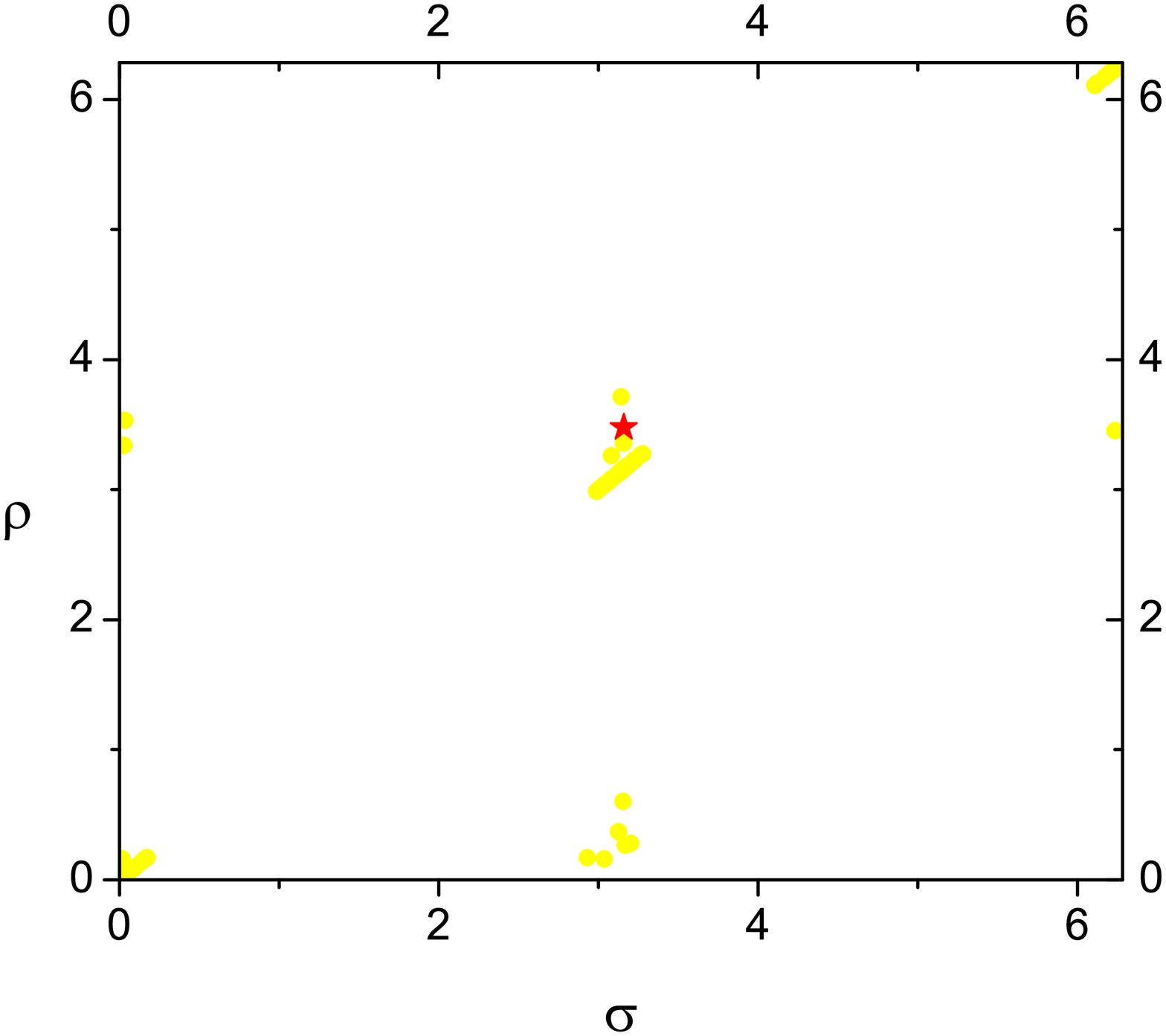,height=48mm,width=54mm}
\hspace{-4mm}
\psfig{file=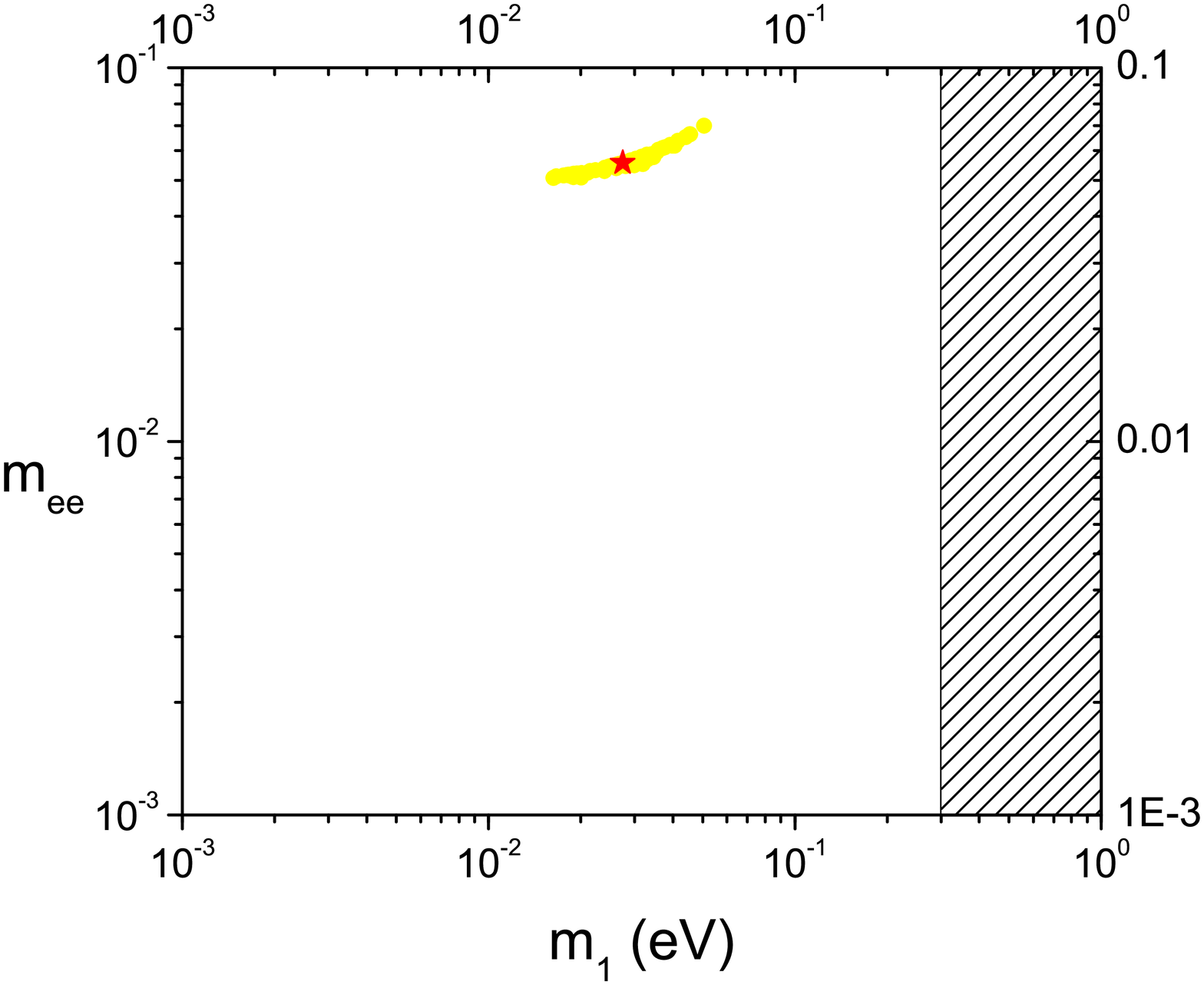,height=48mm,width=54mm}
\end{center}
\caption{Case $V_L=I$, IO. Scatter plot of points in the parameter space that satisfy
the condition $\eta_B>5.9\times 10^{-9}$
for $\a_2=5$ (yellow circles) and $\a_2=4.65$ (red star). In the top left panel the
lower bound on $T_{RH}$ (cf. eq.~(\ref{TRHmin})) is also indicated for the same values of $\alpha_2$
but with different symbols: $\a_2=5$ (grey squares), $\a_2=4.65$  (blue star).
The three mixing angles are in degrees, the three phases in radiant.}
\label{constrIO}
\end{figure}

We repeated the same scan performed in the case of NO
and the results are shown in figure~\ref{constrIO},
the analogous of the figure~\ref{constrNO} for the NO case.
One can see that IO is only very marginally allowed. For $\a_2=5$,
there is only a small region at large values of $m_1=(0.02-0.05)\,{\rm eV}$. Extending the analysis
in Ref. \cite{SO10},
this is explained by the fact that the wash-out parameter $K_{1\tau}$ turns out to be
\be
K_{1\tau}\simeq \frac{1}{3}\frac{\left(m_2-m_1\right)^2}{(2m_2+m_1)}\cdot 10^3\,{\rm eV}\propto
m_{\rm atm}\, ,
\ee
while in the NO case $K_{1\tau}$ was proportional to $m_{\rm sol}$. This constrains $\ve_{2\tau}\propto m_1$ to be as large as possible, thus ruling out  small values of $m_1$.

It is interesting to notice that in this case the allowed values for $\theta_{23}$
lie in the second octant and correspond to the largest ones compatible with the current
experimental limits. The allowed values of the effective neutrino mass fall in
a narrow range, $m_{ee}=(0.05-0.07)\,{\rm eV}$.
Therefore, IO will be in any case fully tested from cosmology and $\b\b 0\n$ experiments
during next years. We will see that this conclusion will hold also allowing  $V_L\neq I$.
As usual, in the plots the red star corresponds to the minimum value of $\a_2$
for which we have found a solution, $\alpha_2=4.65$. The corresponding set of values
indicates approximately where the asymmetry has a maximum for a fixed $\a_2$ value.

\begin{figure}
\begin{center}
\psfig{file=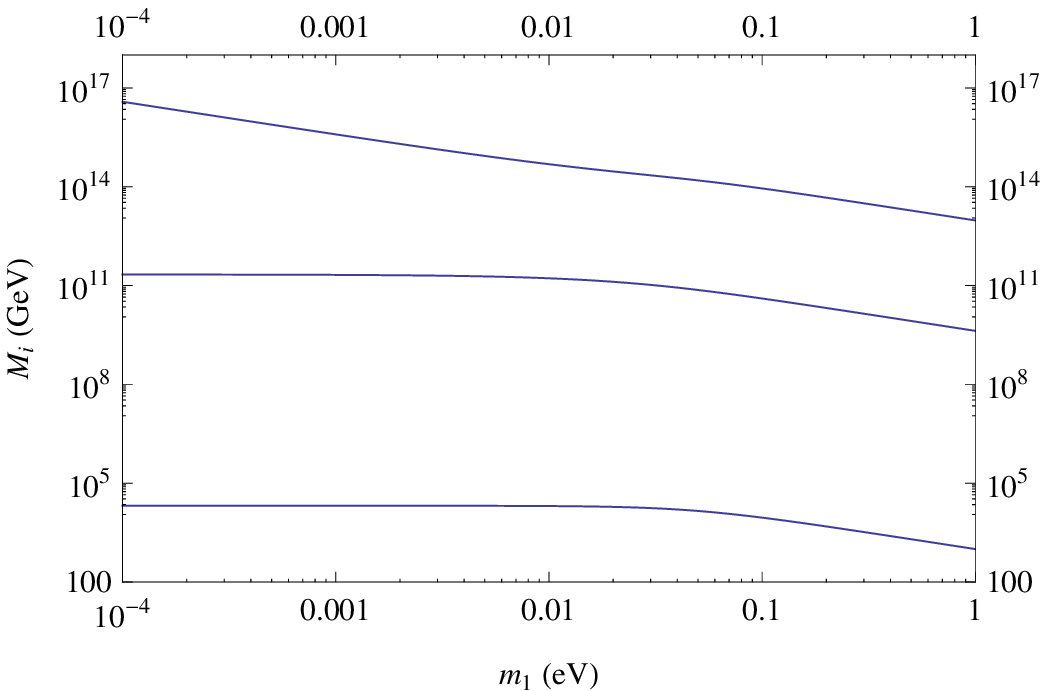,height=48mm,width=50mm}
\hspace{-3mm}
\psfig{file=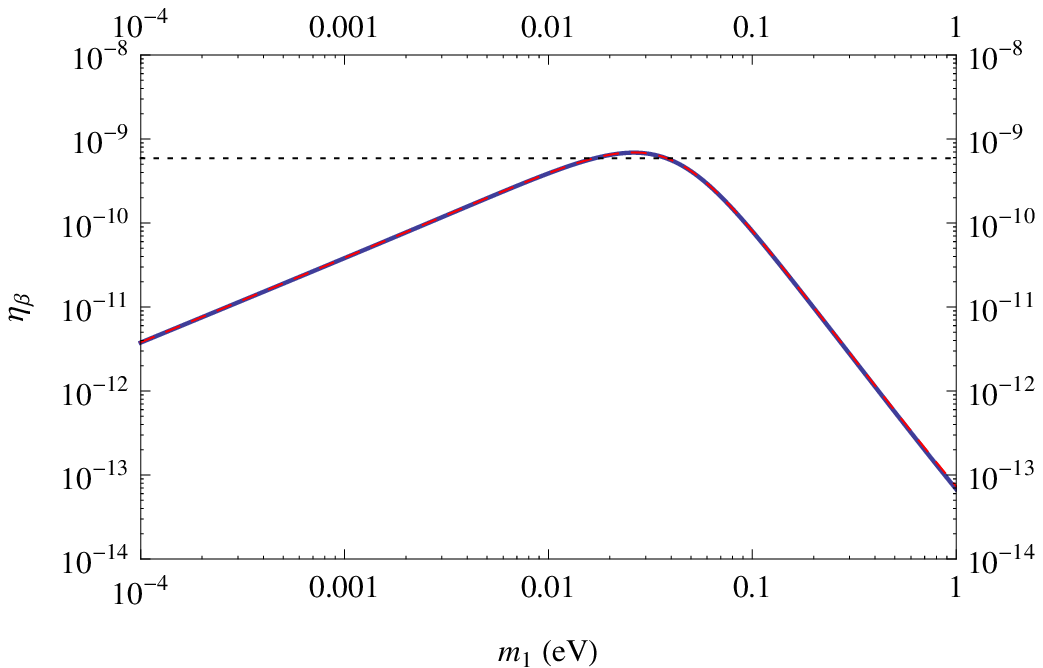,height=48mm,width=50mm}
\hspace{-3mm}
\psfig{file=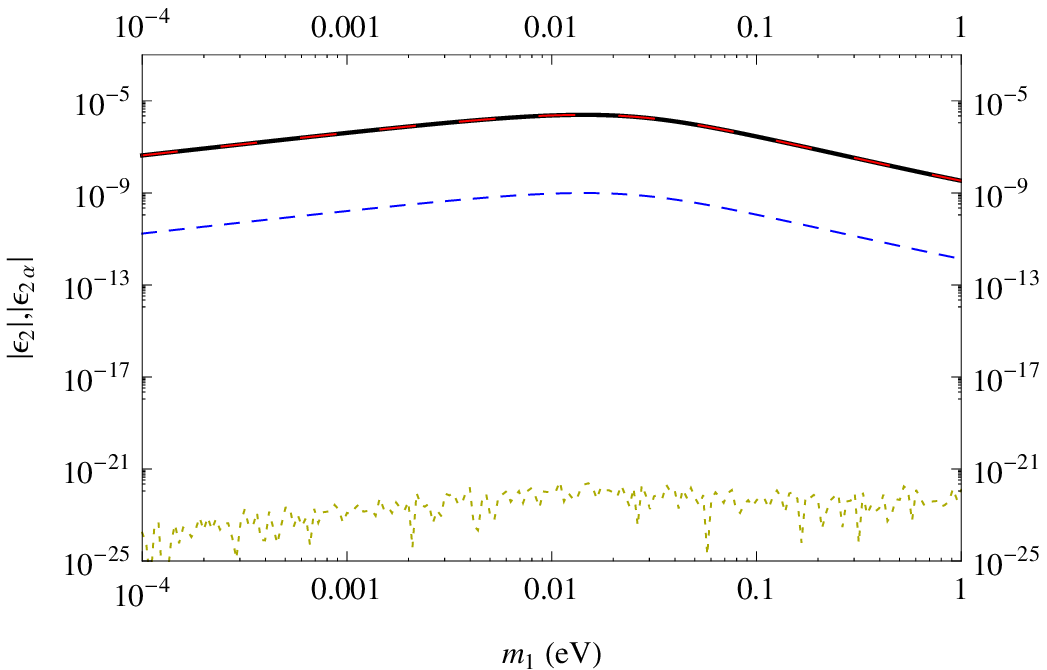,height=48mm,width=50mm} \\ \vspace*{4mm}
\psfig{file=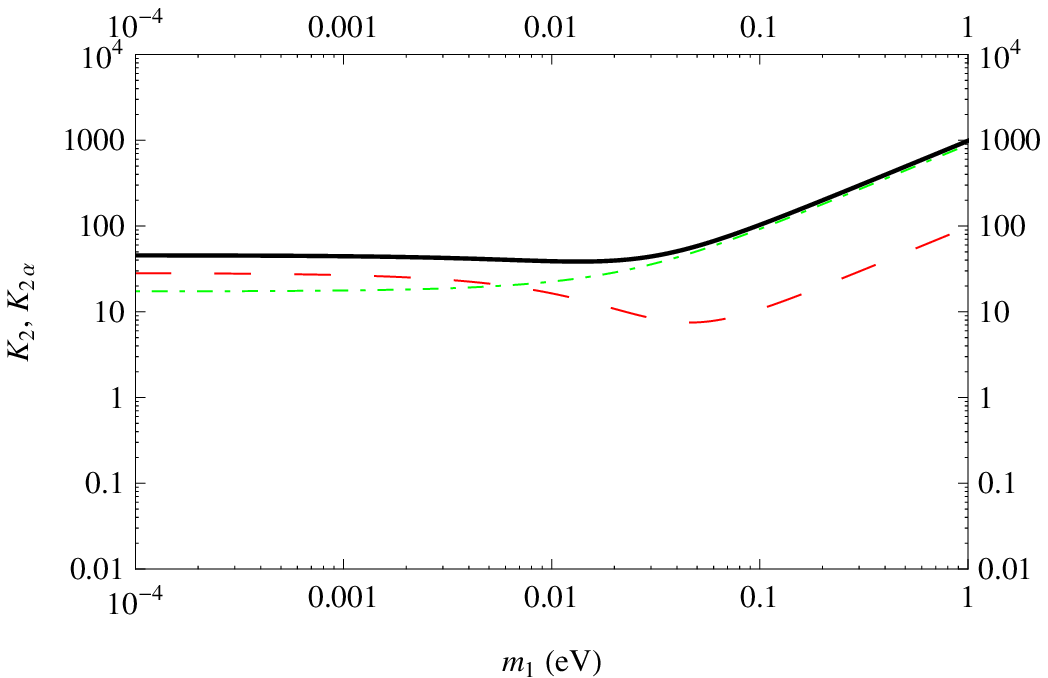,height=48mm,width=54mm}
\hspace{1mm}
\psfig{file=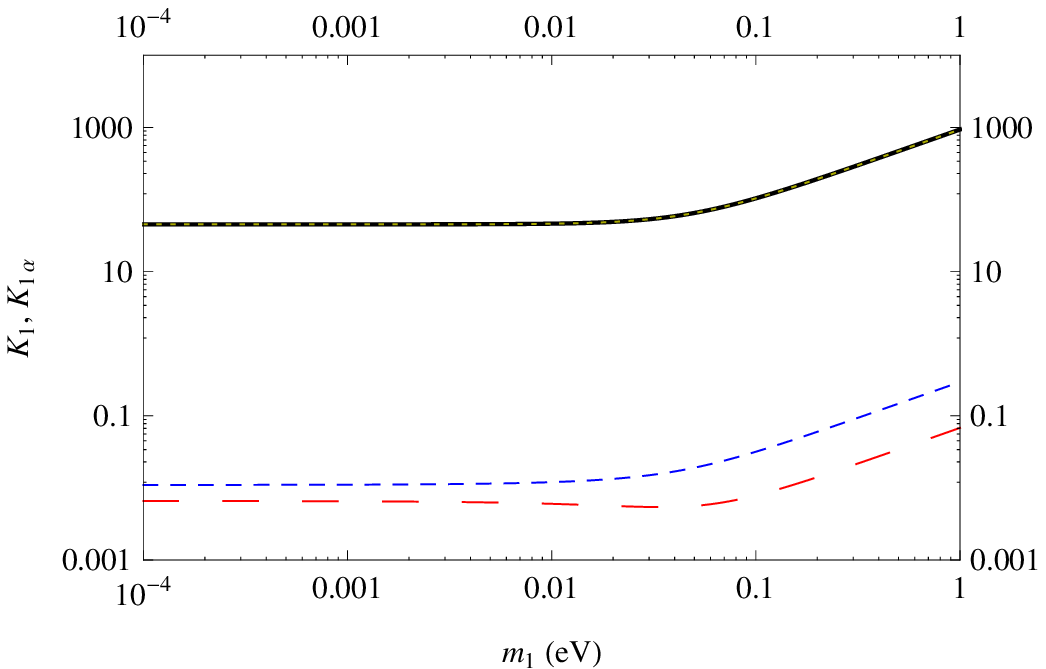,height=48mm,width=54mm}
\end{center}
\caption{Case $V_L=I$, IO. Plot of all relevant quantities versus $m_1$
for the set of values ($\theta_{13}=0.32^{\circ}$,$\theta_{23}=52.03^{\circ}$,$\theta_{12}=32.16^{\circ}$,
$\rho=3.16$,$\sigma=3.48$,$\delta=2.47$)
corresponding to the red star in the previous figure ($\a_2=4.65$).}
\label{IO}
\end{figure}

For this choice of values, in figure~\ref{IO}, we show the plots of
the RH neutrino masses, of the asymmetry $\eta_B$, of the $C\!P$ asymmetries $\ve_2$, $\ve_{2\a}$,
of $K_1, K_{1\a}$ and of $K_2, K_{2\a}$ versus $m_1$.
One can see how the heaviest RH neutrino mass $M_3$ decreases with $m_1$ much faster
and at $m_1 \simeq 0.001\,{\rm eV}$ one has $M_3\simeq 10^{17}\,{\rm GeV}$. Therefore,
as one can see from the central top panel,
the $C\!P$ asymmetries are this time strongly suppressed at $m_1\simeq 10^{-3}\,{\rm eV}$.
On the other hand, in the range $m_1\simeq (0.02-0.05)\,{\rm eV}$ the $C\!P$ asymmetries
are large enough that successful leptogenesis is still possible. Notice, that this kind of solutions
are a sort of modification of the solution obtained at large $m_1$ values for NO, simply shifted
at somehow larger values. The asymmetry is therefore strongly depending on the initial conditions
($K_{2\t}\simeq 1$). The first kind of solution, at small $m_1$ values, is completely absent.

Therefore, though IO is strongly disfavoured, it is not completely ruled out,
a conclusion somehow very similar to that one obtained from completely independent
arguments \cite{albright}. In this case, however, leptogenesis provides
quite a precise quantitative test.

We can conclude this section saying that these results confirm and complete those shown in
\cite{SO10}. In particular it is confirmed that
there are viable solutions corresponding to the different points shown in the figures
falling in the currently experimentally allowed ranges of the parameters,. The model is therefore
not ruled out. A further step is now to understand whether the model is predictive, excluding
regions of the parameter space that future experiments can test. From the figures, as we have discussed,
it is clear that assuming $V_L=I$ such excluded regions exist and therefore one obtains
interesting constraints. However, it is important to go beyond
the simple condition $V_L=I$ in order to test the stability of the constraints for
variations of $V_L$. This is the main objective of the next sections.

%%%%%%%%%%%%%%%%%%%%%%%%%%%%%%%%%%
\section{The case $V_L = V_{CKM}$}
%%%%%%%%%%%%%%%%%%%%%%%%%%%%%%%%%%

We now study how the constraints change when a misalignment between the
physical basis where $m_D$ is diagonal and the flavour basis,
where the charged lepton mass matrix is diagonal, is considered,
corresponding to  $V_L\neq I$. Since $V_L$ is unitary, we can  parameterize
it similarly to the leptonic mixing matrix introducing three mixing angles,
one Dirac-like phase and two Majorana-like phases,
\begin{equation}\label{Umatrix}
V_L=
\left( \begin{array}{ccc}
c^L_{12}\,c^L_{13} & s^L_{12}\,c^L_{13} & s^L_{13}\,e^{-{\rm i}\,\d_L} \\
-s^L_{12}\,c^L_{23}-c^L_{12}\,s^L_{23}\,s^L_{13}\,e^{{\rm i}\,\d_L} &
c^L_{12}\,c^L_{23}-s^L_{12}\,s^L_{23}\,s^L_{13}\,e^{{\rm i}\,\d_L} & s^L_{23}\,c^L_{13} \\
s^L_{12}\,s^L_{23}-c^L_{12}\,c^L_{23}\,s^L_{13}\,e^{{\rm i}\,\d_L}
& -c^L_{12}\,s^L_{23}-s^L_{12}\,c^L_{23}\,s^L_{13}\,e^{{\rm i}\,\d_L}  &
c^L_{23}\,c^L_{13}
\end{array}\right)
\cdot {\rm diag}\left(e^{i\,\rho_L}, 1, e^{i\,\sigma_L}  \right)\, ,
\end{equation}
where we defined $s^L_{ij}\equiv \sin \theta^L_{ij}$ and  $c^L_{ij}\equiv \cos \theta^L_{ij}$.
Therefore, we have now six additional parameters that give much more freedom.
We will not explore the full parameter space but,
in the spirit of $SO(10)$-inspired models, we will allow only small mixing
angles $\theta_{ij}^L$ at the level of the mixing angles in the CKM matrix.

As a first definite example we repeat the analysis performed for
the case $V_L=I$ for a definite case where the $\theta^L_{ij}$ are exactly equal
to the mixing angles in the CKM matrix and therefore we set
$\theta_{13}^L=0.21^{\circ}$, $\theta_{23}^L=2.3^{\circ}$, $\theta_{12}^L=13^{\circ}$,
where the latter is the measured value of the Cabibbo angle.

\subsection{Normal ordering}

For NO the results are shown in Figure 6.
\begin{figure}
\begin{center}
\psfig{file=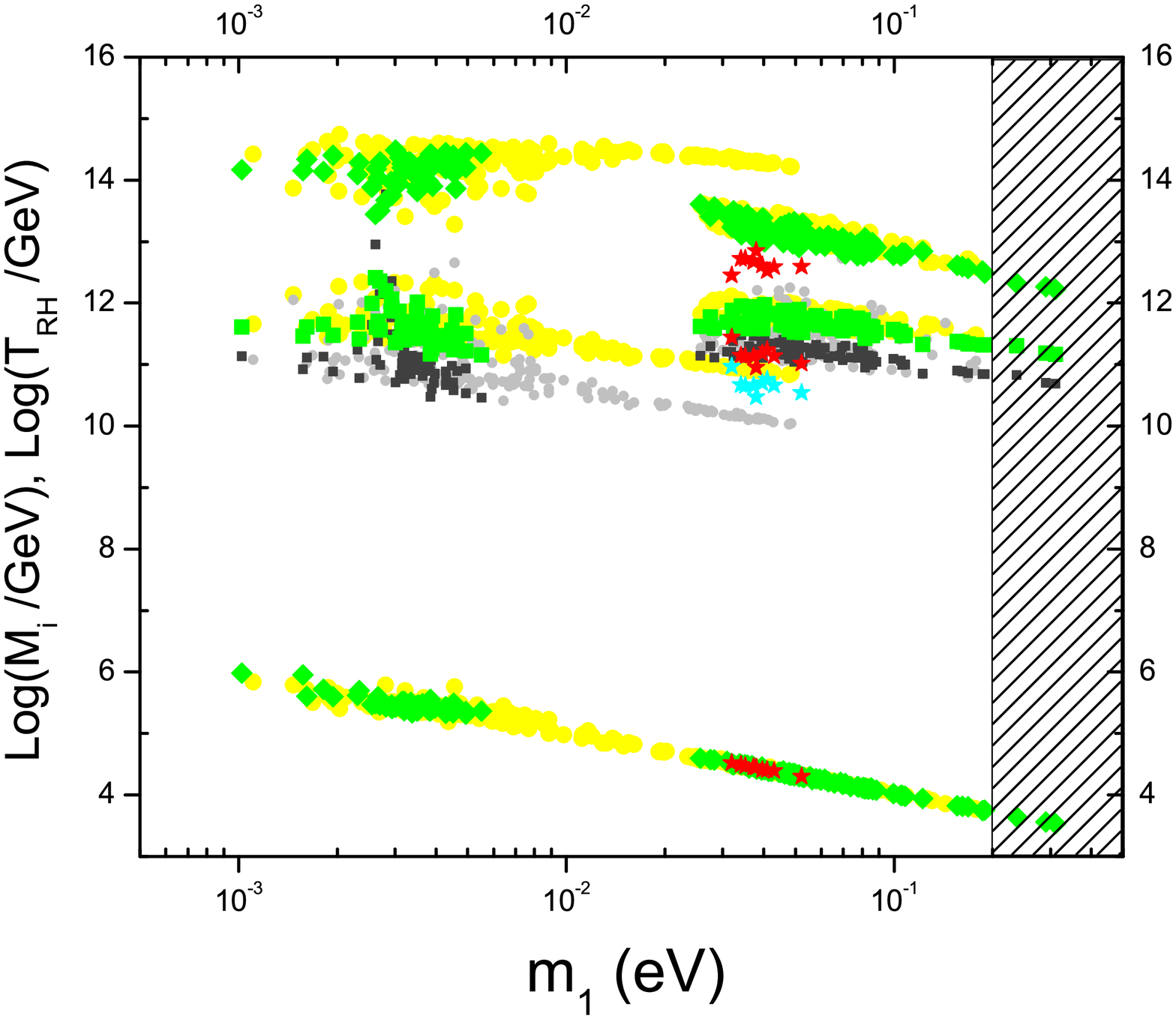,height=48mm,width=54mm}
\hspace{-4mm}
\psfig{file=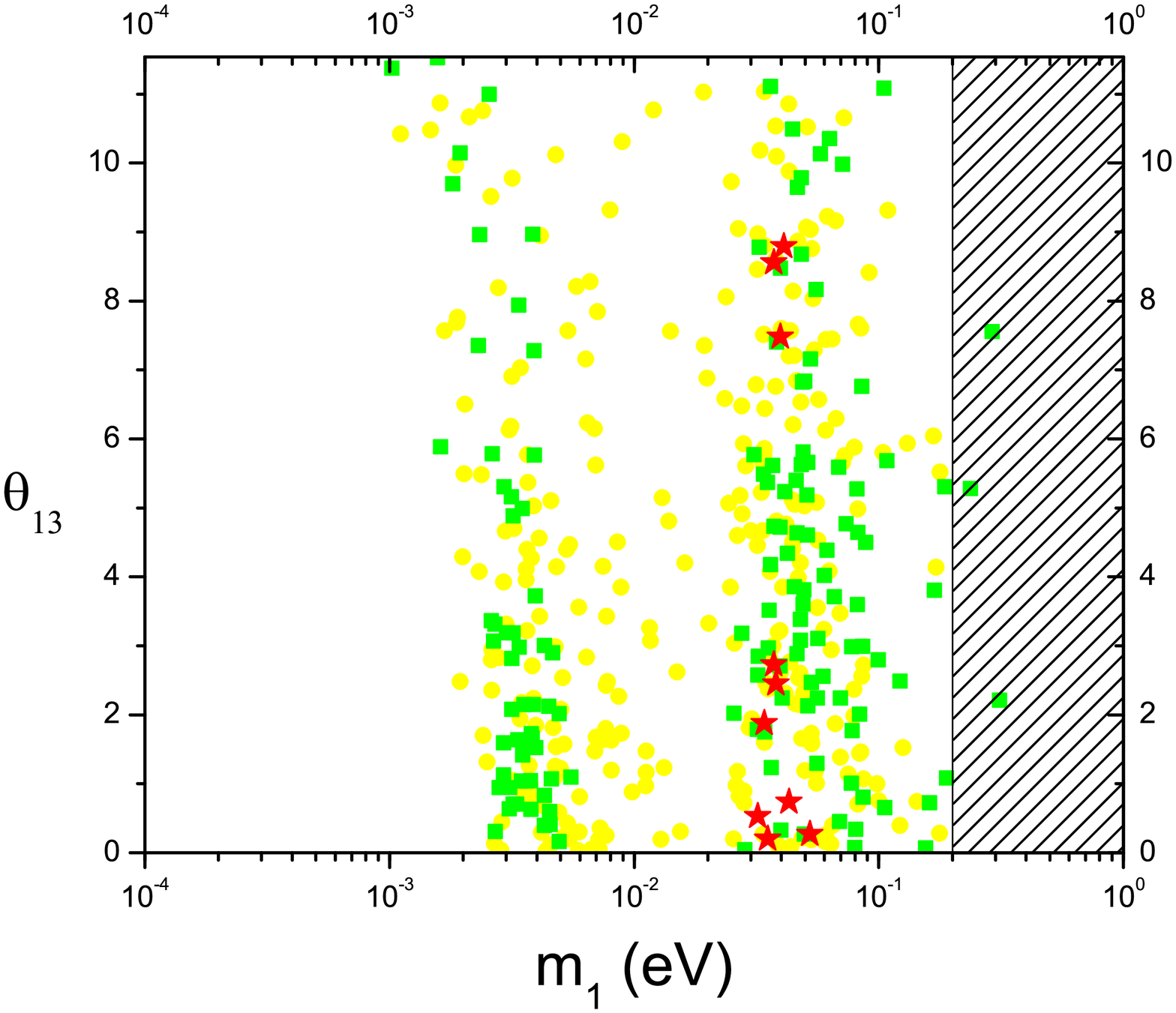,height=48mm,width=54mm}
\hspace{-4mm}
\psfig{file=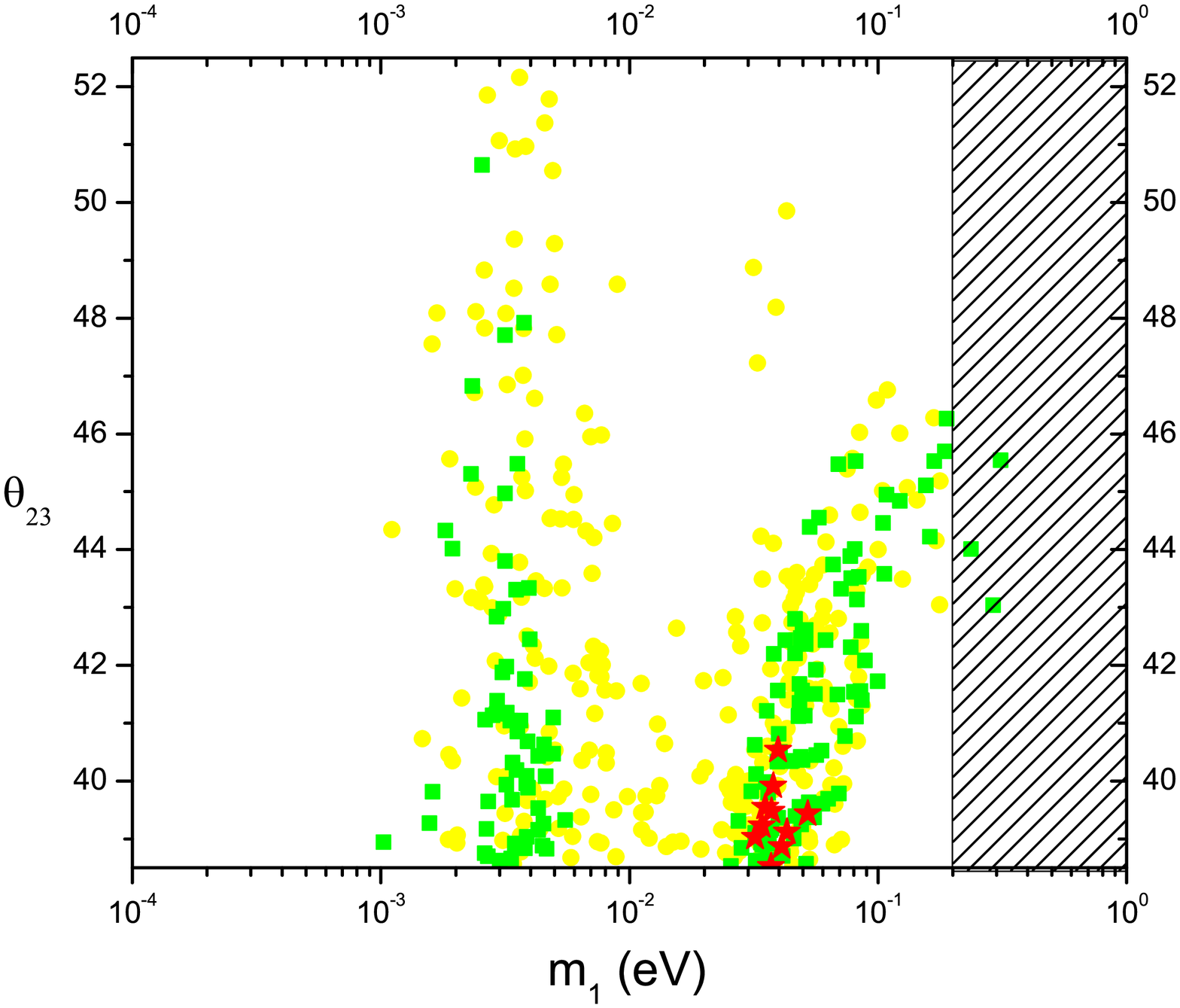,height=48mm,width=54mm} \\
\psfig{file=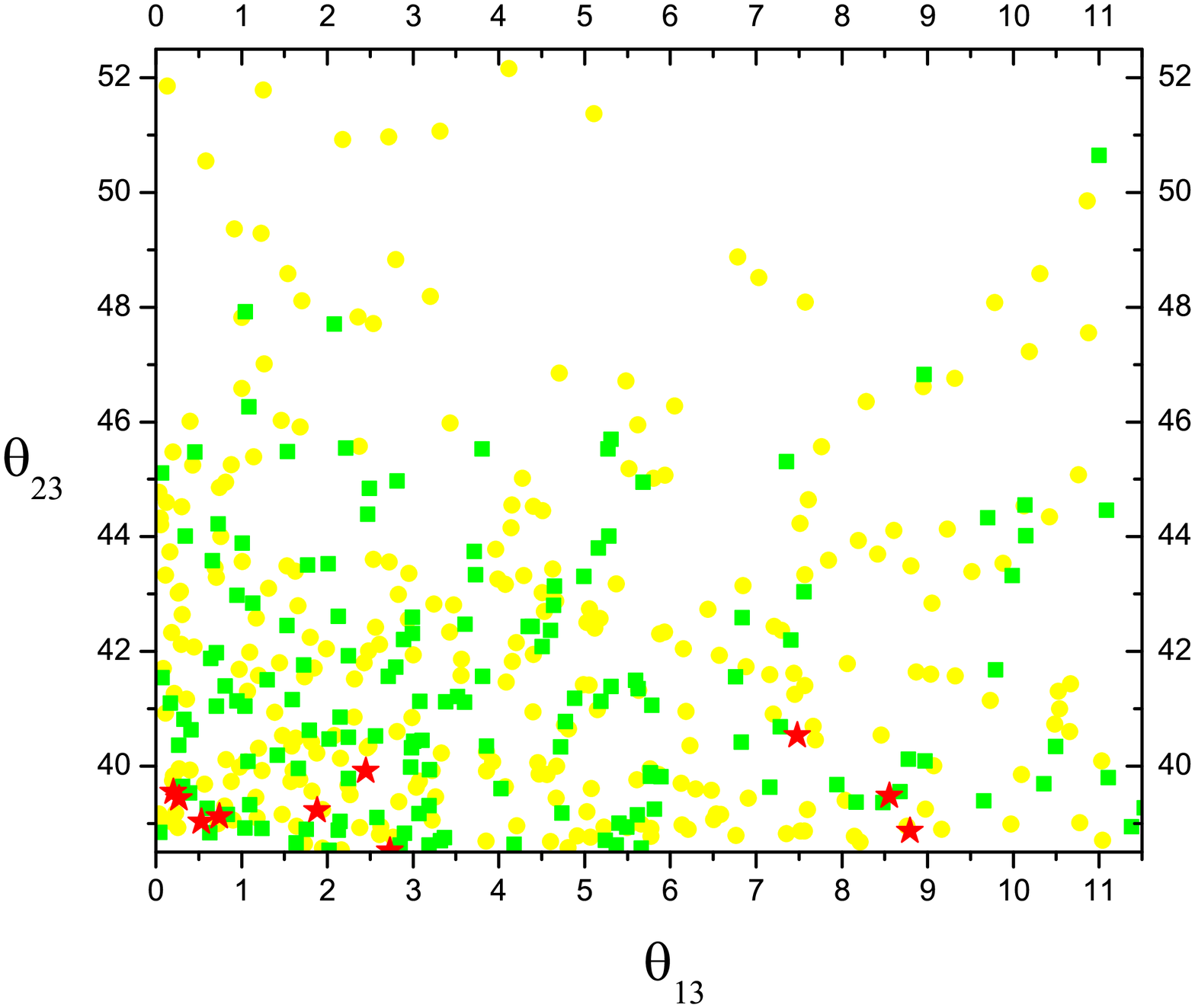,height=48mm,width=54mm}
\hspace{-4mm}
\psfig{file=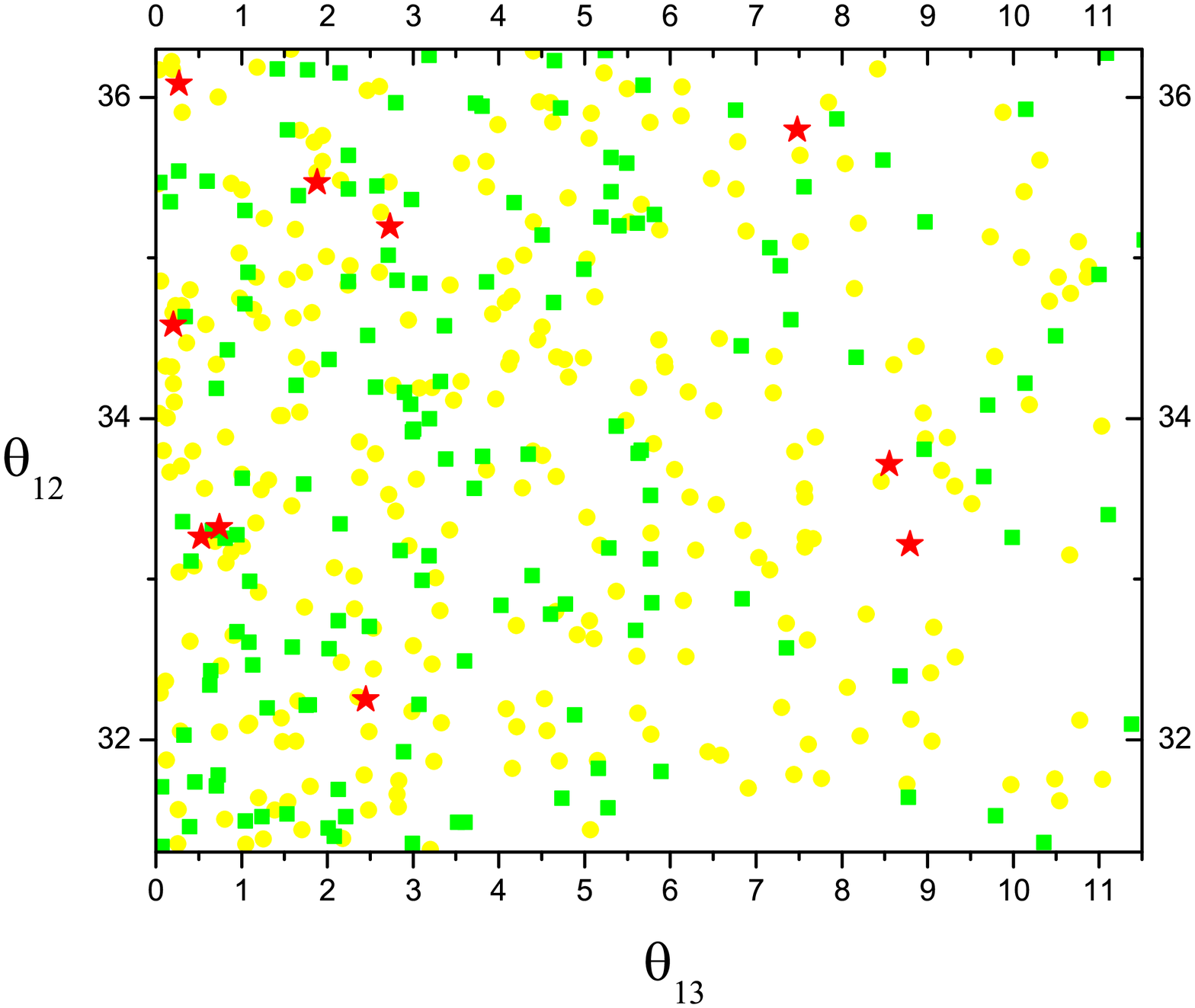,height=48mm,width=54mm}
\hspace{-4mm}
\psfig{file=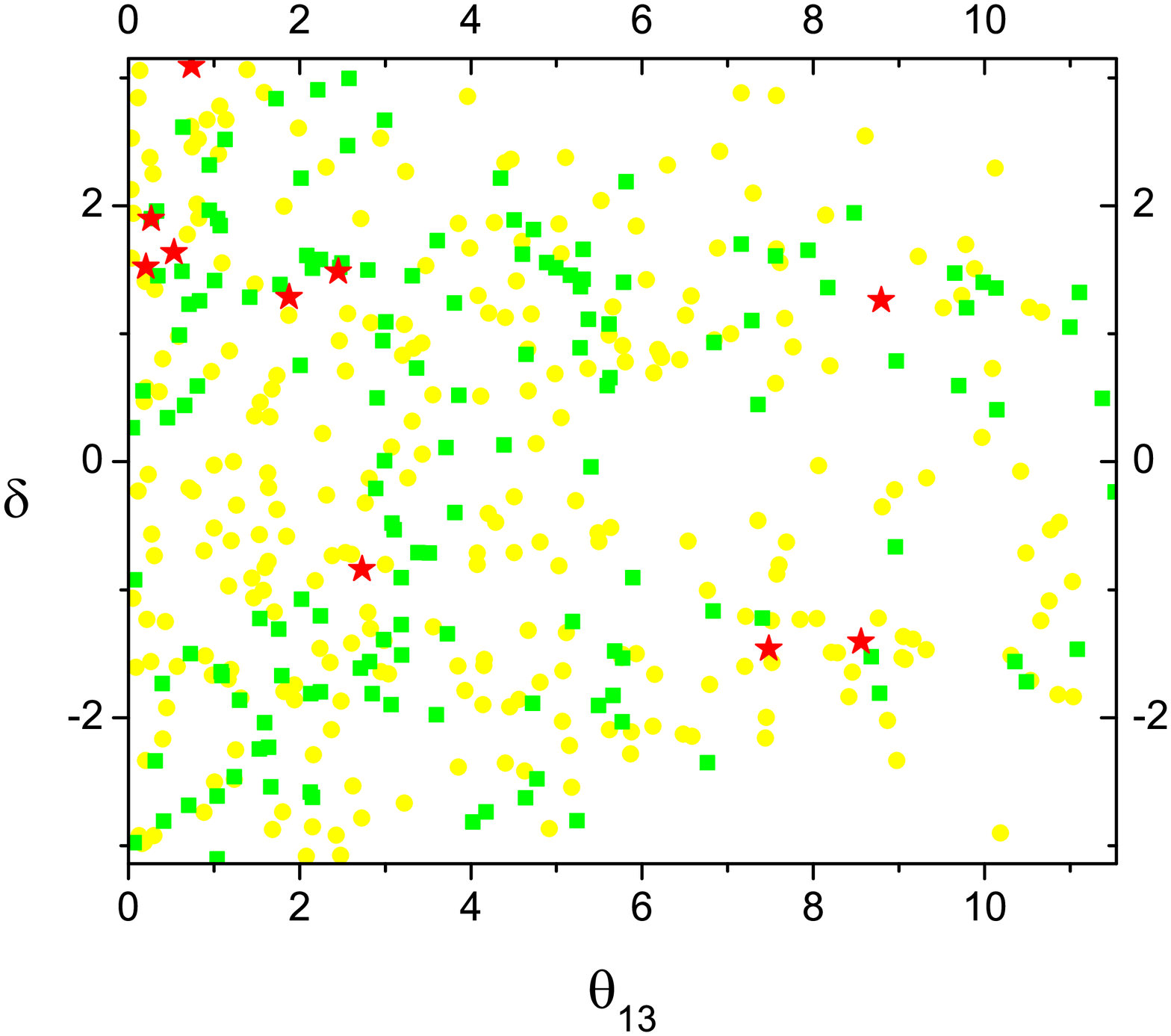,height=48mm,width=54mm} \\
\psfig{file=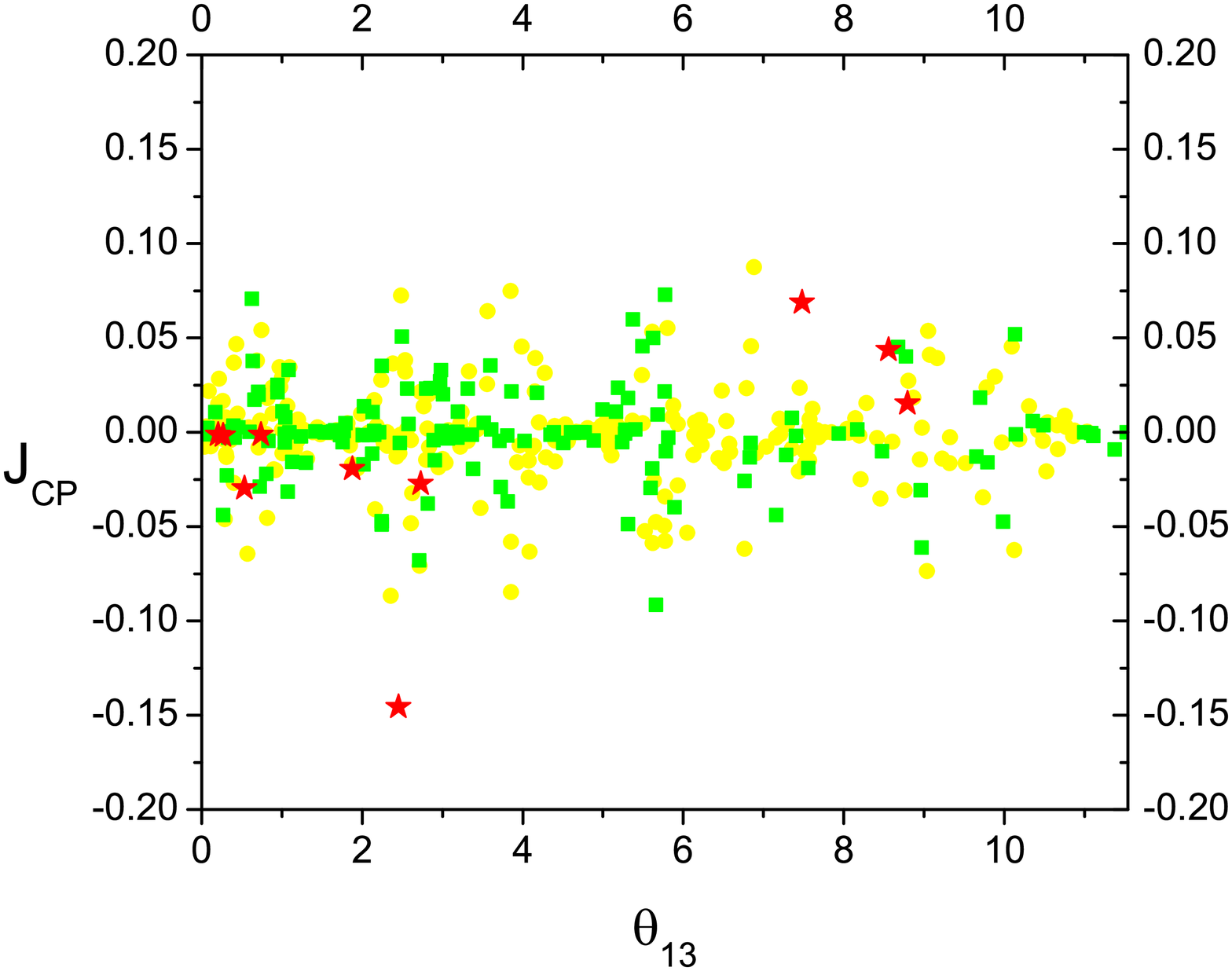,height=48mm,width=54mm}
\hspace{-4mm}
\psfig{file=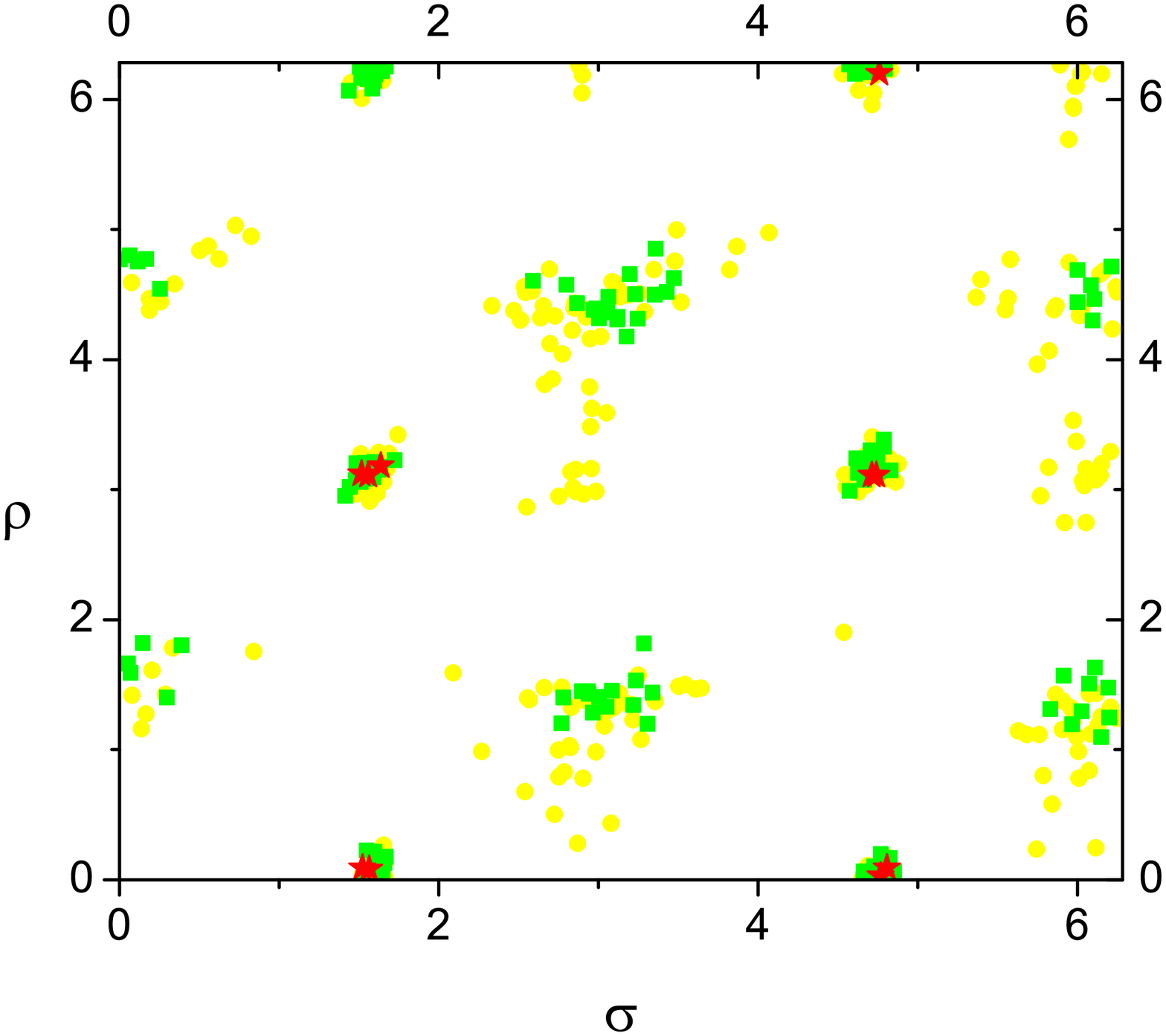,height=48mm,width=54mm}
\hspace{-4mm}
\psfig{file=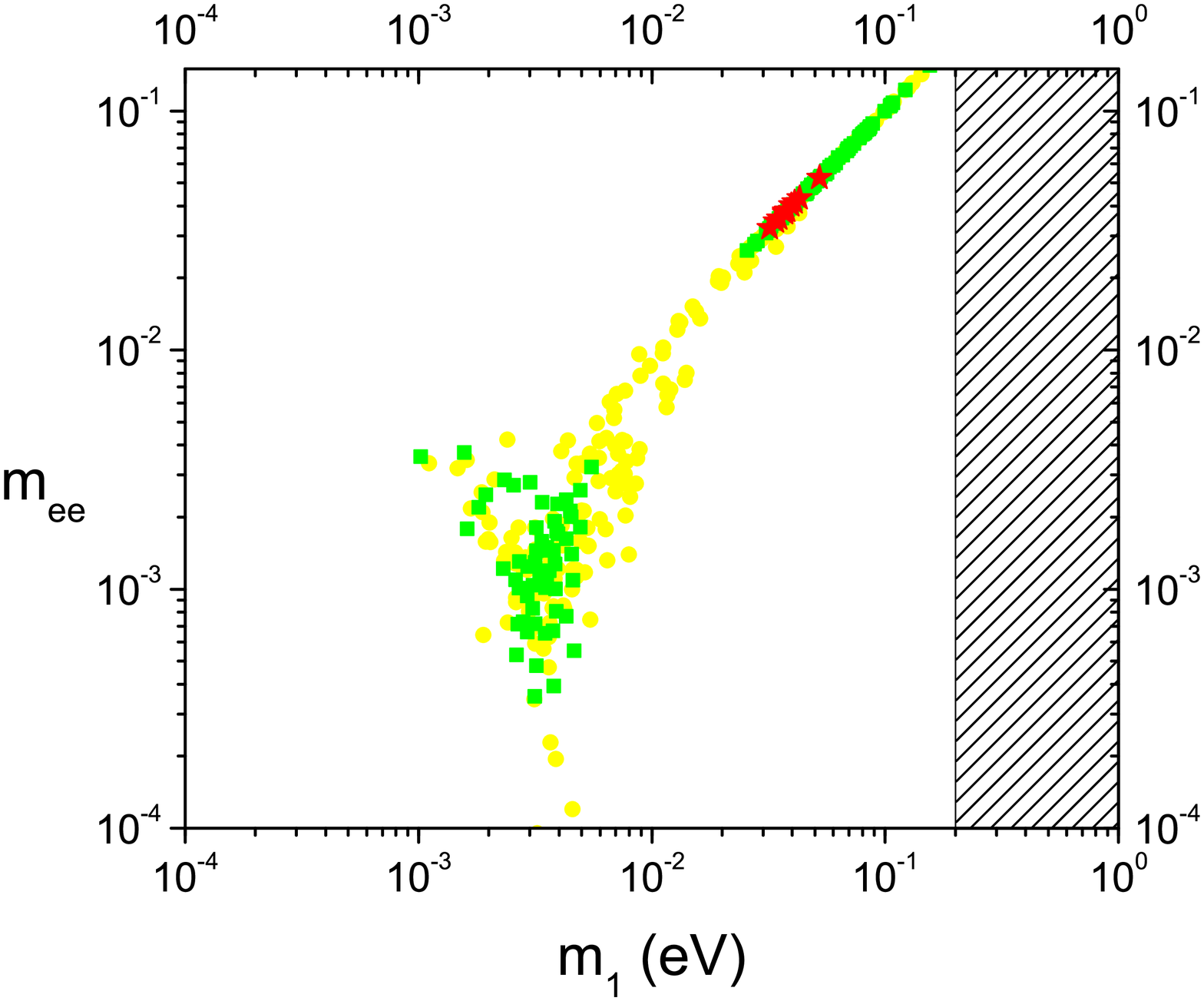,height=48mm,width=54mm}
\end{center}
\caption{Case $V_L=V_{CKM}$, NO. Scatter plot of points in the parameter space that satisfy
the condition $\eta_B>5.9\times 10^{-9}$ for $\a_2=5$ (yellow circles),
$\a_2=4$ (green squares) and $\a_2=1$ (red stars). In the top-left panel, the same convention of Fig.~2 is adopted
to indicate $T_{RH}$.}
\end{figure}
There is a first result to highlight: $\a_2$  values as low as $\a_2=1$
are now allowed. This is an interesting result in connection with the study
of realistic $SO(10)$ models. At the same time this result also implies slightly
lower values of $M_2$ and consequentially of the minimum value of $T_{\rm reh}$
that can be now as low as $\simeq 10^{10}\,{\rm GeV}$,
as it can be noticed in the top-left panel in Fig.~6.
In this case we have more generally calculated the minimum reheat temperature as
\be\label{TRHmin2}
T_{RH}^{\rm min}\simeq {\rm min}\left[{M_2\over z_B(K_{2\t})-2},{M_2\over z_B(K_{2e+\m})-2}\right] \, ,
\ee
considering that in the case the asymmetry at the production can be either
tauon dominated or $e+\m$ dominated. This is because the third kind of solution
that was highly suppressed in the case $V_L=I$, the right panel in Fig.~1, becomes
now viable and is $e+\m$ dominated, as we will discuss soon in more detail.

Notice that if we compare the allowed points for $\a_2=4$ with those
found for $V_L=I$, the constraints on the low energy neutrino parameters
are now less stringent. In particular an allowed region for values $m_1\simeq 0.003\,{\rm eV}$
is also found for very small values of $\theta_{13}$.
Indeed, in  the case $V_L=I$, and for small values of
 $m_1$, the suppression of  the  wash out value  $K_{1\tau}$ imposed   a lower bound on $\theta_{13}$.
 By choosing $V_L=V_{CKM}$ introduces the possibility of getting vanishing $K_{1\tau}$ even for zero $\theta_{13}$ angles. Extending the analysis of Ref. \cite{SO10}, one finds indeed that one configuration where $K_{1\tau}$ is
 smaller than unity is attained if $\rho=0$ (mod $2\pi$) and
 $\cos\sigma=-[1/(12\,\theta_{12}^L)]\,[m_{\rm sol}^4/(m_2^3\, m_3)]\sim -(5/12)(m_{\rm sol}/m_{\rm atm})\sim -10^{-1}$.
 This implies $\sigma\simeq \pi/2$ (mod $2\pi$), as confirmed by our numerical results.
Including in the
analysis the atmospheric neutrino mixing angle, as one can see from
the panel with the constraints in the $\theta_{13}-\theta_{23}$ plane,
only values $\theta_{23}\lesssim 48^{\circ}$ for $\theta_{13}\lesssim 10^{\circ}$
are allowed for $\a_2 \lesssim 4$. Notice that, for $\a_2 \leq 4$, the allowed region in  $m_1,\theta_{13},\theta_{23}$
only marginally overlaps, at small values of $\theta_{23}$,
with the region for the case $V_L=I$.
This means that a measurement of these three quantities can distinguish between the two
cases, $V_L=I$ and $V_L=V_{CKM}$, and not all combinations of these three quantities
seem to be possible. We will be back on this point in the next section.

In Figure 7 we  plotted the relevant quantities for three particular
choices of the parameters, as indicated in the figure caption, corresponding to
the three kinds of solutions found for  $V_L=V_{CKM}$.
\begin{figure}
\begin{center}
\psfig{file=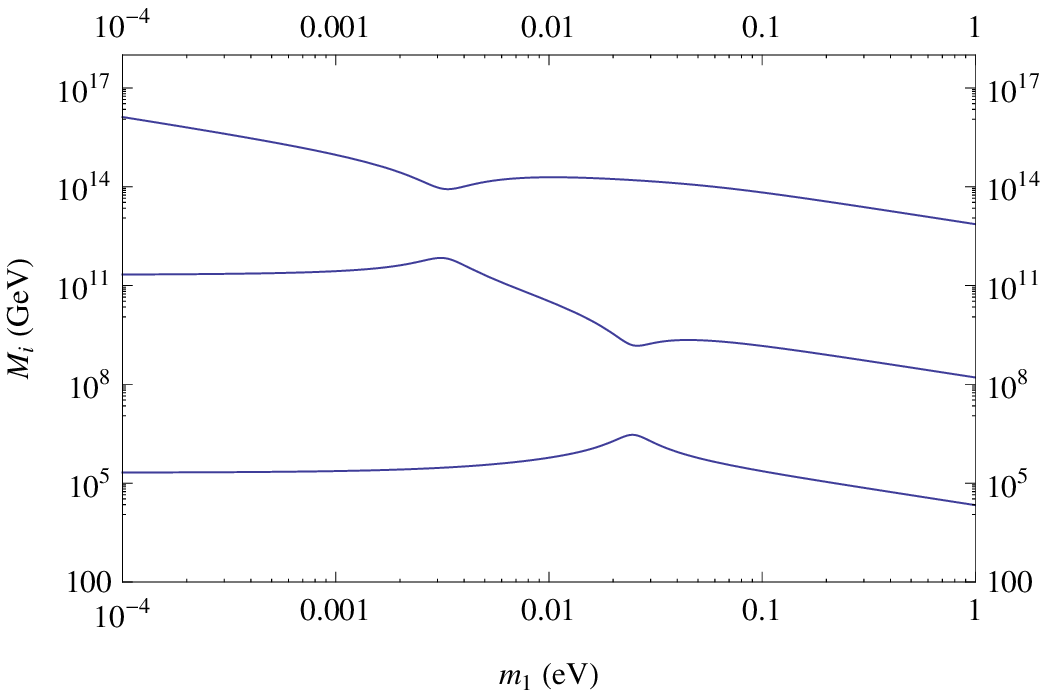,height=38mm,width=45mm}
\hspace{3mm}
\psfig{file=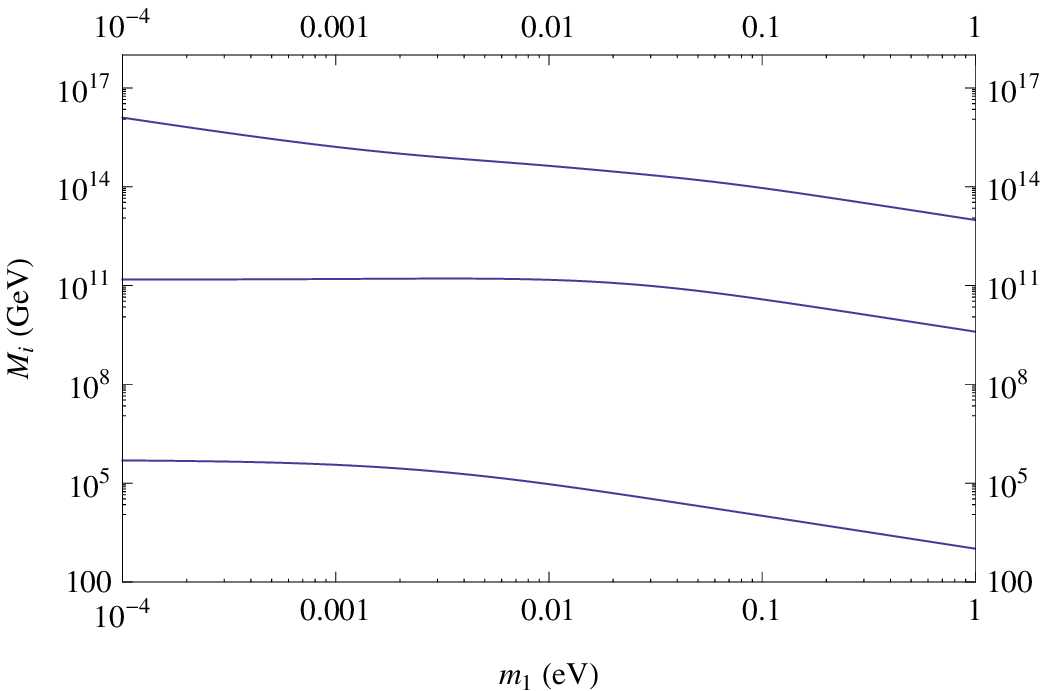,height=38mm,width=45mm}
\hspace{3mm}
\psfig{file=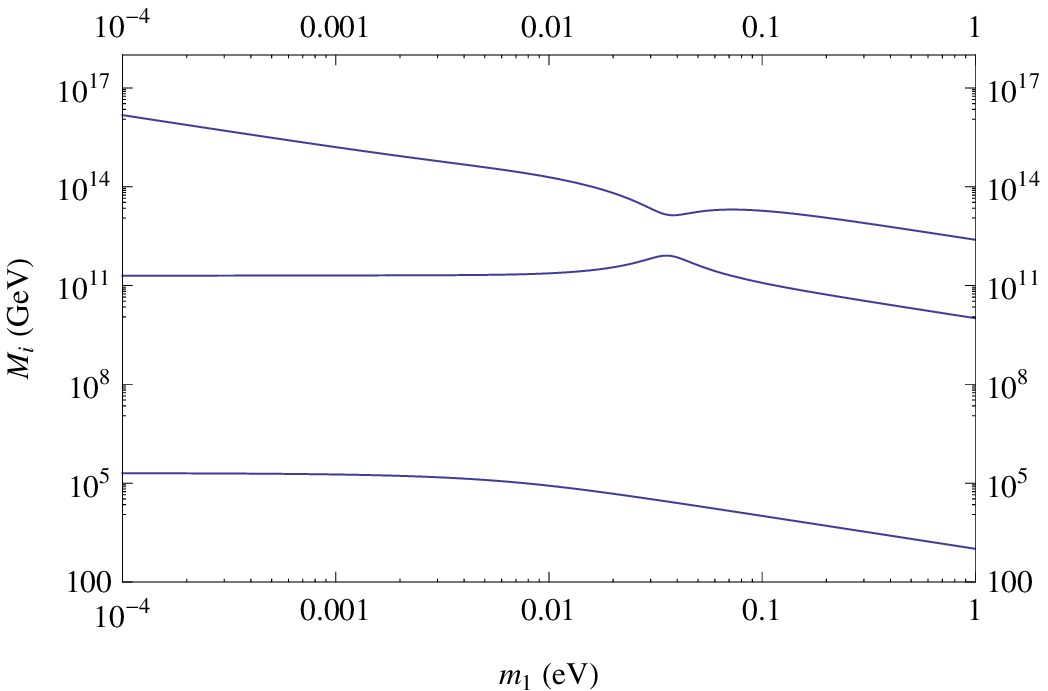,height=38mm,width=45mm} \\
\psfig{file=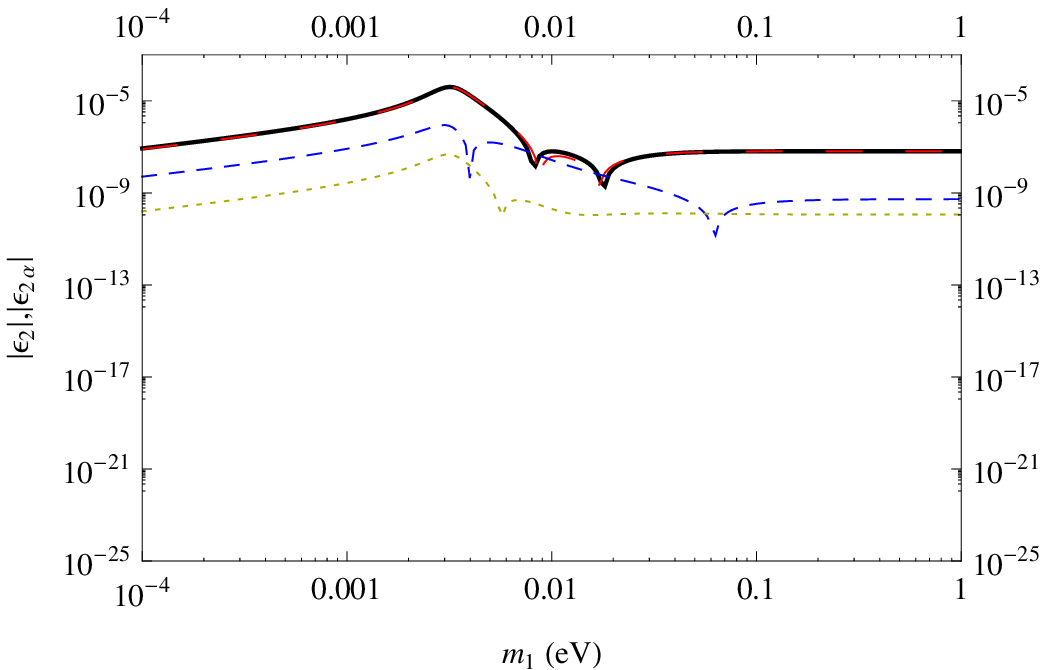,height=38mm,width=45mm}
\hspace{3mm}
\psfig{file=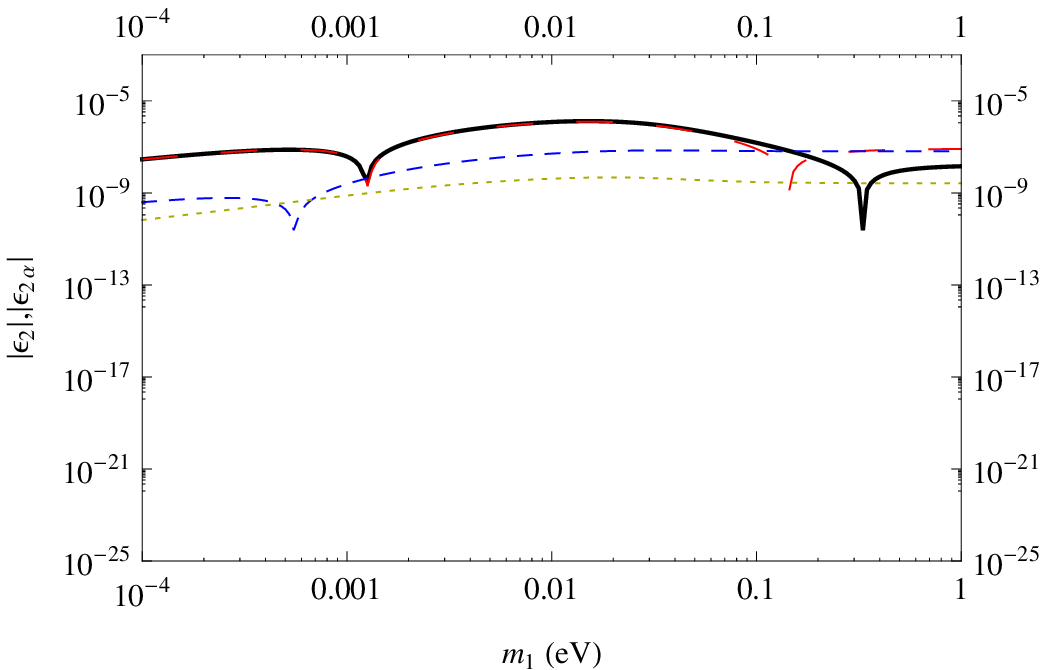,height=38mm,width=45mm}
\hspace{3mm}
\psfig{file=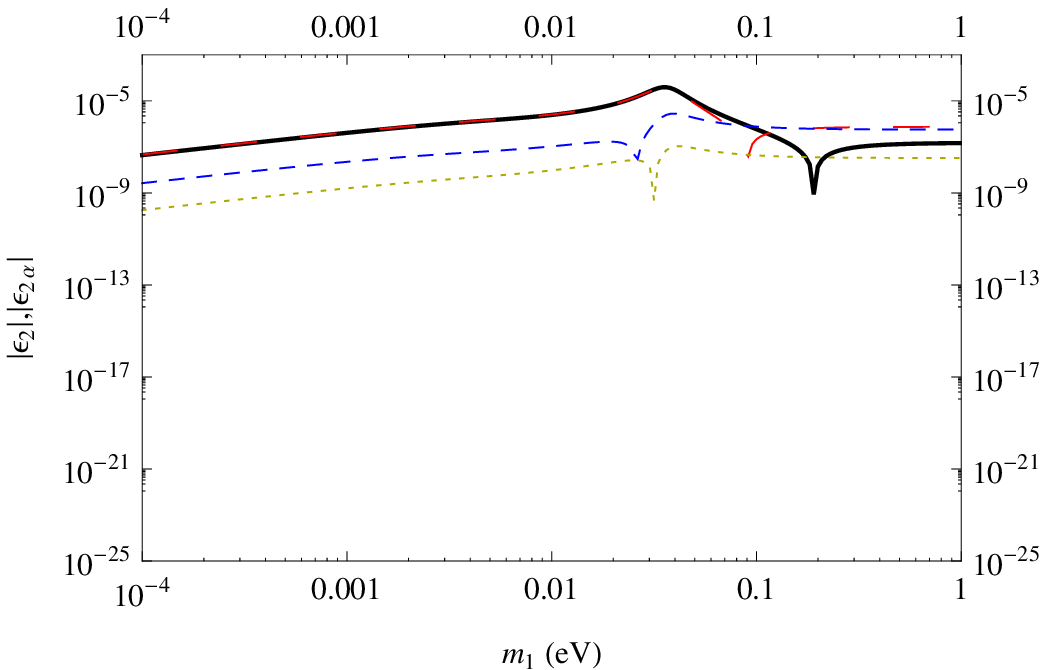,height=38mm,width=45mm} \\
\psfig{file=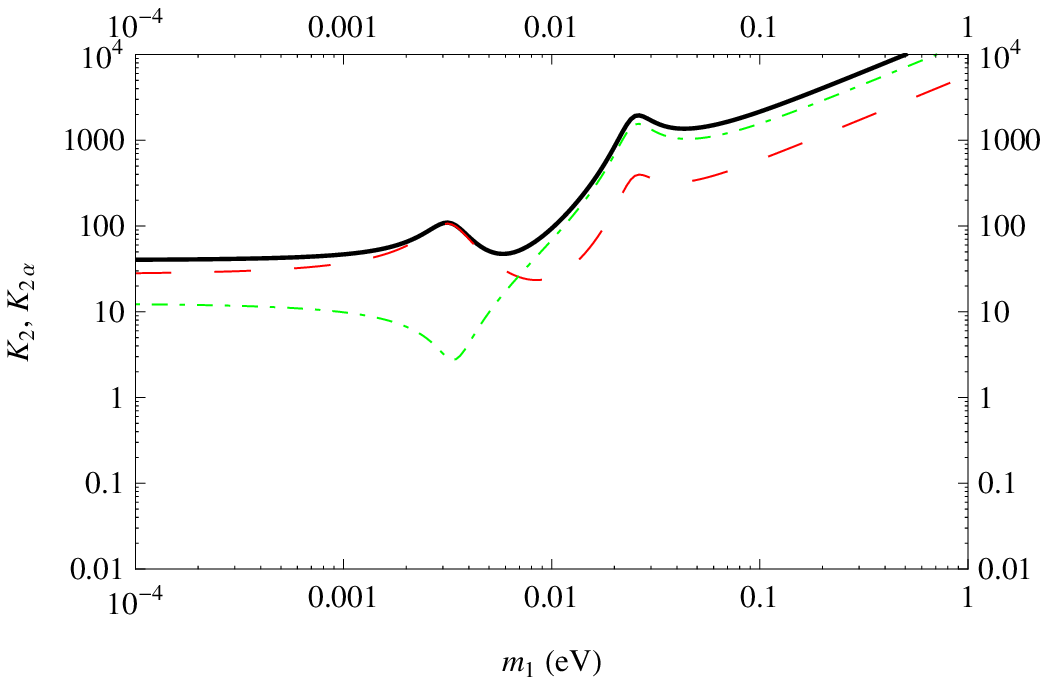,height=38mm,width=45mm}
\hspace{3mm}
\psfig{file=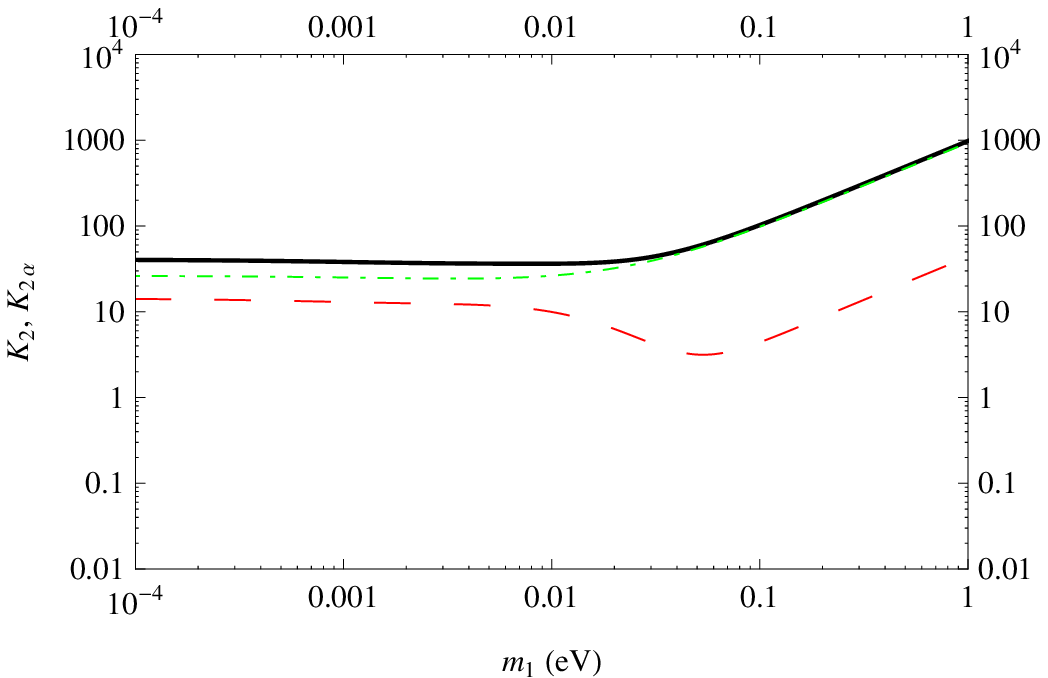,height=38mm,width=45mm}
\hspace{3mm}
\psfig{file=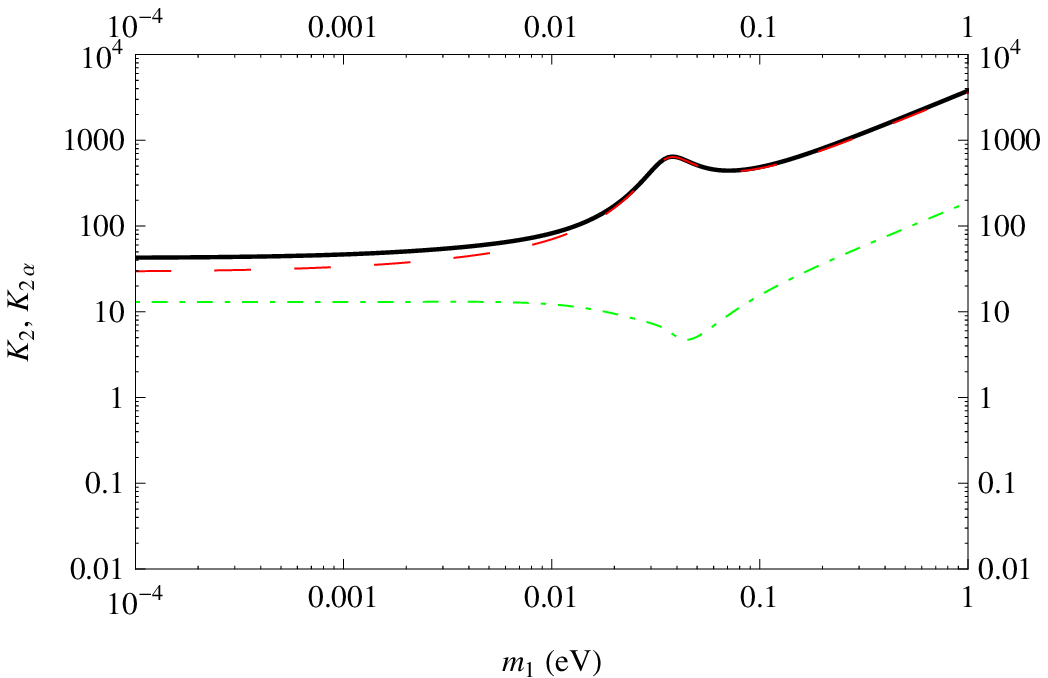,height=38mm,width=45mm} \\
\psfig{file=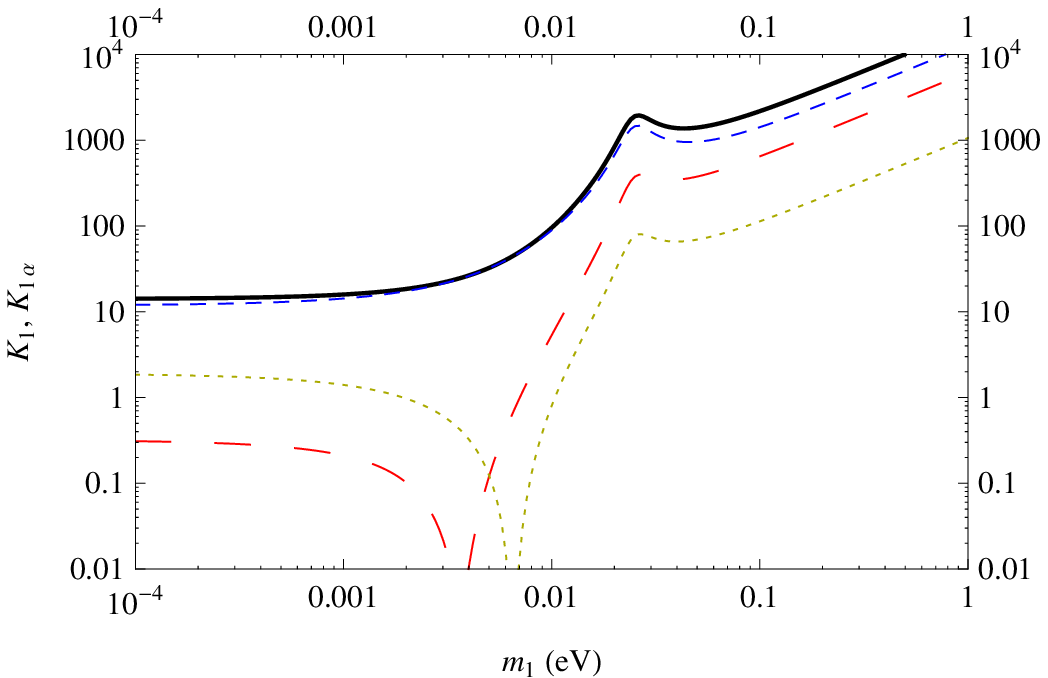,height=38mm,width=45mm}
\hspace{3mm}
\psfig{file=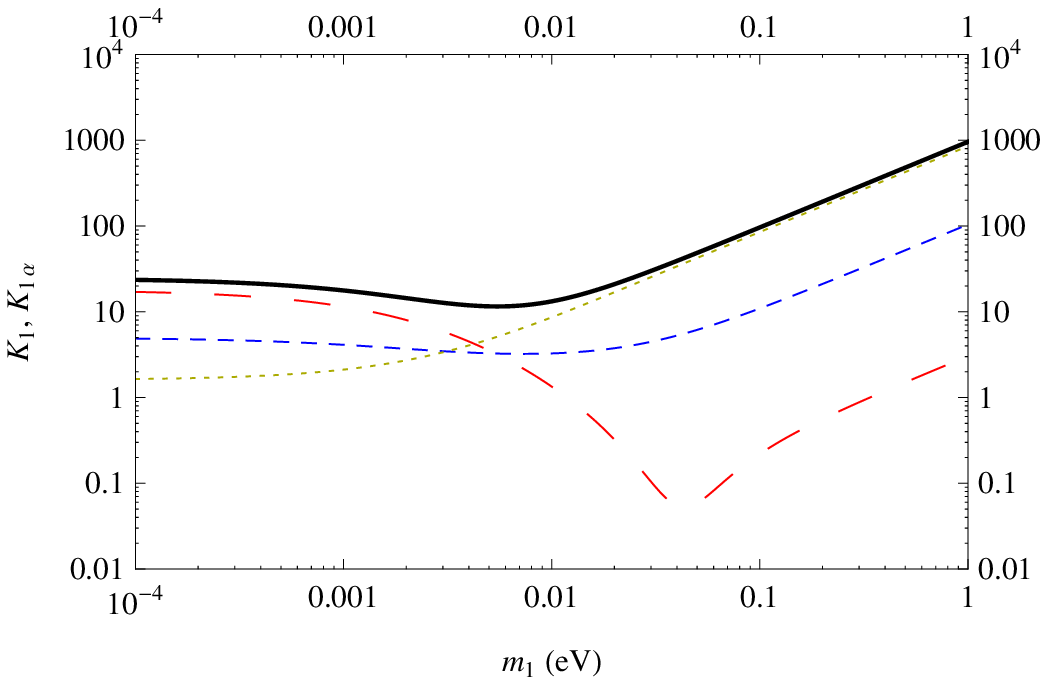,height=38mm,width=45mm}
\hspace{3mm}
\psfig{file=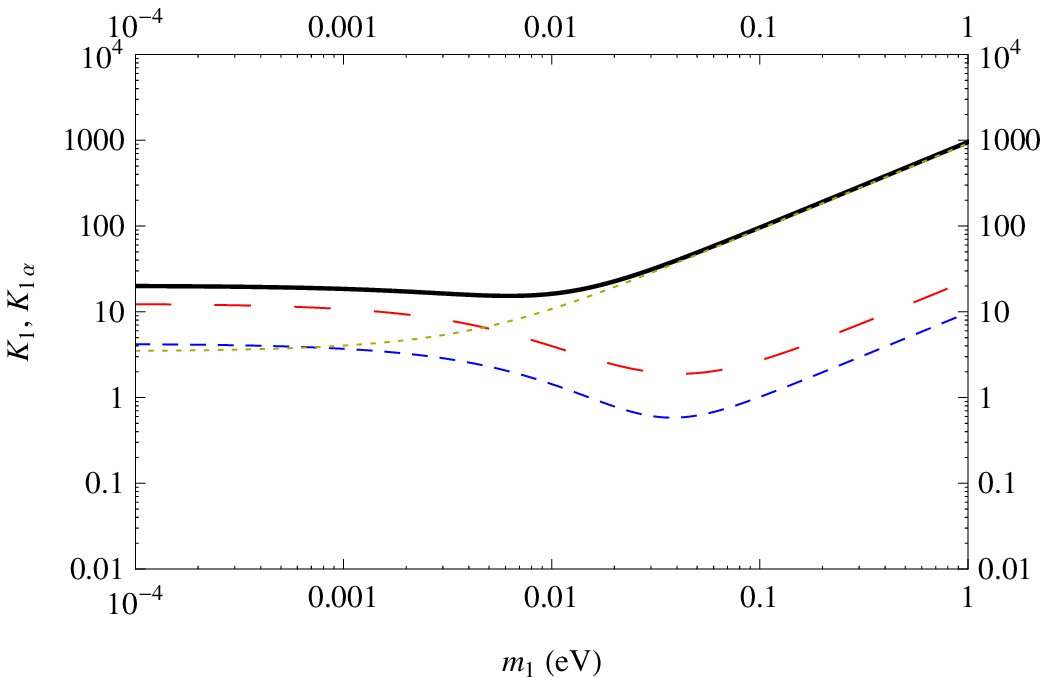,height=38mm,width=45mm} \\
\psfig{file=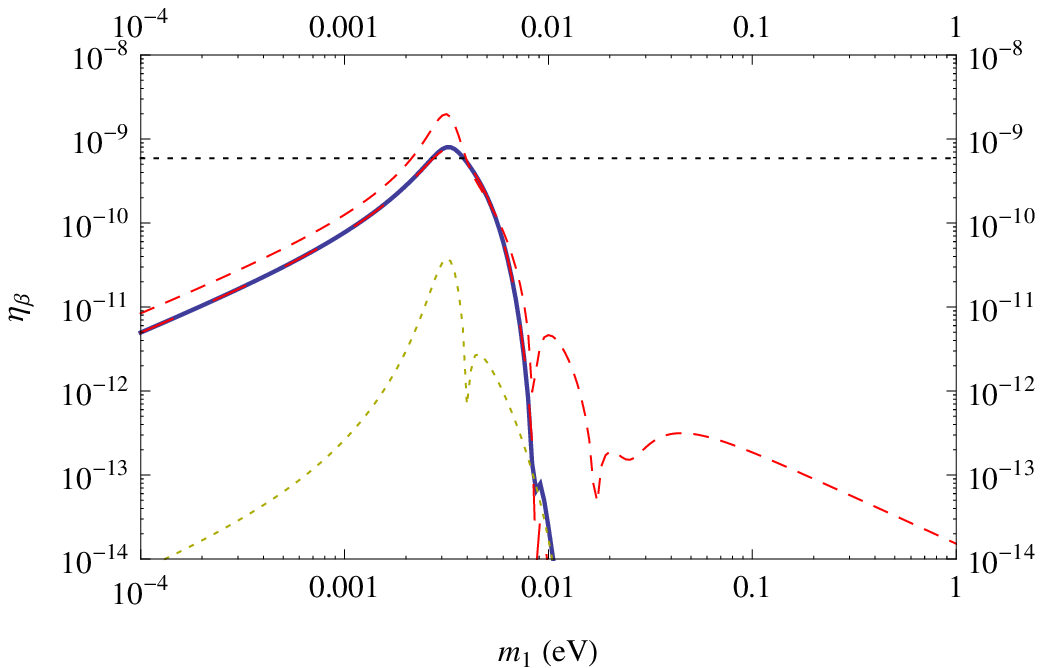,height=38mm,width=45mm}
\hspace{3mm}
\psfig{file=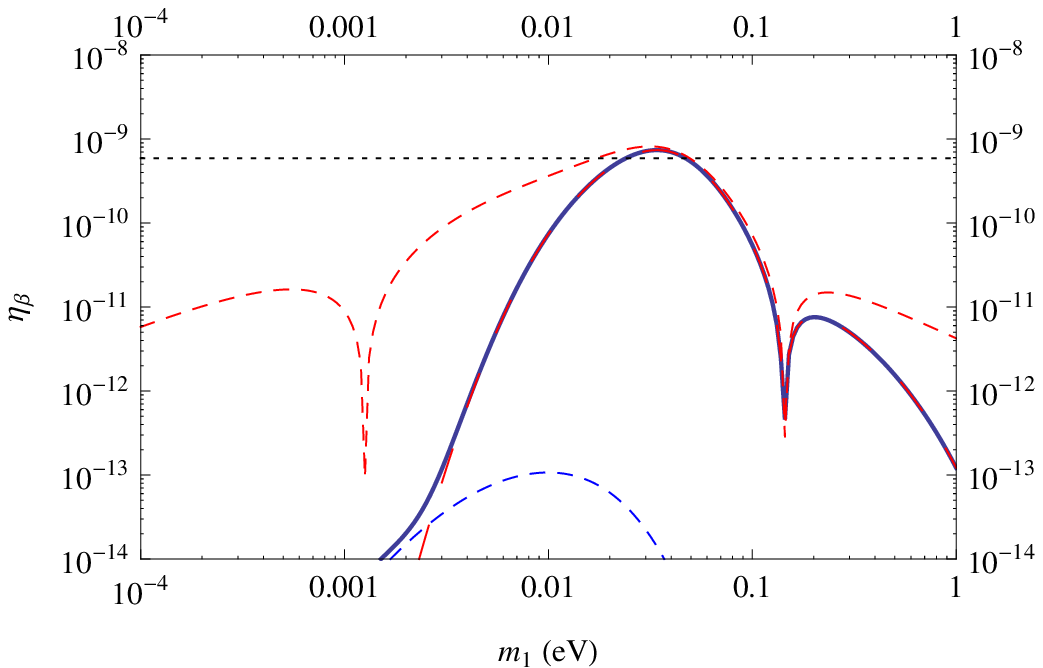,height=38mm,width=45mm}
\hspace{3mm}
\psfig{file=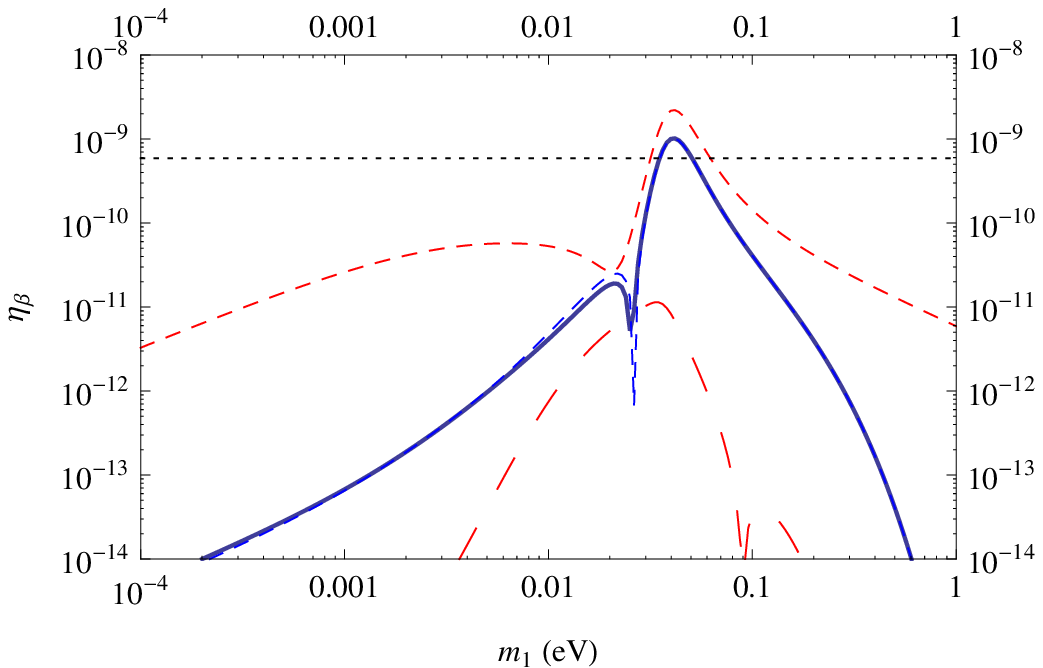,height=38mm,width=45mm}
\end{center}
\vspace{-8mm}
\caption{Case $V_L=V_{CKM}$, NO. Plots of the relevant quantities for three
choices of the involved parameters. The long-dashed red lines correspond to $\a=\t$,
the dashed blue lines to $\a=\m$ and the short-dashed dark yellow lines to $\a=e$.
Left panels:
$\a_2=4$, $\theta_{13}=1.7^{\circ}$, $\theta_{12}=33.6^{\circ}$,
$\theta_{23}=41.8^{\circ}$, $\d=2.84$, $\rho=$1.53 $\sigma=3.24$, $\rho_L=0.12$, $\sigma_L=2.56$;
central panels: $\a_2=5$, $\theta_{13}=3.3^{\circ}$, $\theta_{12}=35.6^{\circ}$,
$\theta_{23}=40.4^{\circ}$, $\d=-1.06$, $\rho=2.87$, $\sigma=6.0$, $\rho_L=3.13$, $\sigma_L=3.25$;
right panels:
$\a_2=4$, $\theta_{13}=4.7^{\circ}$, $\theta_{12}=35.9^{\circ}$,
$\theta_{23}=40.3^{\circ}$, $\d=-1.89$, $\rho=0.065$, $\sigma=4.85$, $\rho_L=5.89$, $\sigma_L=3.69$.}
\end{figure}
These three sets of values correspond to the three kinds of solutions that are
found in the scan plots. The first two sets, corresponding to the left and central
panels, give a tauon dominated asymmetry, while the third set, corresponding to the right panels,
yields a muon dominated asymmetry. Notice that these three kinds of
solutions are the same three kinds, with slight modifications, found for the
case $V_L=I$. However, one can see that this time the third kind of solution,
where the final asymmetry is muon dominated,
also yields successful leptogenesis.
The major difference that explains this result, is that for $V_L=V_{CKM}$ the flavoured $C\!P$
asymmetries $\ve_{2\a}$ are not as hierarchical as in the case $V_L=I$, as it can be clearly
seen in the three panels showing the $C\!P$ asymmetries in Figure 7.

\begin{figure}
\begin{center}
\psfig{file=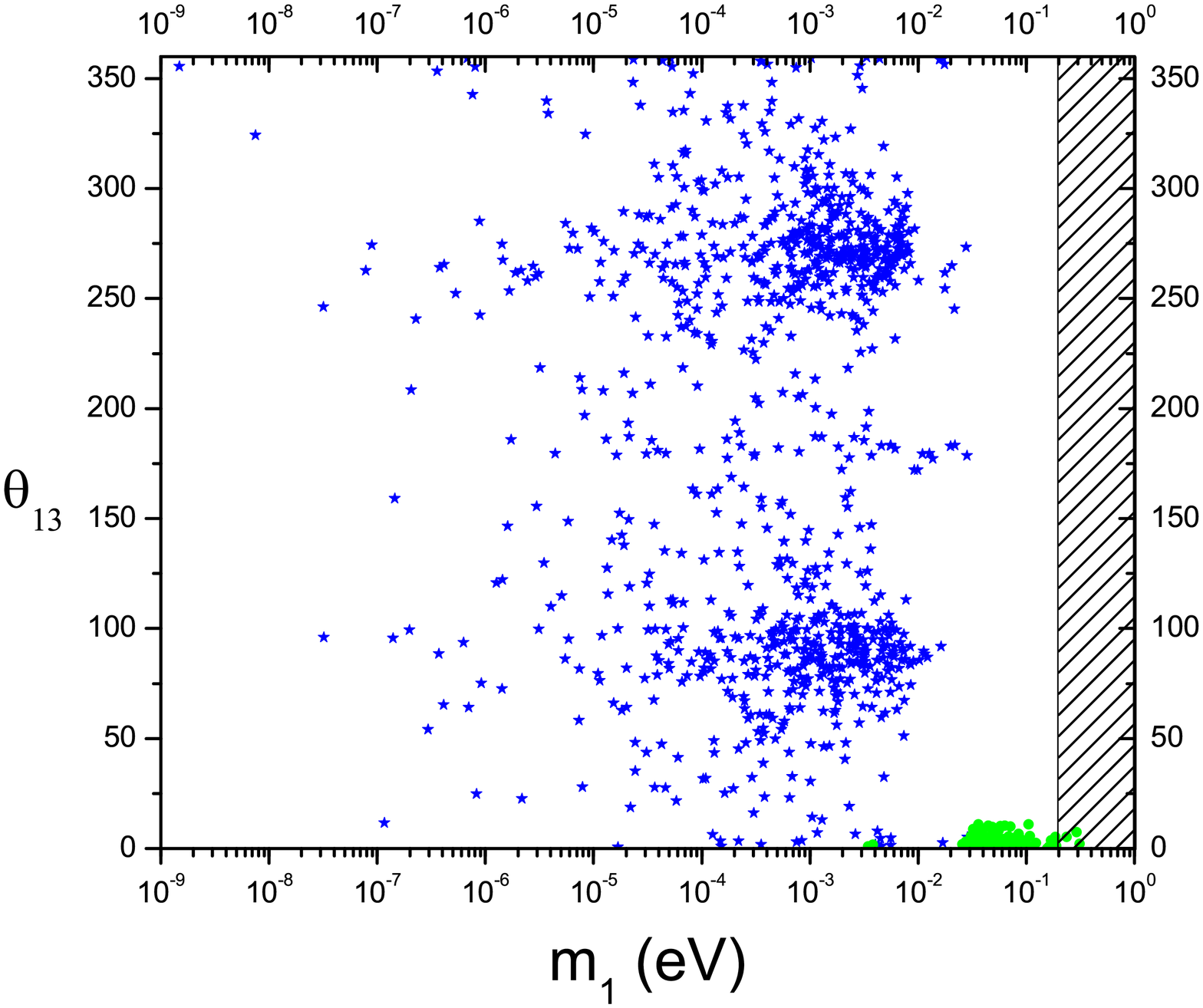,height=48mm,width=54mm}
\hspace{-4mm}
\psfig{file=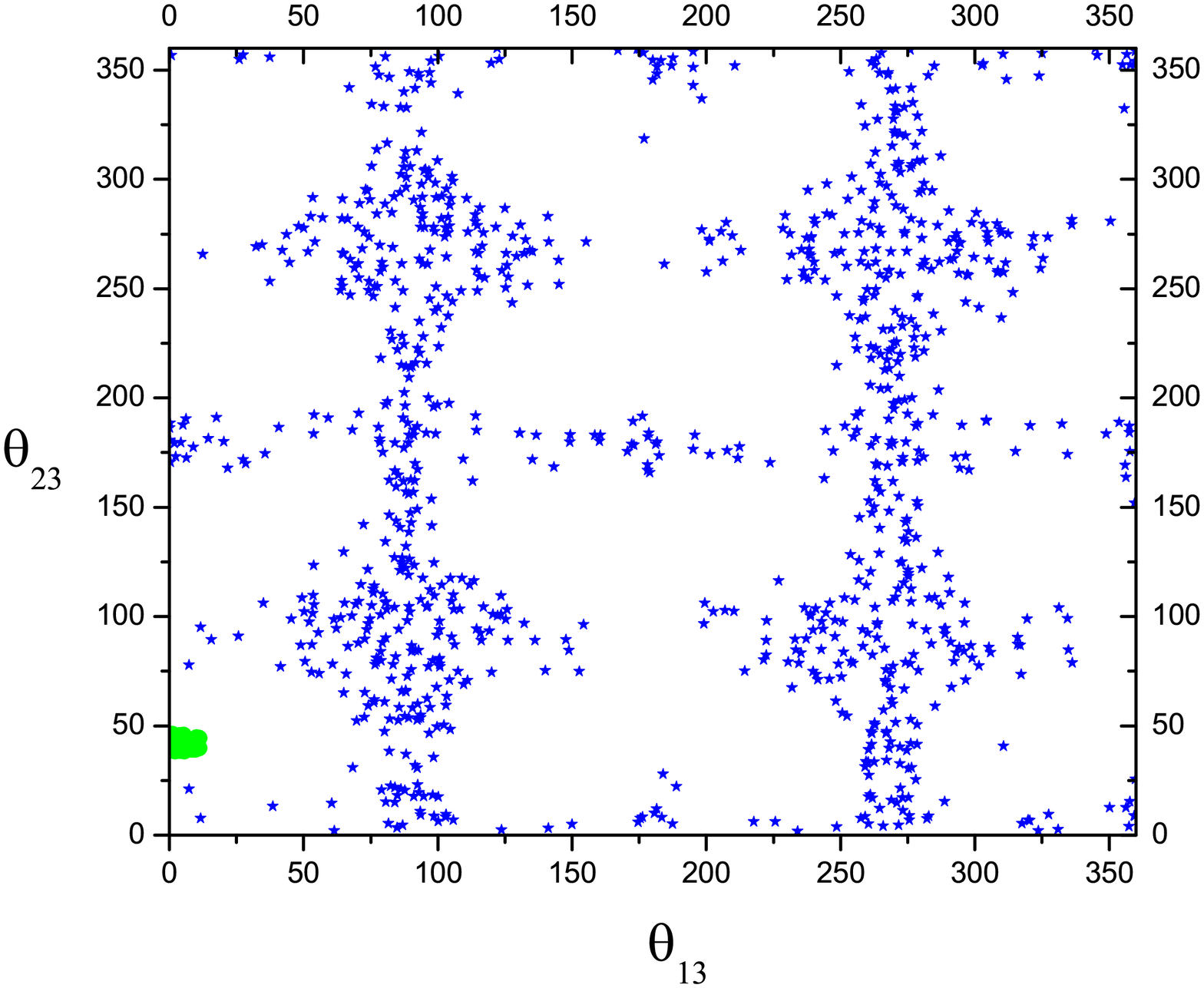,height=48mm,width=54mm}
\hspace{-4mm}
\psfig{file=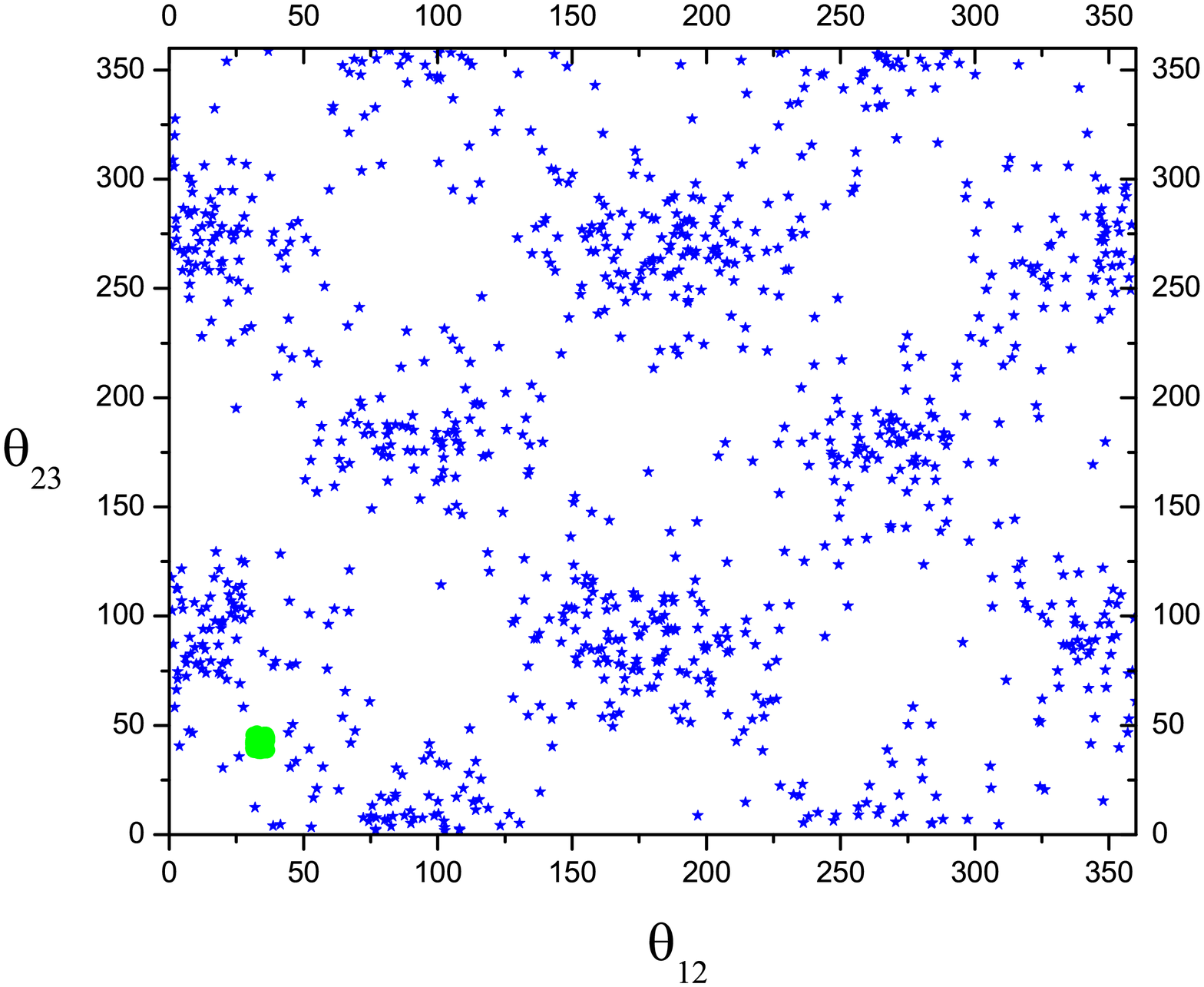,height=48mm,width=54mm}
\end{center}
\caption{Case $V_L=V_{CKM}$, NO. Constraints on the mixing angles without making use of the current
experimental information from neutrino oscillation experiments.}
\label{thijarb2}
\end{figure}

We have also repeated, as for the case $V_L=I$, the exercise to leave
 the mixing angles completely free, without imposing any experimental constraint
 finding the results shown in Fig.~\ref{thijarb2}.
One can see that in this case the points found when the current experimental
constraints are imposed (the green points) fall in more marginally allowed regions,
also for $\theta_{13}$. This might suggest that $V_L=I$ seems to be
a more attractive case than $V_L=V_{CKM}$.

\subsection{Inverted ordering}

Finally, we also present in Figure~9 the constraints obtained for IO.
Even though there is again a remarkable
suppression of the allowed regions compared to NO, they are
somehow less restrictive than for $V_L=I$. In particular
now a broader range of values for $m_1$ is allowed and $\theta_{23}$
can be as low as $\simeq 45^{\circ}$ for $\a_2 \leq 4$.
\begin{figure}
\begin{center}
\psfig{file=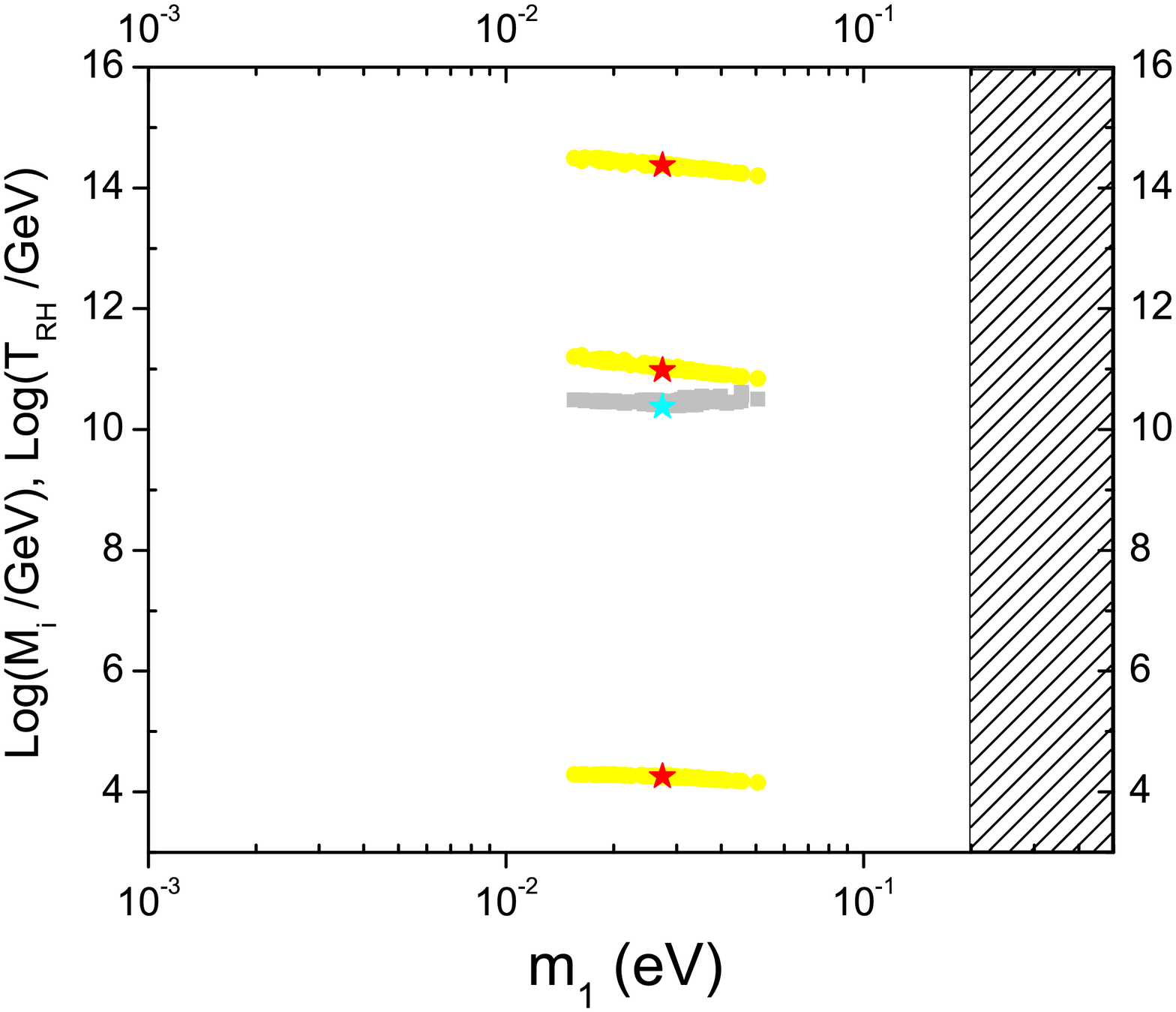,height=48mm,width=54mm}
\hspace{-4mm}
\psfig{file=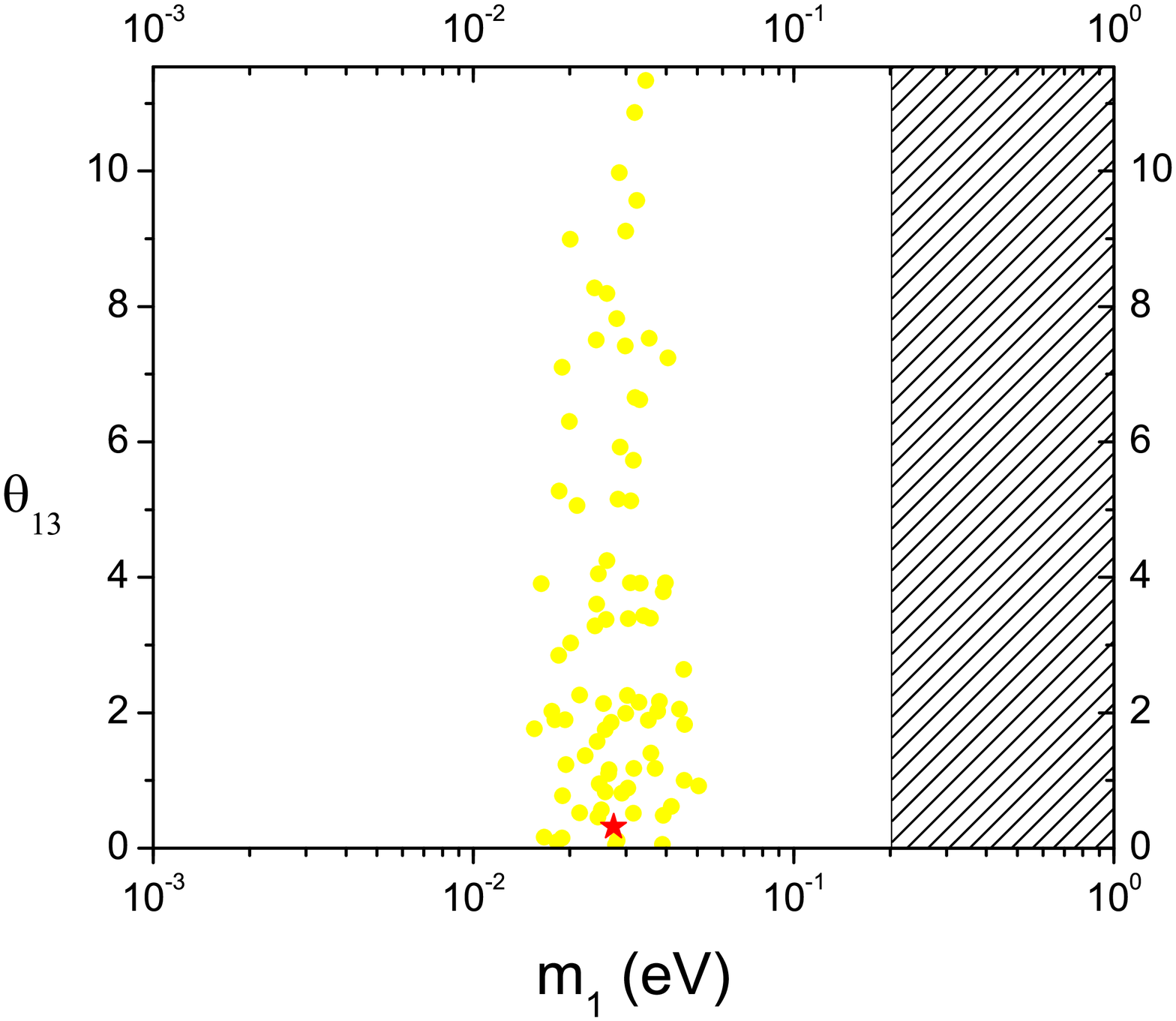,height=48mm,width=54mm}
\hspace{-4mm}
\psfig{file=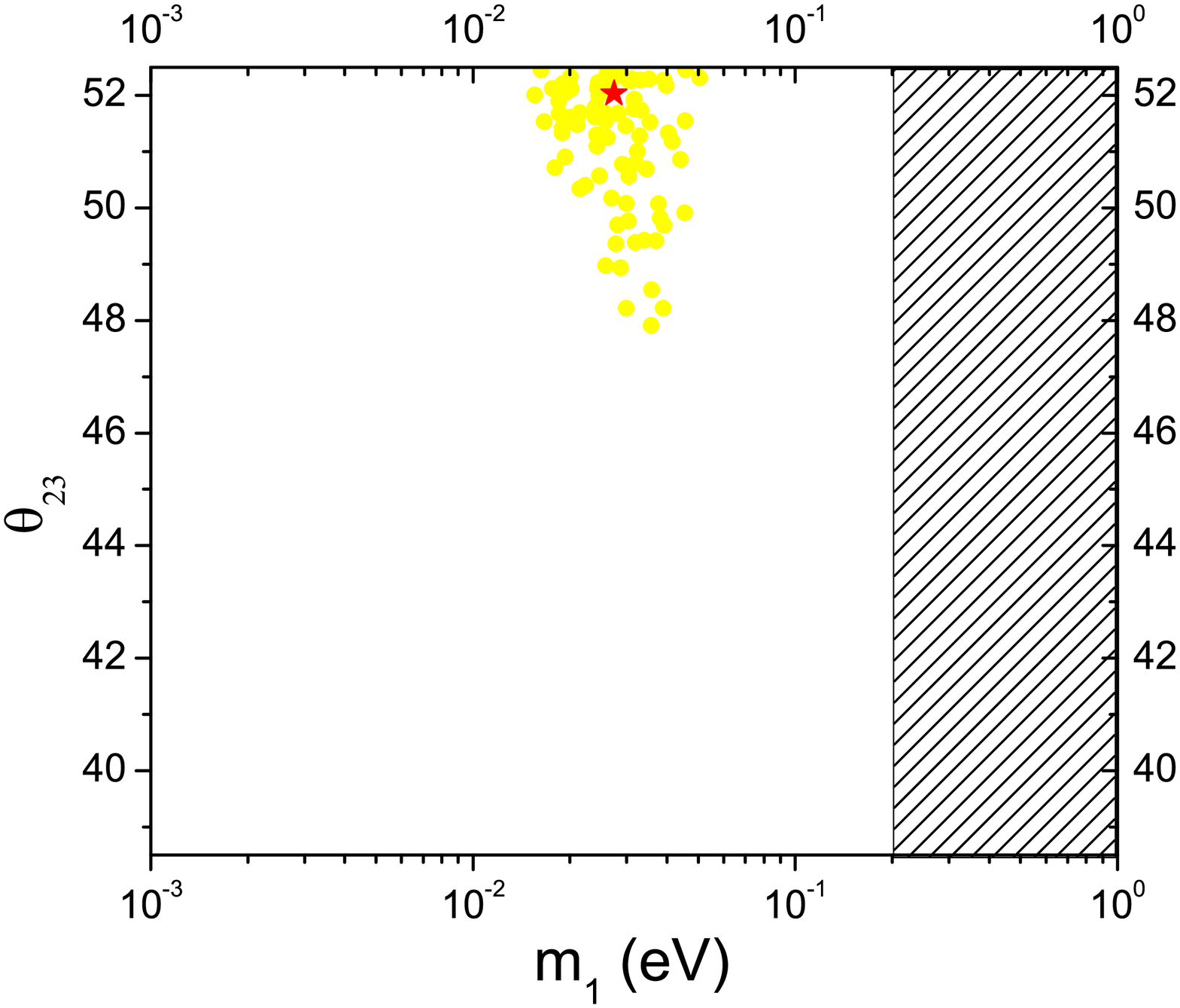,height=48mm,width=54mm} \\
\psfig{file=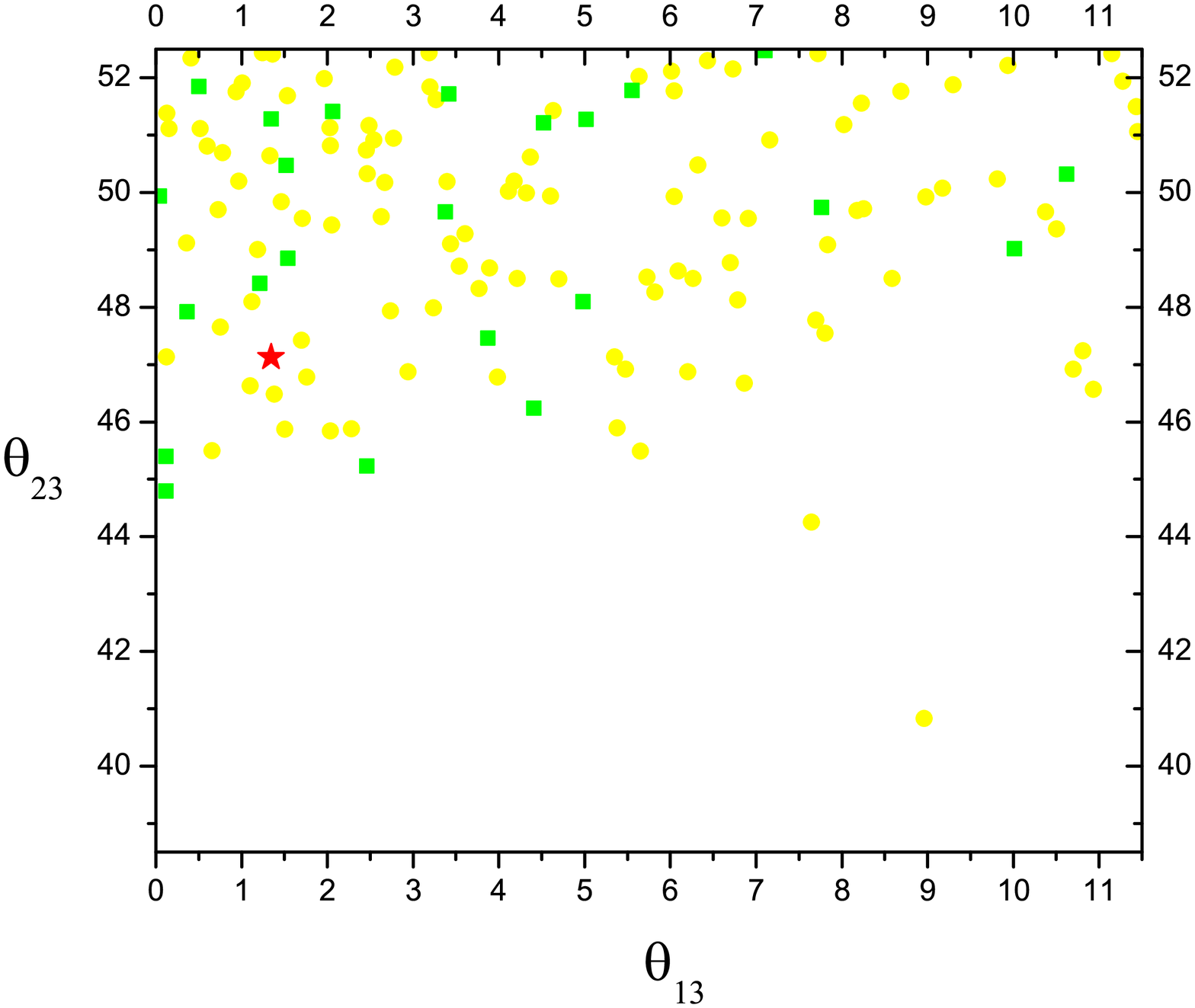,height=48mm,width=54mm}
\hspace{-4mm}
\psfig{file=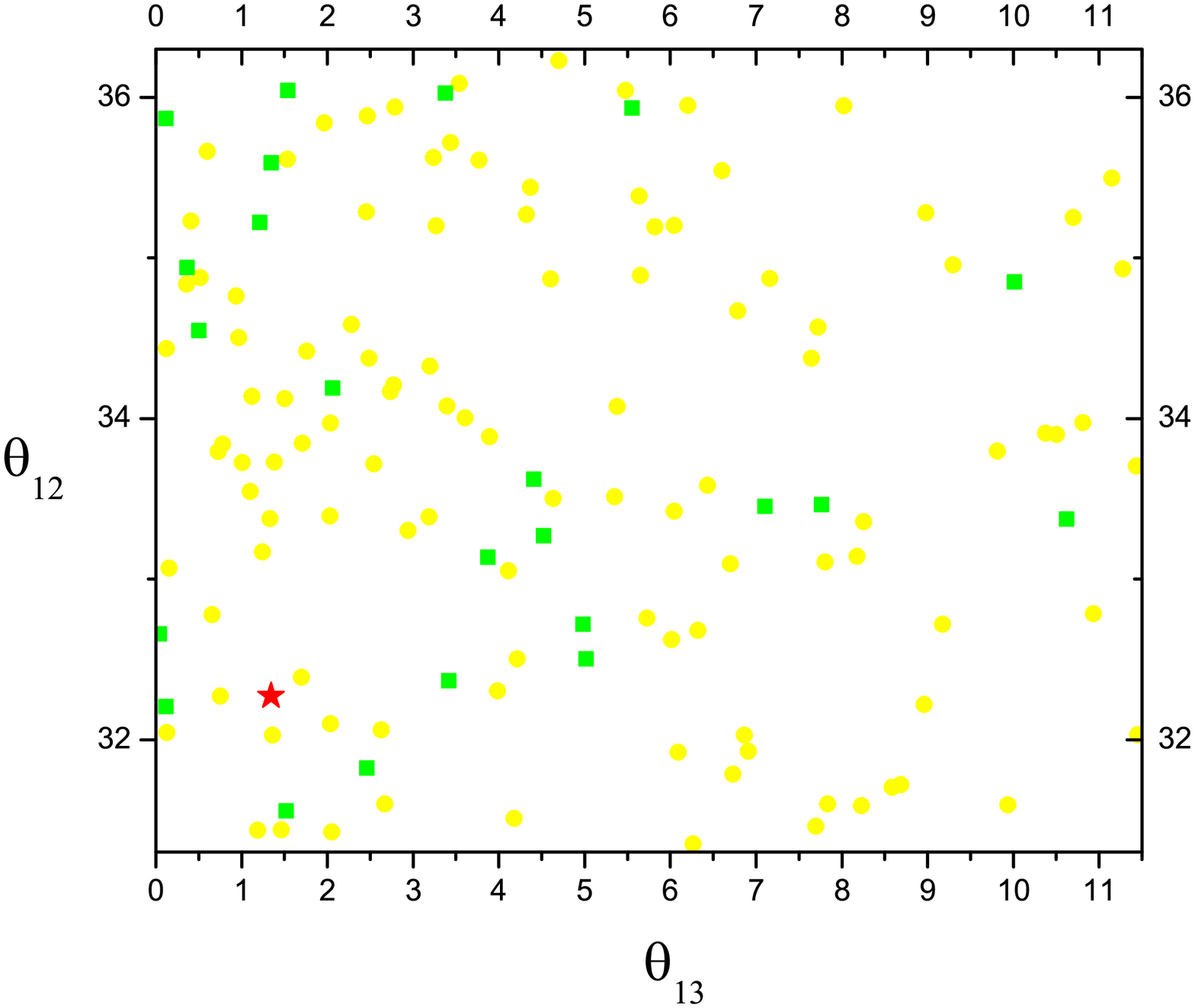,height=48mm,width=54mm}
\hspace{-4mm}
\psfig{file=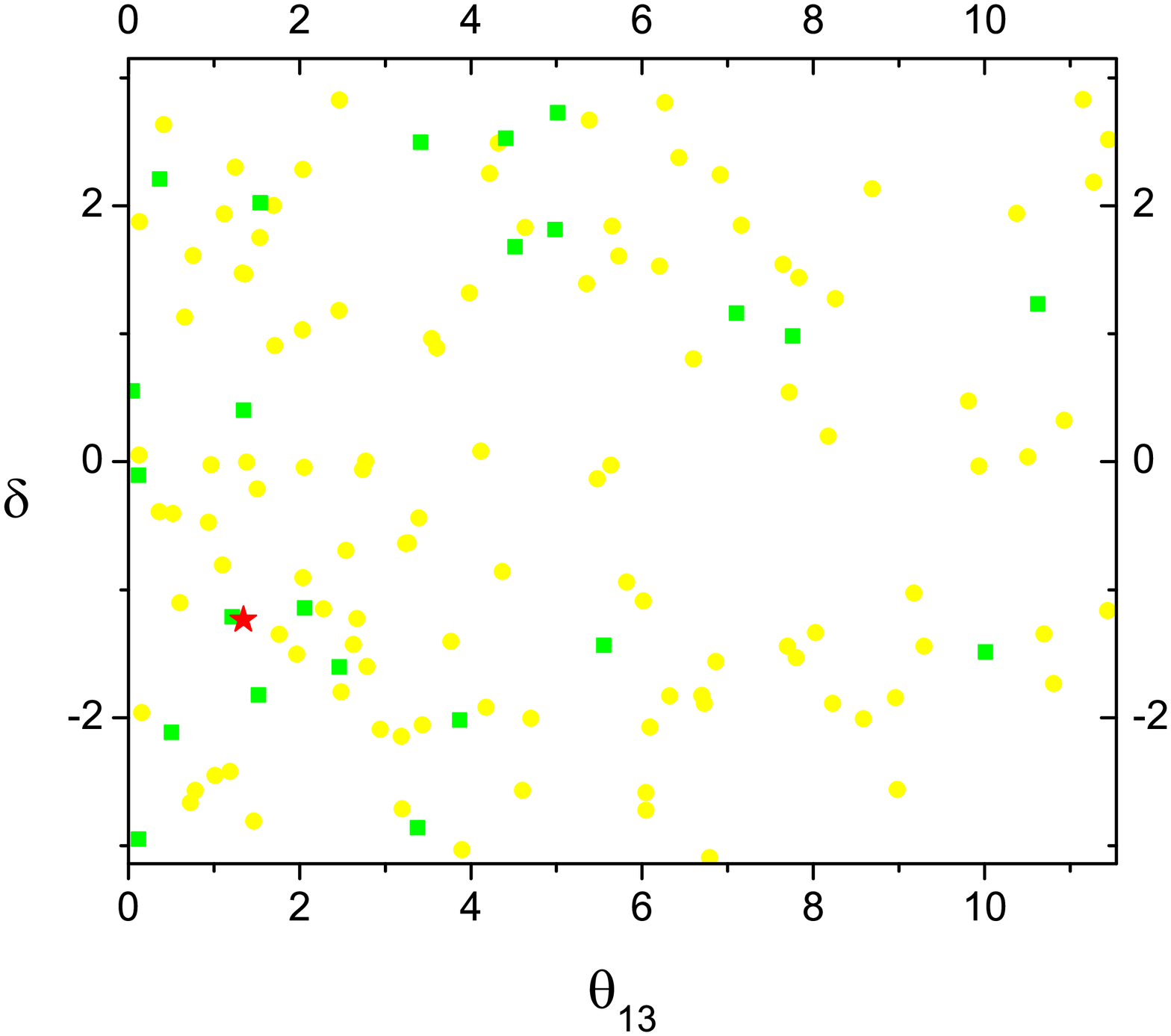,height=48mm,width=54mm} \\
\psfig{file=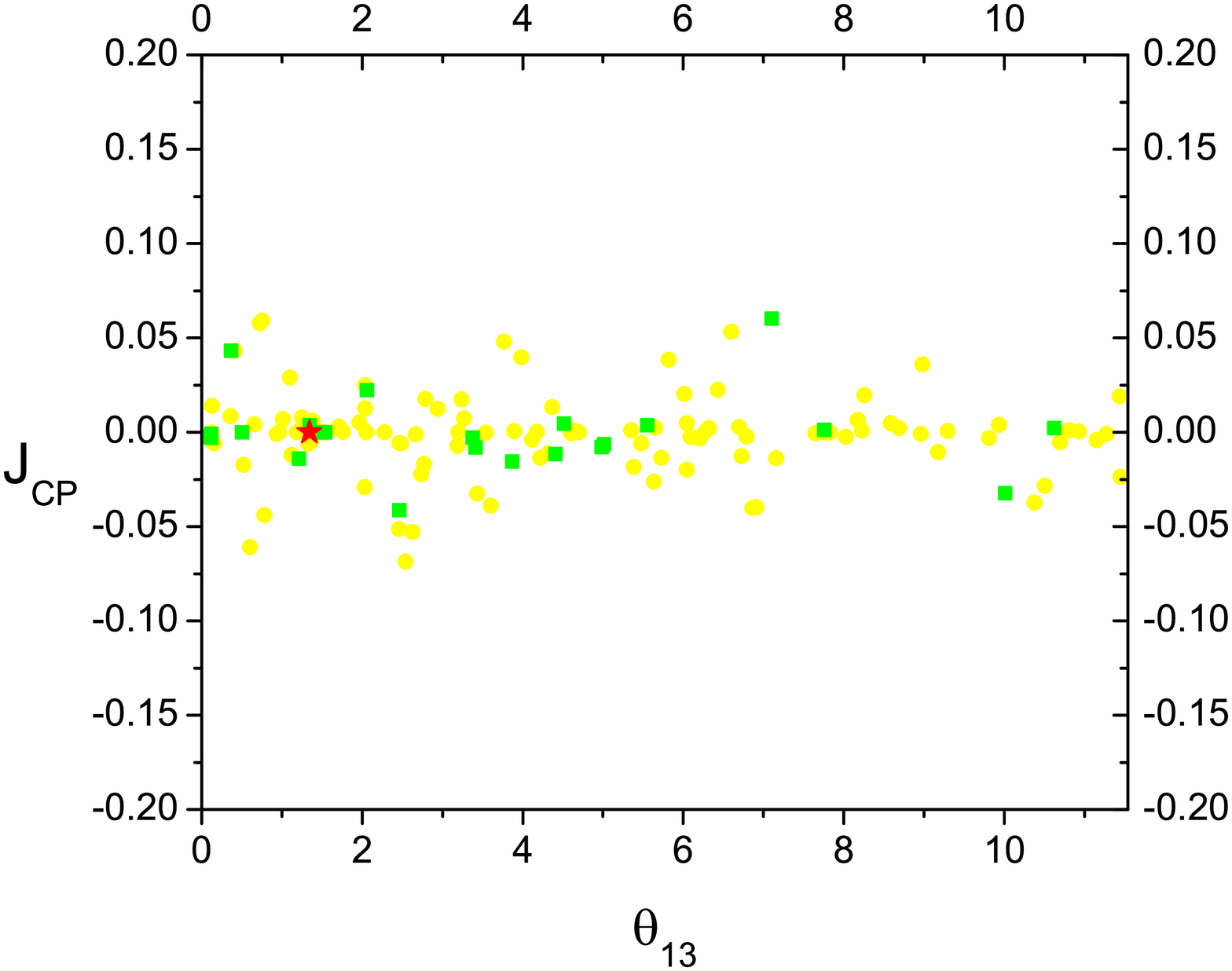,height=48mm,width=54mm}
\hspace{-4mm}
\psfig{file=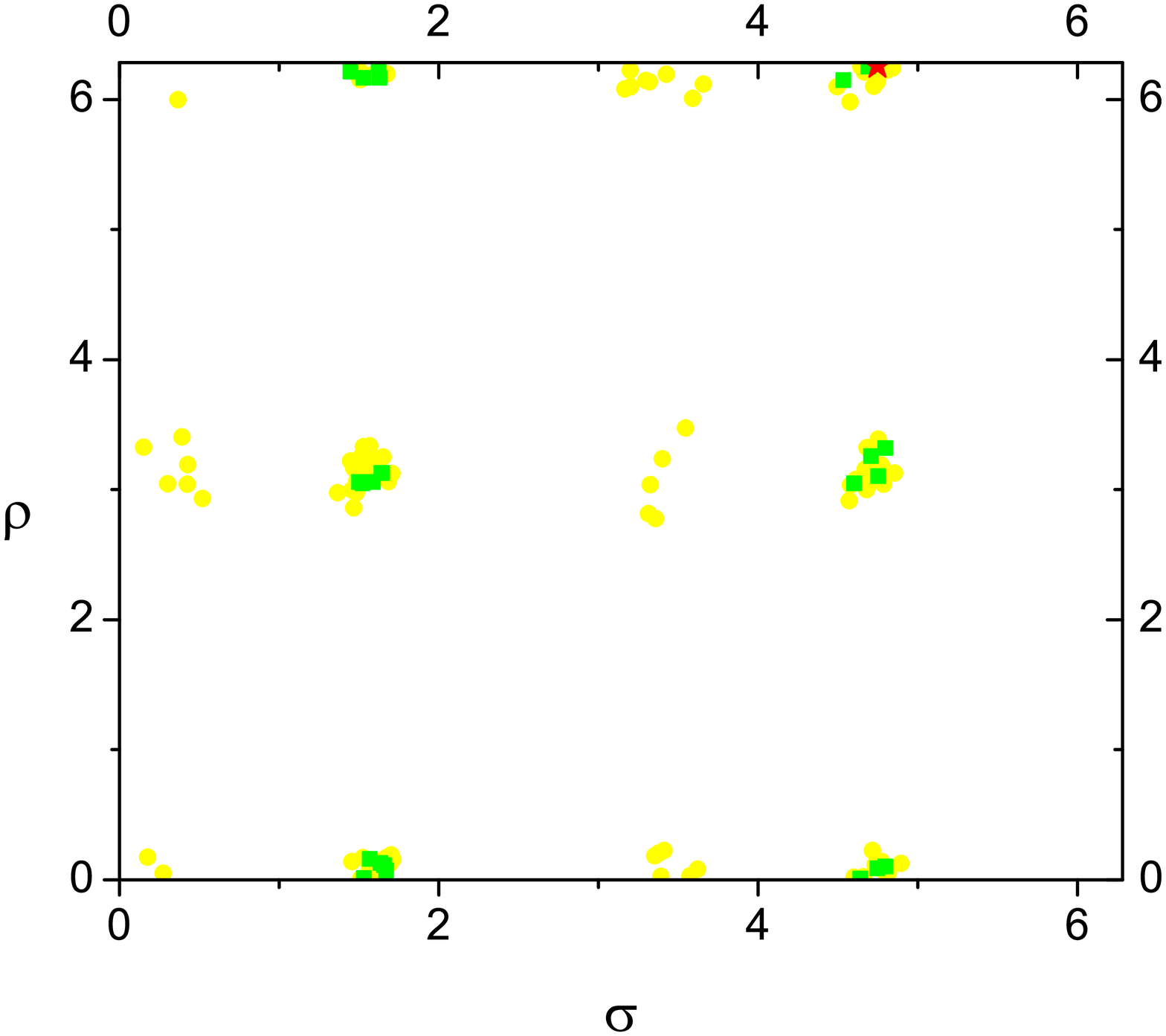,height=48mm,width=54mm}
\hspace{-4mm}
\psfig{file=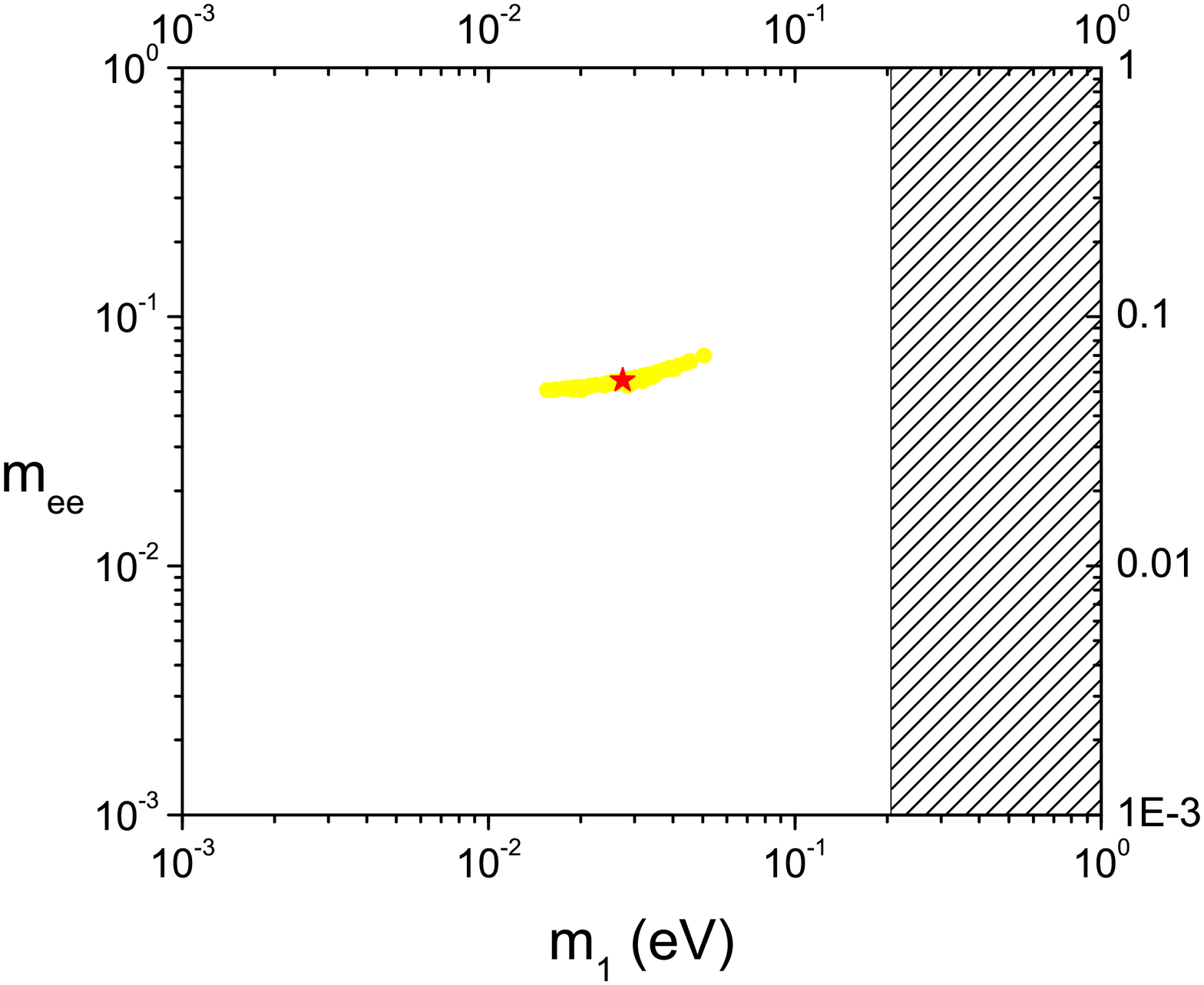,height=48mm,width=54mm}
\end{center}
\caption{Case $V_L=V_{CKM}$, IO. Scatter plot of points in the parameter space that satisfy
the condition $\eta_B>5.9\times 10^{-9}$ for $\a_2=5$ (yellow circles),
$\a_2=4$ (green squares) and $\a_2= 2$ (red star).}
\end{figure}
This is also confirmed by the fact that lowest allowed value is now $\a_2=2$, much lower
than in the case $V_L=I$ (it was $\a_2=4.65)$. However, it is still fair to say
that the IO case is only marginally allowed and certainly
disfavoured compared to the NO case.

%%%%%%%%%%%%%%%%%%%%%%
\section{Global scans}
%%%%%%%%%%%%%%%%%%%%%%

The two specific cases that we discussed, $V_L=I$ and $V_L=V_{CKM}$,
suggest an interesting sensitivity of $SO(10)$-inspired leptogenesis to slight deviations
of $V_L$ from the identity. This sensitivity was
absent in the results found in $N_1$-dominated leptogenesis \cite{branco}.
In this  way it seems that one could even gain some information
on $V_L$ from low energy neutrino experiments.
However, there is a potentially dangerous aspect of such a sensitivity: if for a slight variation of $V_L$
the entire space of low energy neutrino parameters becomes accessible, then
any chance to test $SO(10)$-inspired leptogenesis is lost.
On the other hand, from a comparison of the results obtained for the two definite cases,
$V_L=I$ and $V_L=V_{CKM}$, one can understand that this does not happen.

One can still suspect that for a continuous variation of the
parameters in $V_L$, such that $V_L$ changes from $V_L=I$ to $V_L=V_{CKM}$,
new solutions appear so that any point in the space of the low energy neutrino parameters
can be obtained for a proper choice of $V_L$.

In this section we study this issue.
We perform a global continuous scan of the parameters for $V_L$ between $V_L=I$
and $V_L=V_{CKM}$. Obviously a precise limit $V_L=V_{CKM}$ for such a global
scan is somehow arbitrary. It should be therefore taken as
a working assumption defining $SO(10)$-inspired leptogenesis, even more
than the condition $\a_i={\cal O}(1)$ that, as we stressed many times,
should not be regarded as a very restrictive assumption.
Clearly within well defined realistic $SO(10)$ models,
more specific conditions on $V_L$ should be obtained. In any case
one expects that if the $V_L$ satisfies the condition $I \leq V_L \leq V_{CKM}$,
then the allowed values for the low energy parameters
should fall in  the allowed regions   for  $SO(10)$-inspired leptogenesis.

Therefore, in this Section we present
the constraints on the low energy neutrino parameters
for a continuous variation of the values of the mixing
angles $\theta_{ij}^L$ in the range $0 \leq \theta_{ij}^L \leq \theta_{ij}^{CKM}$
(i.e. for $I \leq V_L \leq V_{CKM}$). More explicitly the shown scatter plots
are obtained for the low energy neutrino parameters scanned over exactly the same ranges as
for the case $V_L=I$. The three angles in $V_L$ are scanned over the
ranges $0 \leq  \theta_{13}^L \leq 0.2^{\circ}$,  $0 \leq  \theta_{13}^L \leq 2.5^{\circ}$,
$0\leq \theta_{12}^L \leq 13^{\circ}$, while the three phases are scanned over $[0,2\pi]$.
In order to determine the allowed regions,
we have followed the same strategy as in the case $V_L=I$, with a similar
total number of scanned points, ${\cal O}(10^7)$.

\subsection{Normal ordering}

The results for NO are shown in figure~10. One can see how
the allowed regions are approximately given by a super-position of those
 found for $V_L=I$ and $V_L=V_{CKM}$ plus all
intermediate solutions. The result is that now the correlations
among the parameters found in the two special cases seem to disappear.
There are however still interesting non trivial constraints.
What clearly survives is
that the allowed points still cluster within two distinguished ranges of values for $m_1$,
one range at small values, $m_1 \simeq (1-5)\times 10^{-3}\,{\rm eV}$, and one range at high values,
$m_1 \simeq 0.03-0.1 \,{\rm eV}$, a distinction that is sharp for $\alpha_2=4$ (green squares)
while it is softer for $\a_2=5$ (yellow circles).
 \begin{figure}
\begin{center}
\psfig{file=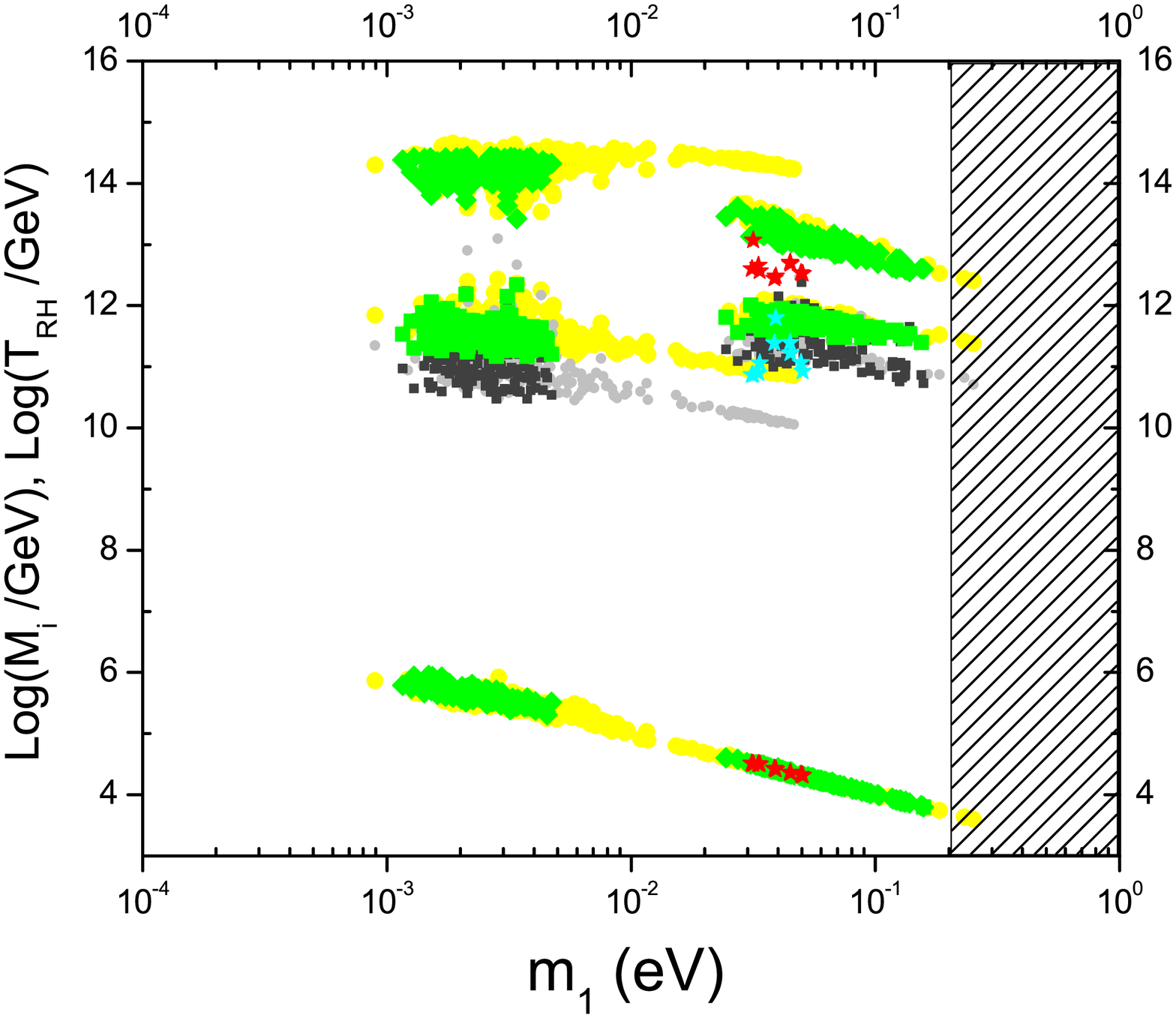,height=48mm,width=54mm}
\hspace{-4mm}
\psfig{file=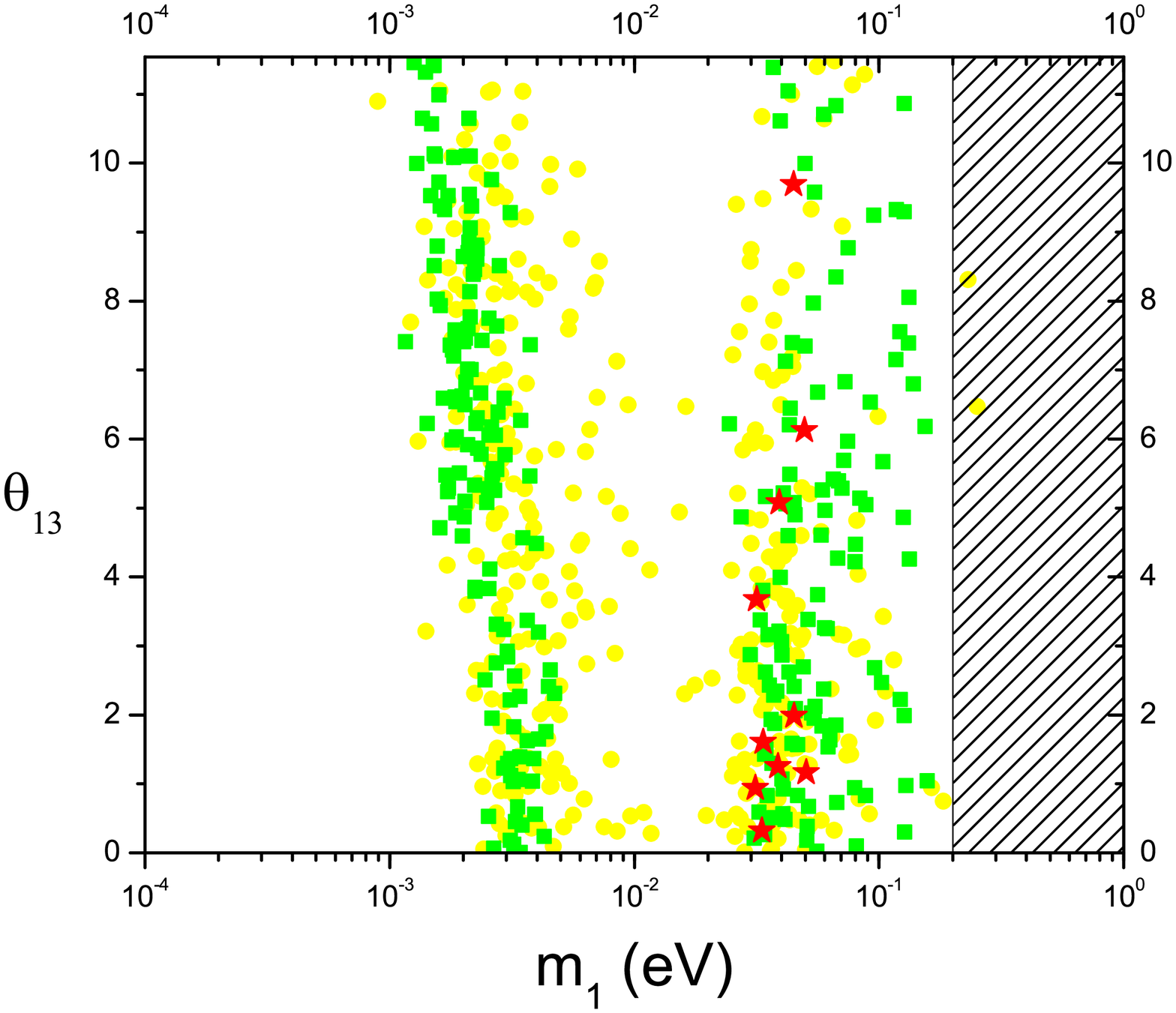,height=48mm,width=54mm}
\hspace{-4mm}
\psfig{file=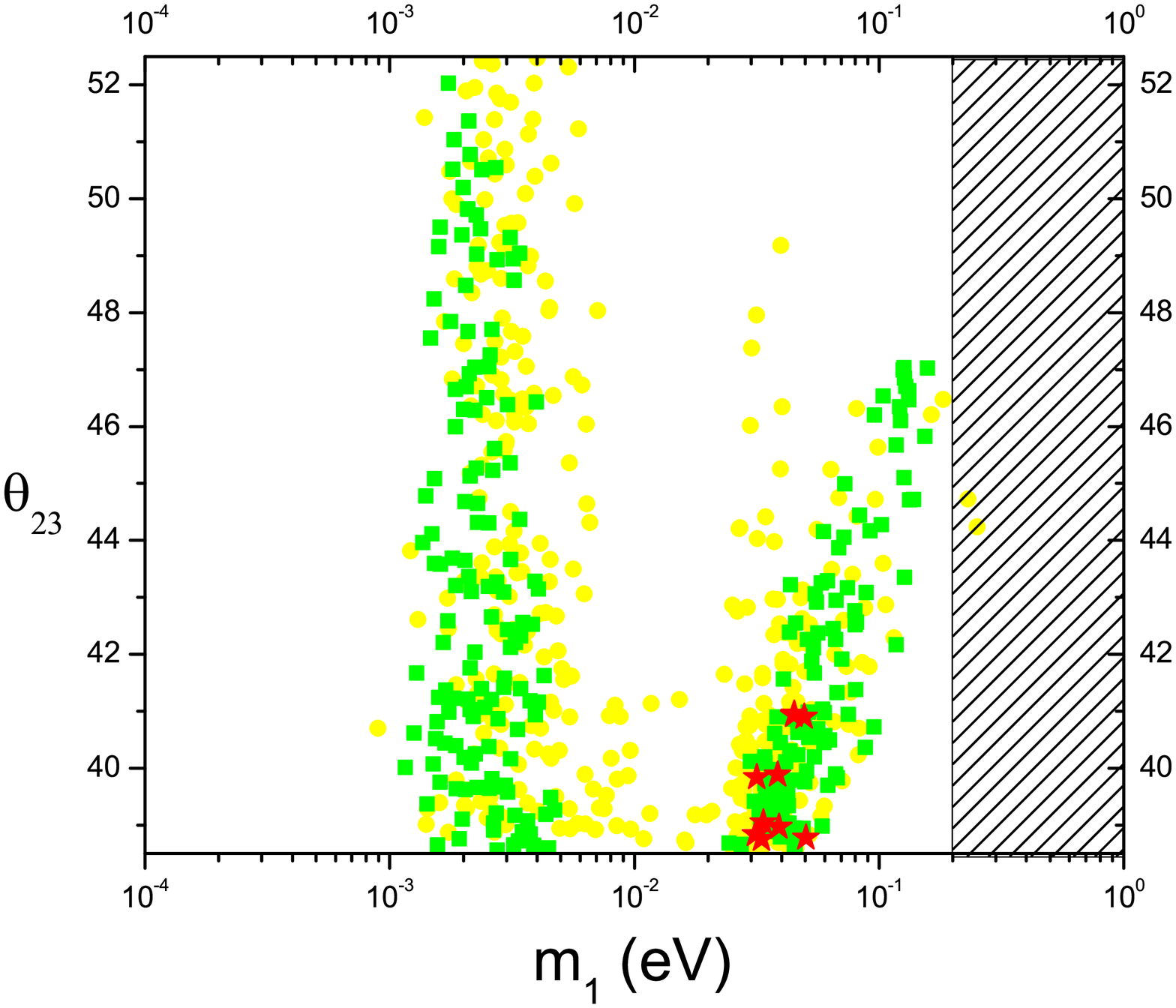,height=48mm,width=54mm} \\
\psfig{file=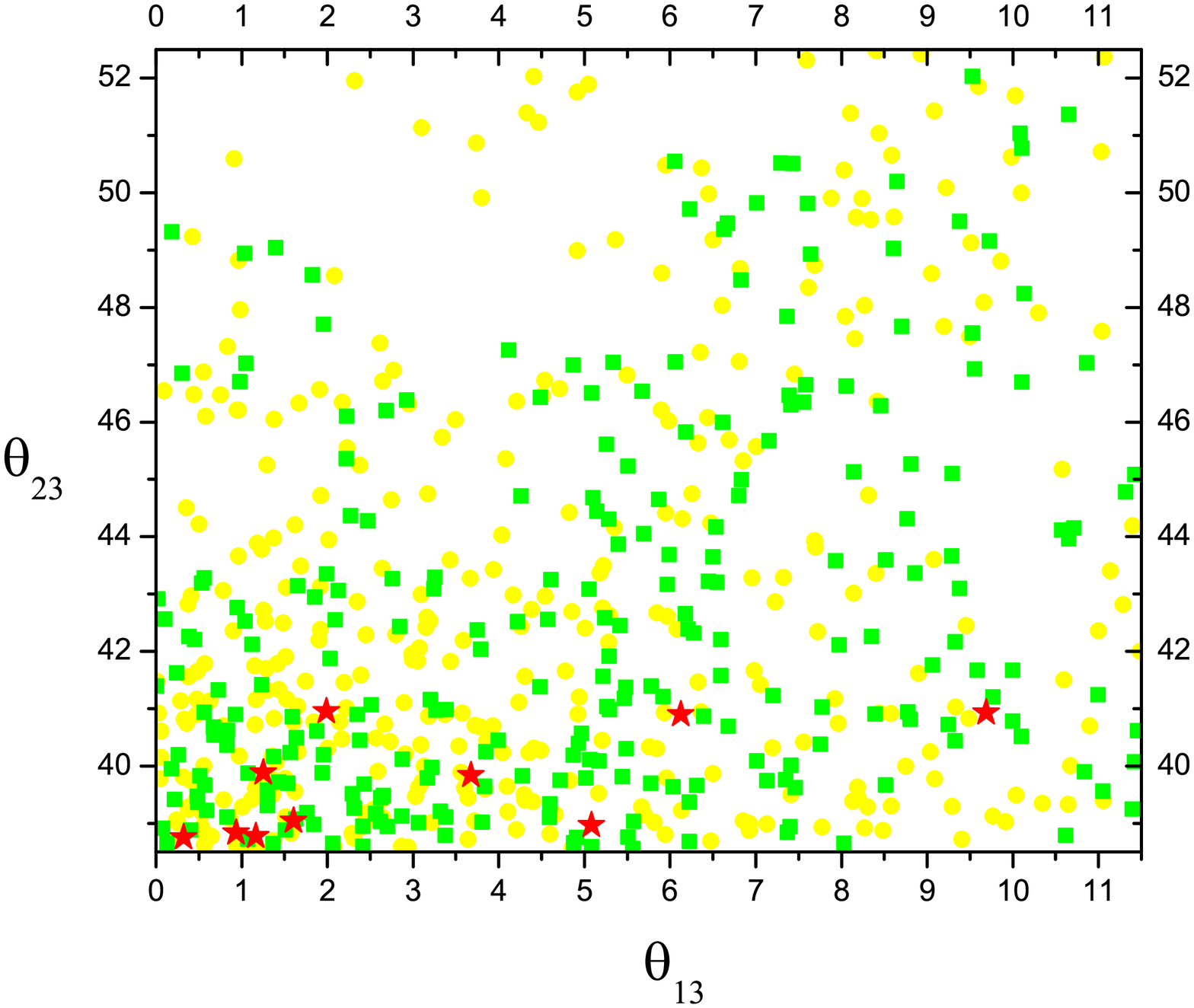,height=48mm,width=54mm}
\hspace{-4mm}
\psfig{file=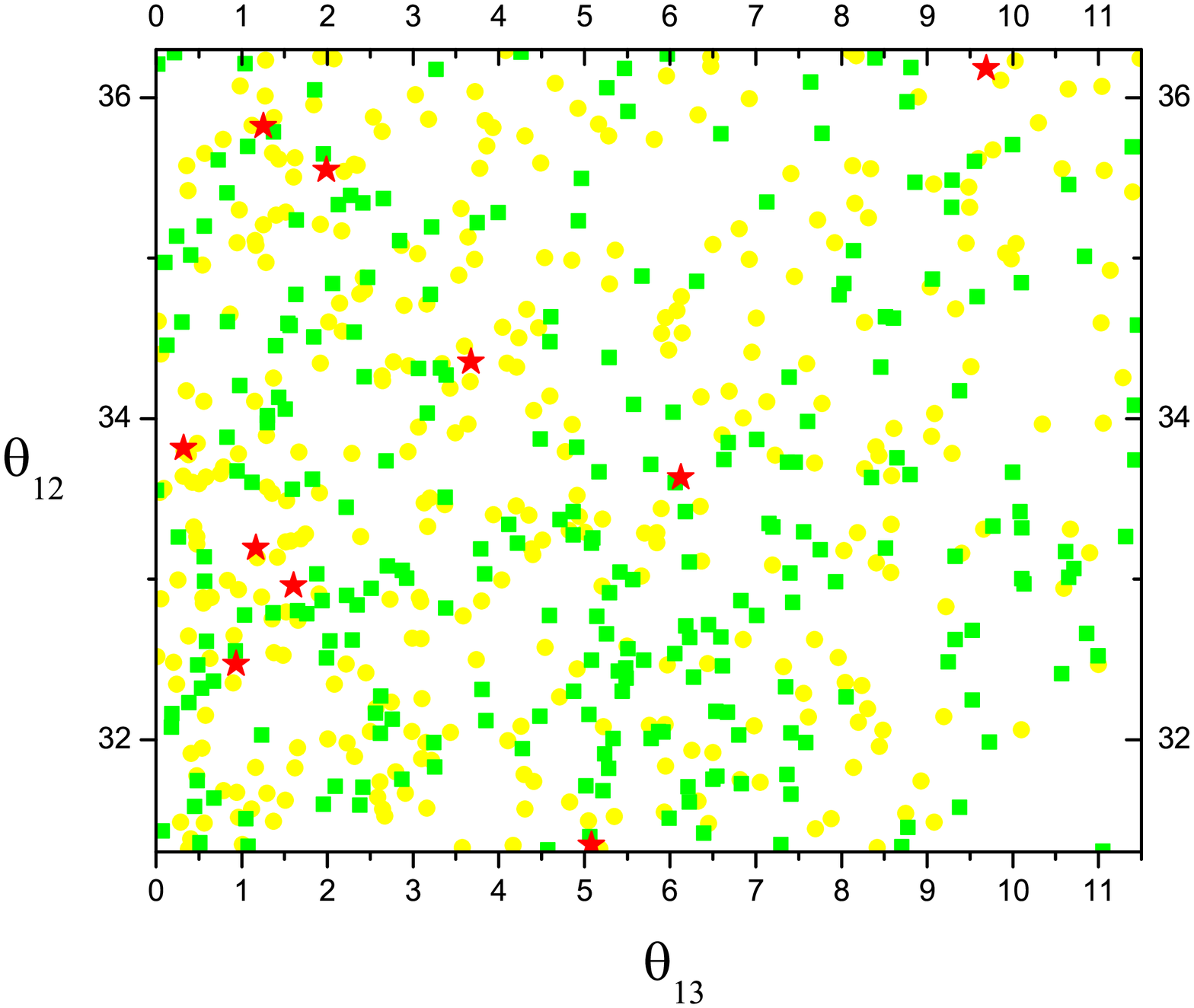,height=48mm,width=54mm}
\hspace{-4mm}
\psfig{file=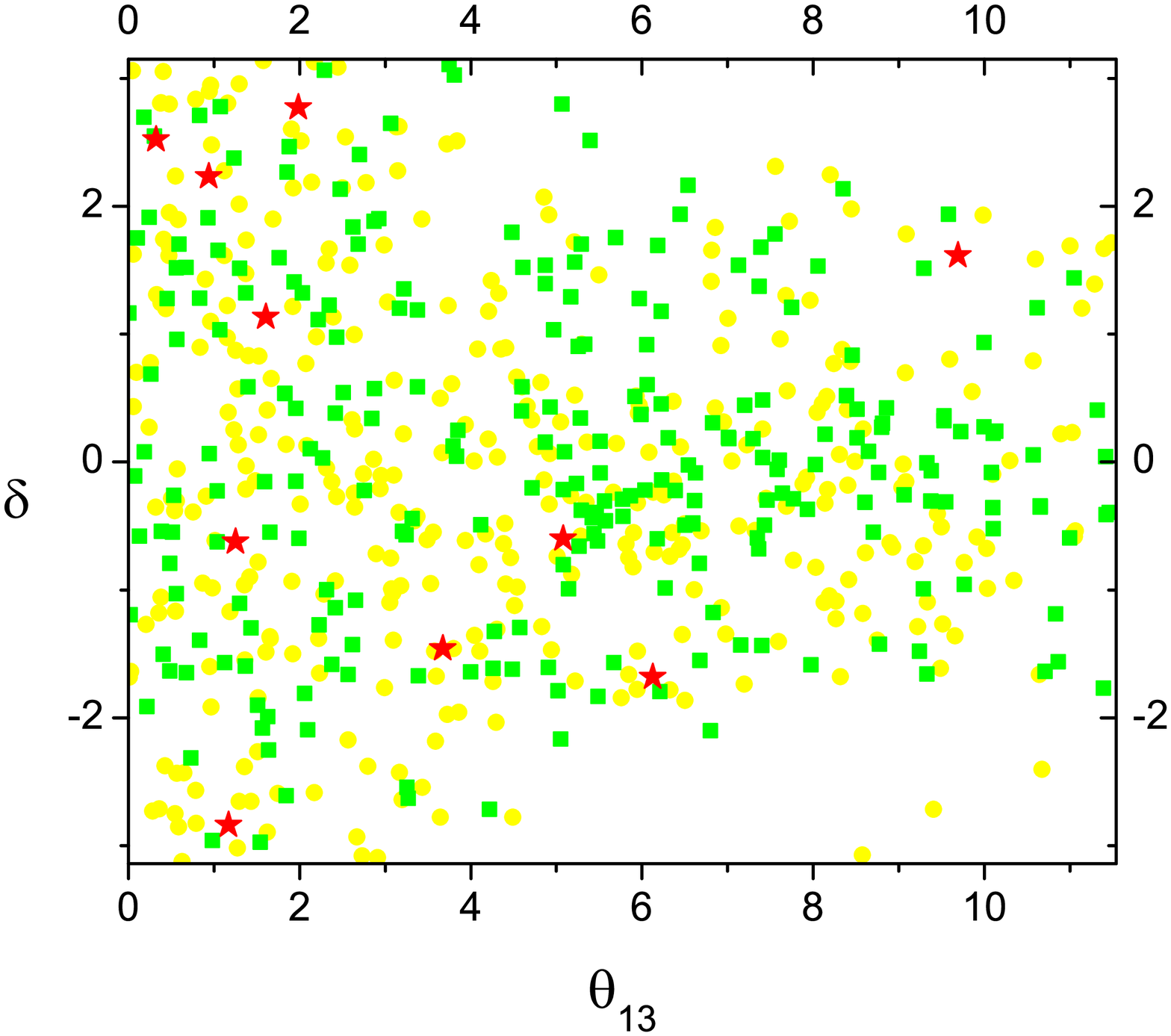,height=48mm,width=54mm} \\
\psfig{file=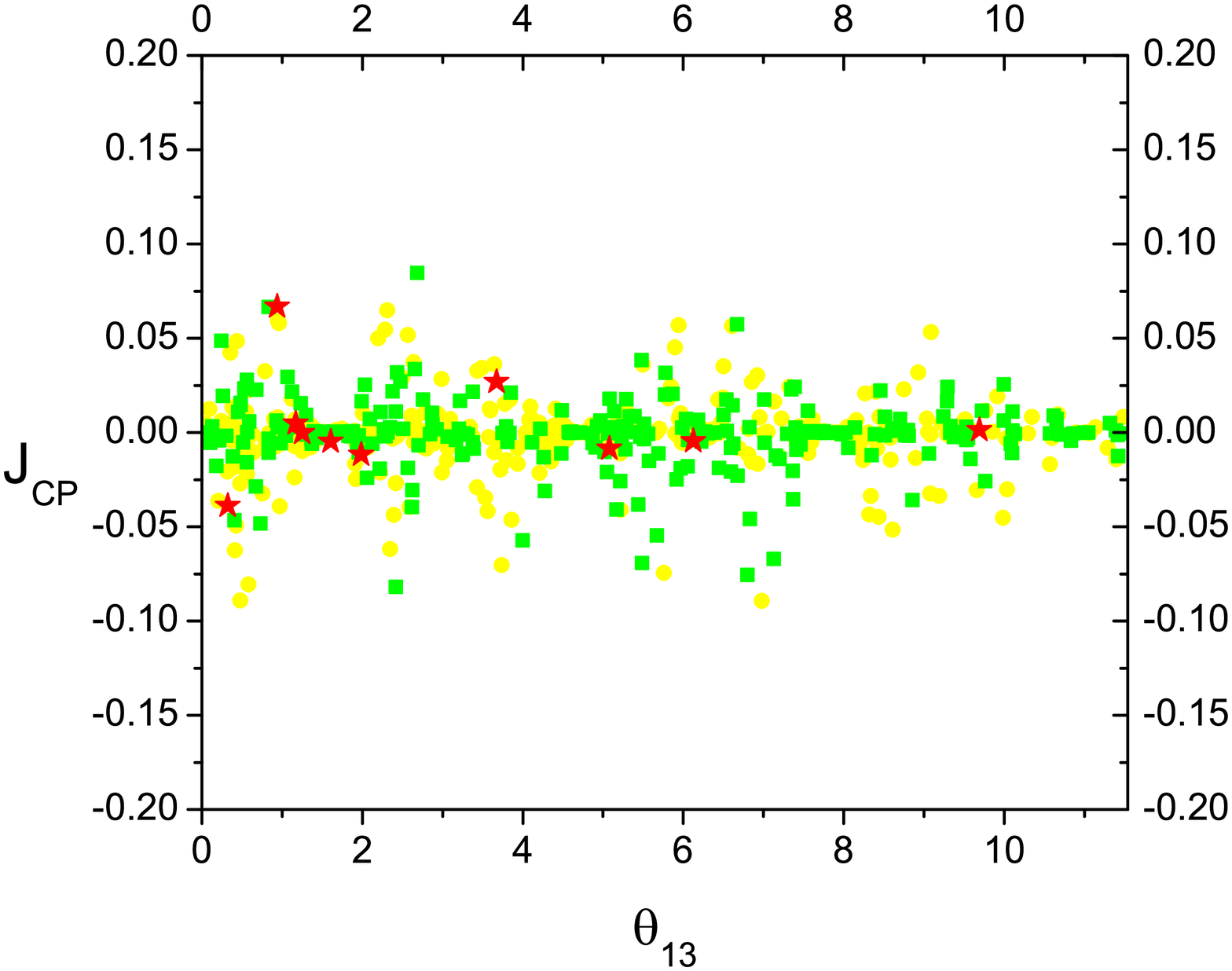,height=48mm,width=54mm}
\hspace{-4mm}
\psfig{file=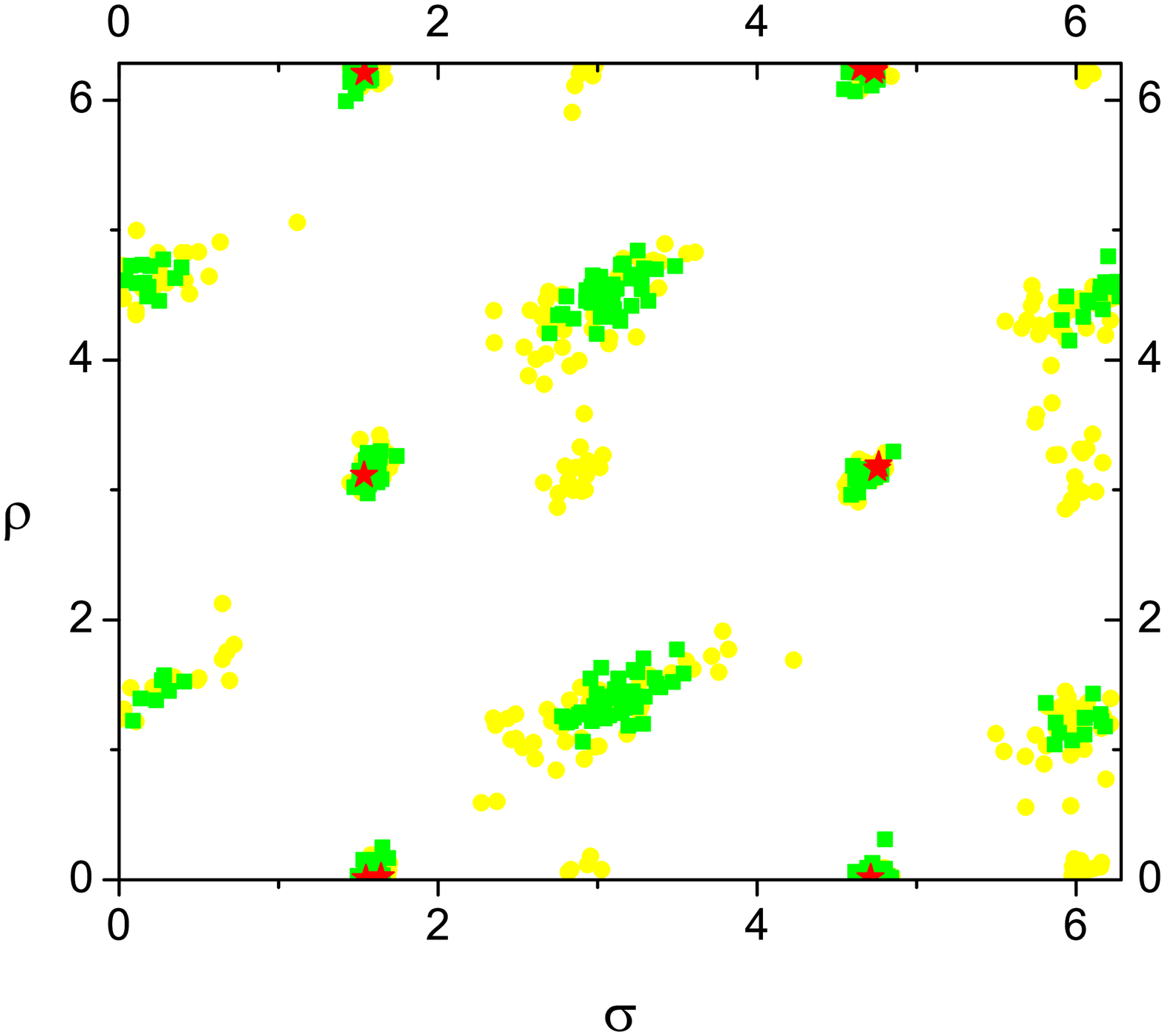,height=48mm,width=54mm}
\hspace{-4mm}
\psfig{file=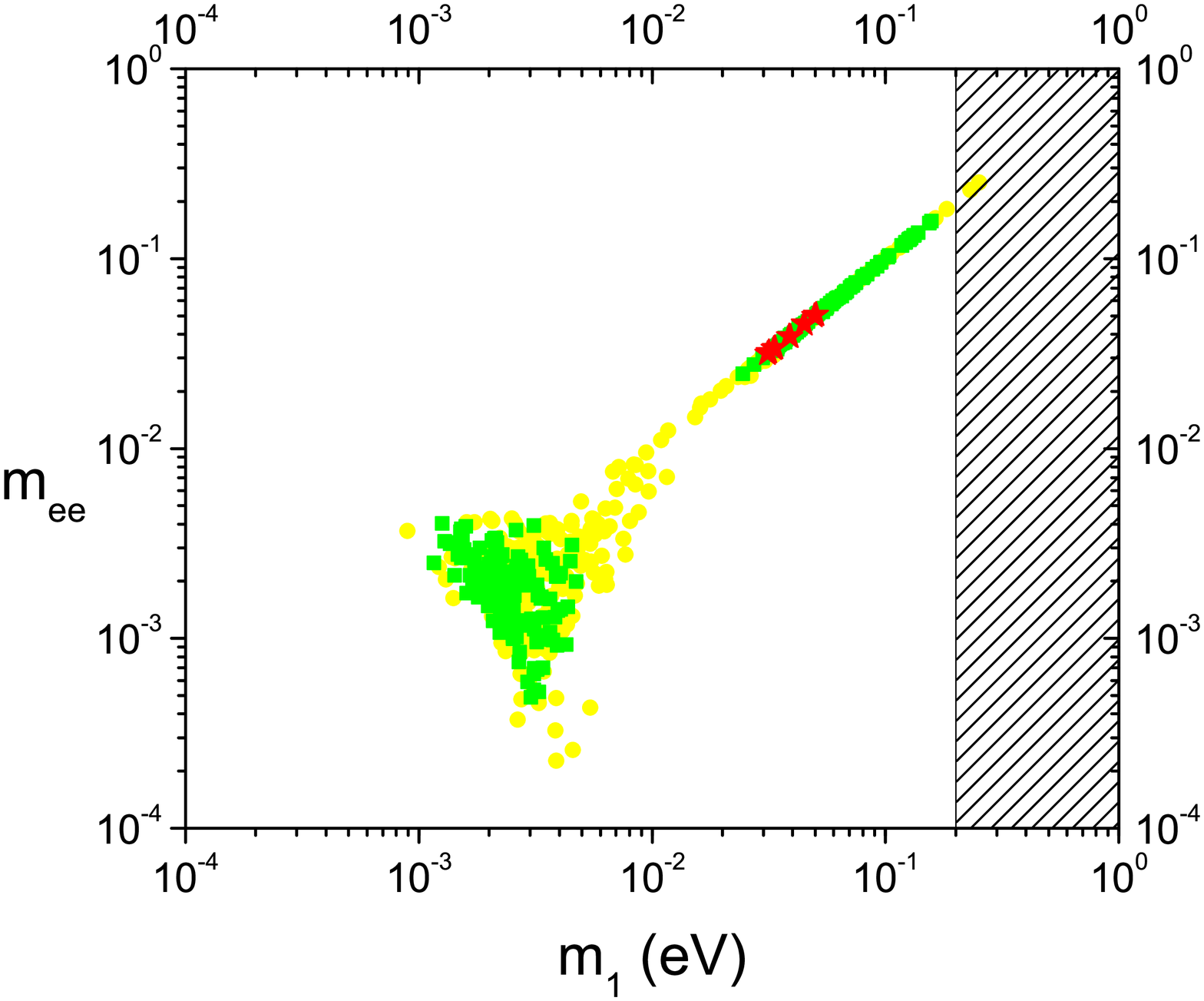,height=48mm,width=54mm} \\
\psfig{file=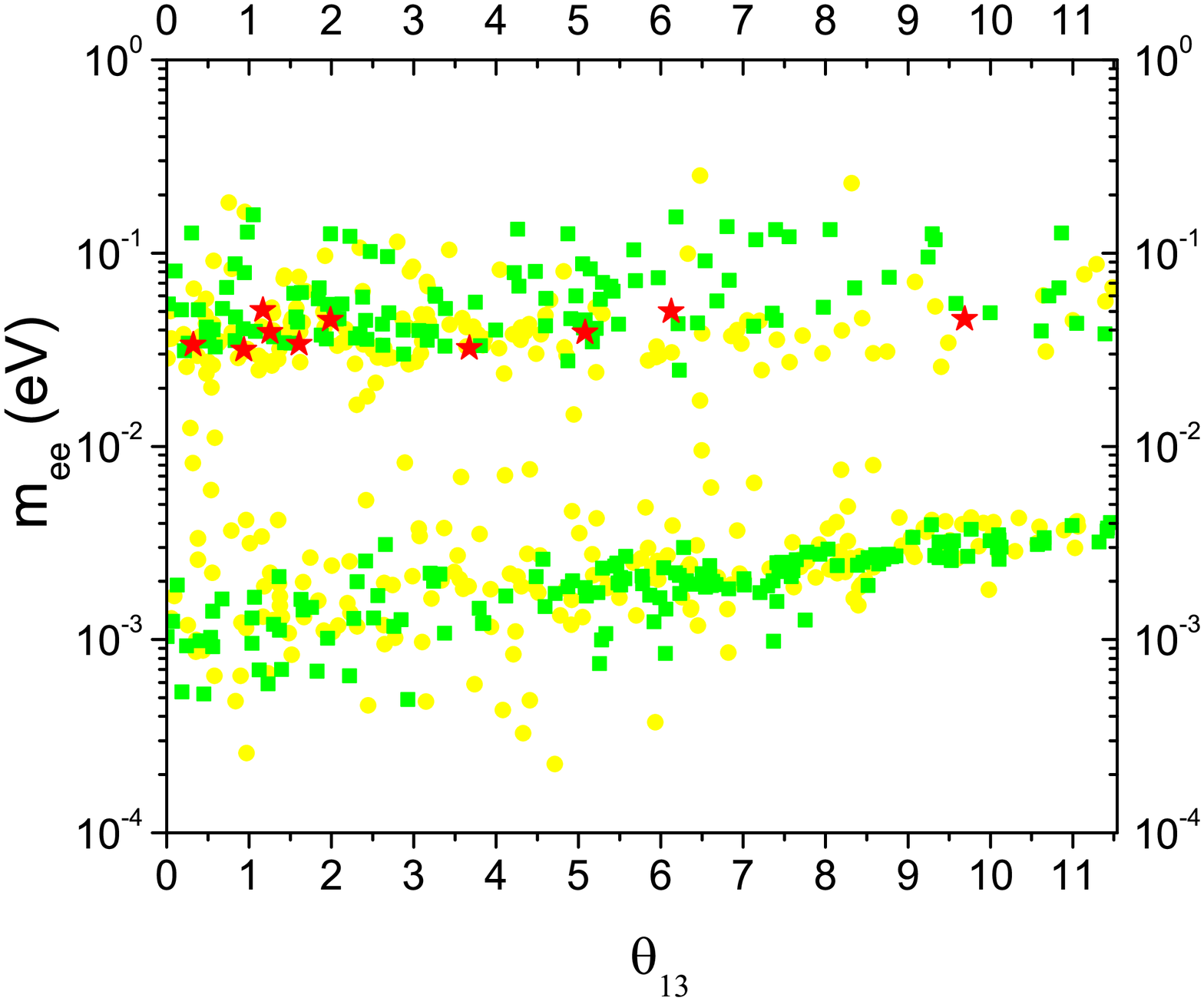,height=48mm,width=54mm}
\end{center}
\caption{Global scan, NO. Scatter plot of points in the parameter space that
satisfy successful leptogenesis ($\eta_B>5.9\times 10^{-9}$), for $\a_2=5$ (yellow circles),
$\a_2=4$ (green squares) and $\a_2=1$ (red stars).}
\end{figure}
At the same time one can see that a global scan actually shows a slight correlation
between $m_1$ and $\theta_{13}$ in the low  $m_1$ range while
the interesting linear dependence between
$\theta_{13}$ and $\theta_{23}$ found for $V_L=I$ seems now to be lost.

However, it should be considered that these plots are projections
on two-parameters planes of an allowed region in a seven-parameter space.
Therefore, only a full multi-parameters analysis would be able
to unreveal correlations involving more than two parameters.
Nevertheless, thanks to
the distinct analysis that we carried out for the two special cases $V_L=I$ and $V_L=V_{CKM}$,
one can catch sight of an interesting correlation among $m_1$, $\theta_{12}$, $\theta_{13}$ and $m_{ee}$.
To this extent, this time we have also plotted the constraints in the plane $\theta_{13}-m_{ee}$,
showing how the lower bound on $m_{ee}$ increases with $\theta_{13}$.

\subsubsection{Low $m_1$ range}

In order to find out whether the linear dependence between $\theta_{13}$ and $\theta_{23}$
found for $V_L=I$ (cf. eq.~(\ref{linear})) still holds for a global scan, we show in Fig.~11 the same constraints
as in Fig.~10 imposing the condition $m_1\lesssim 0.01\,{\rm eV}$,
since the linear dependence was found in that range of values.
 We only show the constraints on the relevant parameters, therefore only those
 in the plane $\theta_{13}-\theta_{23}$, in the plane $\theta_{12}-\theta_{13}$
 and in the plane $\theta_{13}-m_{ee}$.
\begin{figure}
\begin{center}
\psfig{file=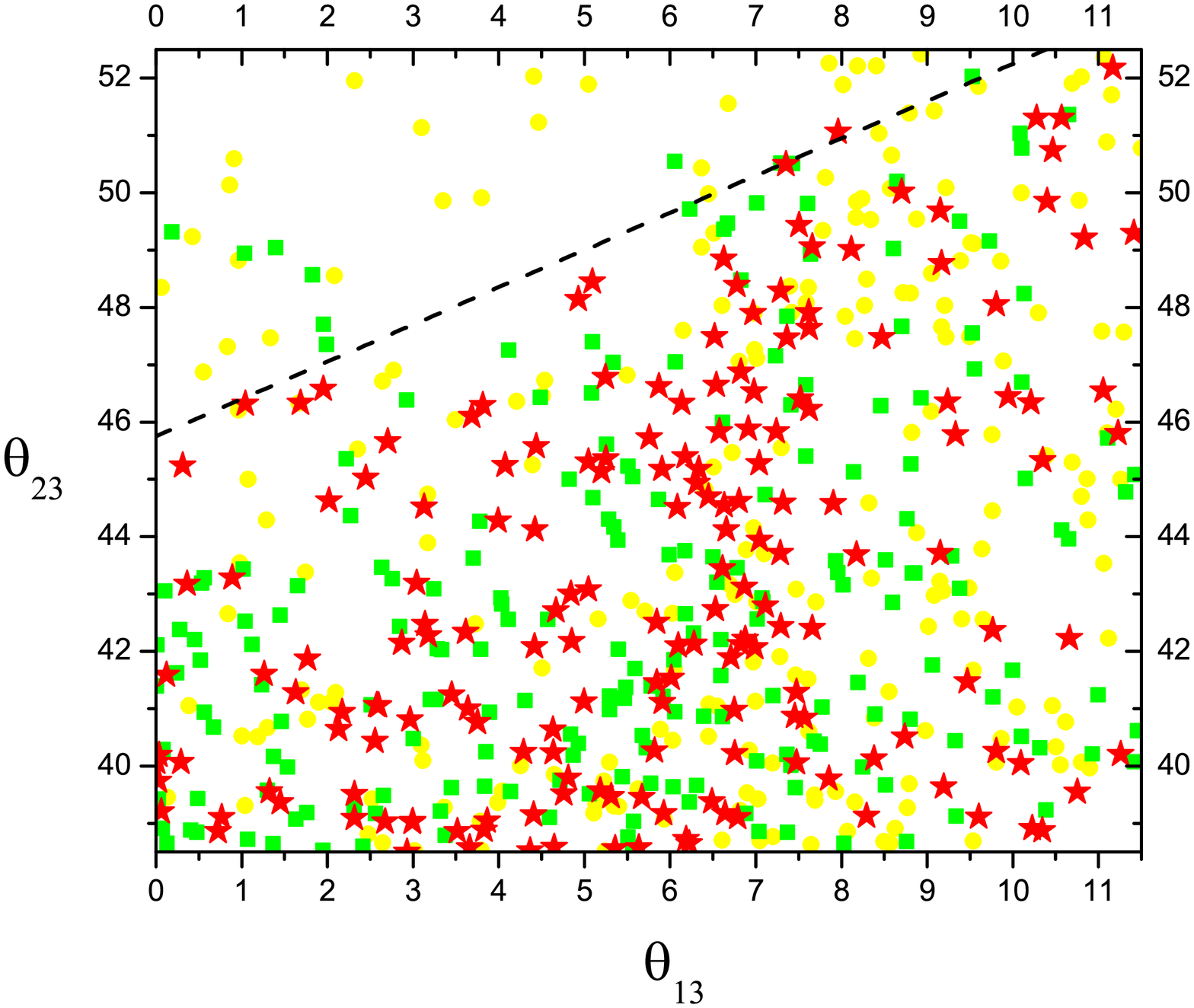,height=48mm,width=54mm}
\hspace{-4mm}
\psfig{file=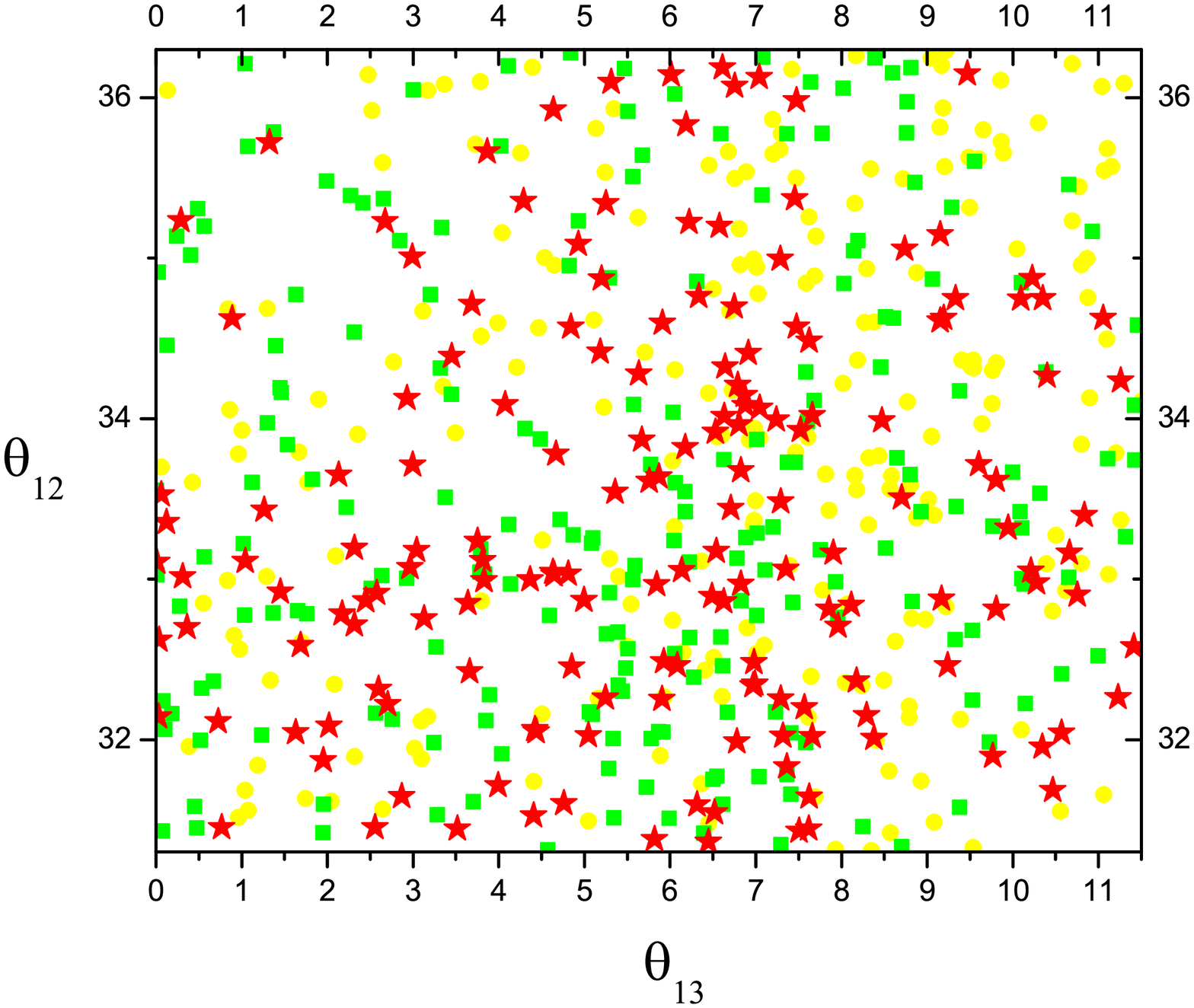,height=48mm,width=54mm}
\hspace{-4mm}
\psfig{file=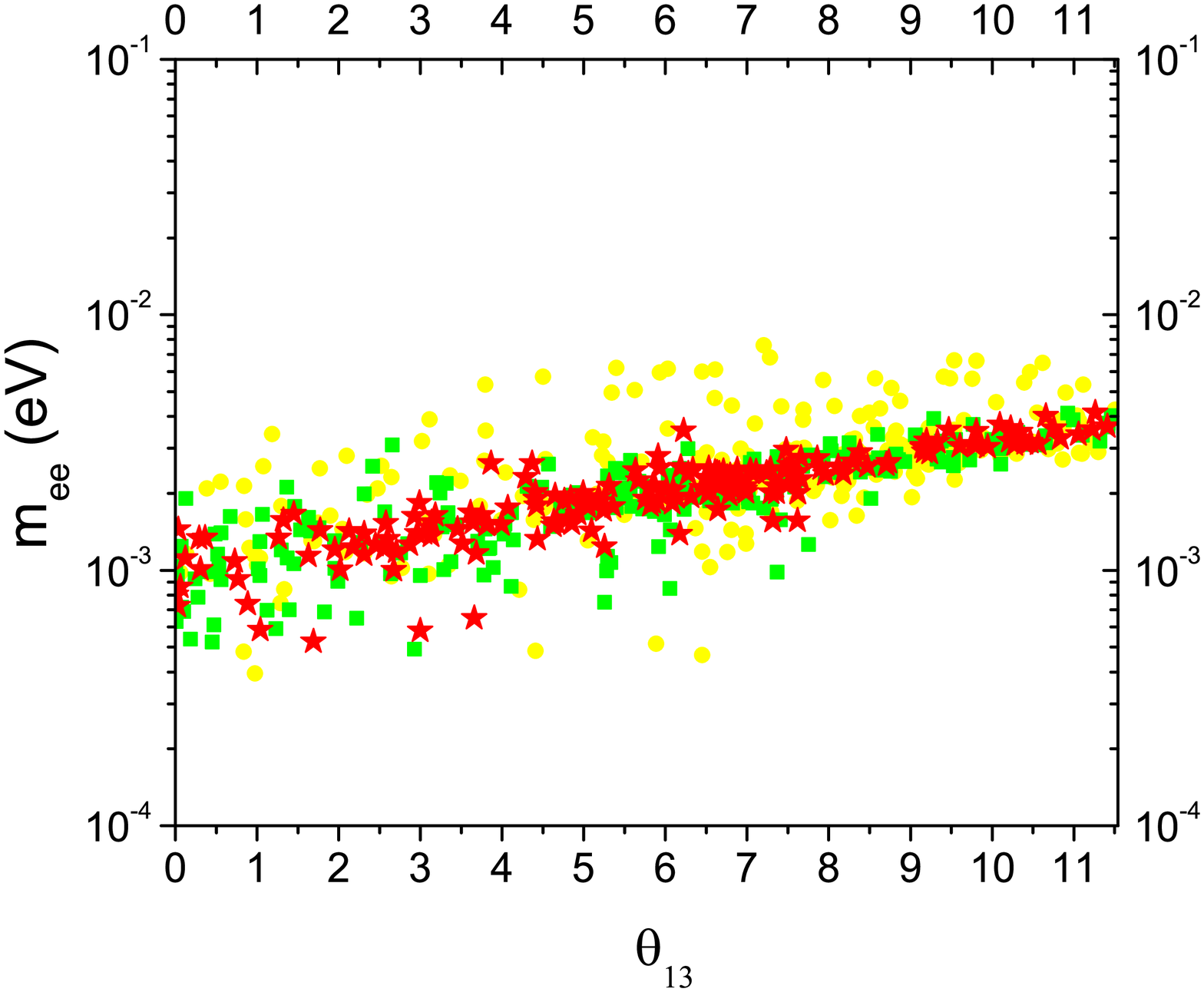,height=48mm,width=54mm}
\end{center}
\vspace{-8mm}
\caption{Global scan, NO, $m_1<0.01 \,{\rm eV}$.
Scatter plot of points in the parameter space that
satisfy successful leptogenesis ($\eta_B>5.9\times 10^{-9}$), for $\a_2=5$ (yellow circles),
$\a_2=4$ (green squares) and $\a_2=3.7$ (red stars).
The region below the dashed line in the left
panel corresponds to the condition eq.~(\ref{th13th23}).}
\end{figure}
This time we could also easily find points for $\a_2=3.7$ (red stars), showing again how
allowing for a $V_L\neq I$ the allowed regions get larger.

One can see that the quite clear linear dependence  eq.~(\ref{linear}) between $\theta_{13}$
and $\theta_{23}$ holding for $V_L=I$, now turns more, for the red star points at $\a_2=3.7$,
into an allowed region below the dashed line showed in the figure and
corresponding approximately to
\be\label{th13th23}
\theta_{23} \lesssim 49^{\circ}+ 0.65\,(\theta_{13}-5^{\circ}) \, .
\ee
This result should be also
understood in terms of the condition $K_{1\tau}\lesssim 1$ (cf. (\ref{pp}))
when a very small $V_L$ is allowed clearly yielding a
dispersion around the linear dependence eq.~(\ref{linear}).
Notice that inside this region there are still sort of sub-regions that
seem to be excluded.

We can summarize these results saying that, at low values of $m_1\lesssim 0.01\,{\rm eV}$,
there is an interesting testable constraints in the plane $\theta_{13}-\theta_{23}$
given by the relation eq.~(\ref{th13th23}). In particular experiments that are already
taking data such as the nuclear reactor
experiment DOUBLE CHOOZ \cite{DC} and the long baseline experiment T2K \cite{T2K}
have the capability of a $3\sigma$ discovery of values  $\theta_{13}\gtrsim 8^{\circ}$.
Our results seem to suggest that  if such high $\theta_{13}$ values will not be found,
then a restricted range of values for $\theta_{23}$ is predicted.
For example, if $\theta_{13}\lesssim 8^{\circ}$ then
$\theta_{23}\lesssim 51^{\circ}$, and
if $\theta_{13}\simeq 6^{\circ}$ then
$\theta_{23}\lesssim 48^{\circ}$.
Such a constraint on $\theta_{23}$ should be also tested during next years
with quite a good accuracy by the T2K experiment \cite{T2K}.
These constraints in
the plane $\theta_{13}-\theta_{23}$ should be considered at this level
indicative, and should also consider that they are quite sensitive to the
value of $\a_2$.

Notice that  at the same time, cosmological observations and/or neutrinoless double beta decay experiments
should also be able to test the condition $m_1<0.01\,{\rm eV}$. It should be therefore appreciated
that this scenario will be tested during next years.

It is also interesting to notice (see right panel in Fig.~11)
that there is a linear dependence between $m_{ee}$ and $\theta_{13}$ as well.
In particular, for $\a_2\leq 4$,
at large values $\theta_{13}\gtrsim 6^{\circ}$ one has $m_{ee}\simeq 10^{-3}\,{\rm eV}$
and even for $\theta_{13}\gtrsim 8^{\circ}$ one has $m_{ee}\simeq 3\times 10^{-3}\,{\rm eV}$.
These values for $m_{ee}$ are below the sensitivity
of future planned experiments ($\gtrsim 0.01\,{\rm eV}$) such as EXO \cite{EXO}.
However, at least, $m_{ee}$  cannot be arbitrary small but has
a lower bound that, for sufficiently large $\theta_{13}$ values, is
$3$ times below the currently planned reachable experimental sensitivity,
a very small value but maybe not completely hopeless.

Within the two-parameter analysis we are presenting, we cannot draw sharper predictions
but is seems quite plausible that from a more involved multi-parameter analysis
precise correlations could emerge, maybe
also involving the solar neutrino angle $\theta_{12}$. In this respect
the central panel in figure~11 suggests that the solar mixing angle
could indeed play also a role and that maybe sharper predictions
in the 3 parameter space $(\theta_{13},\theta_{12},\theta_{23})$
exist.

\subsubsection{Large $m_1$ range}

We can also study how the allowed regions would reduce
requiring large values  $m_1> 0.01 \,{\rm eV}$. The results are
shown in figure~12. One can see that in this case one obtains very clear
\begin{figure}
\begin{center}
\psfig{file=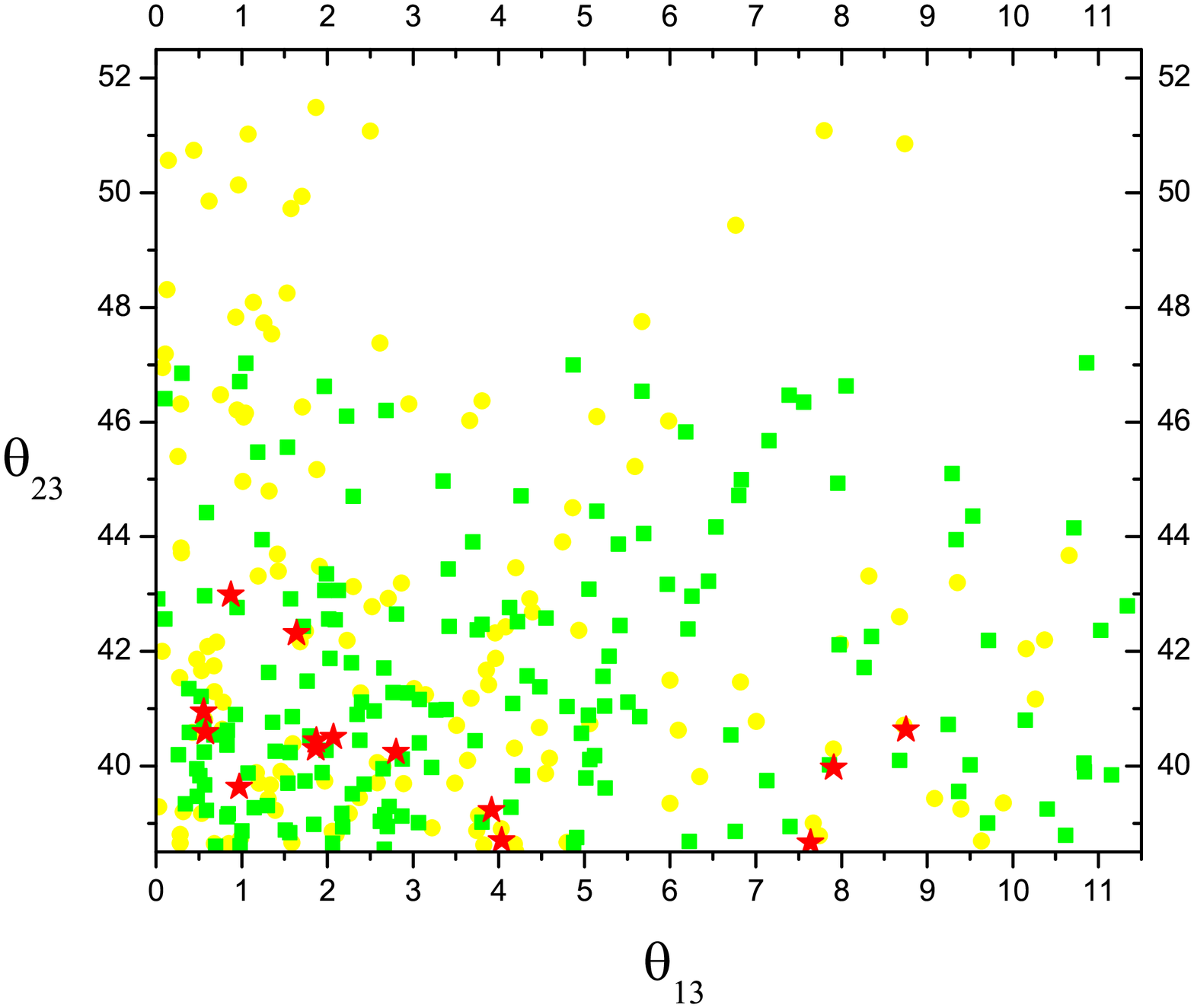,height=48mm,width=54mm}
\hspace{-4mm}
\psfig{file=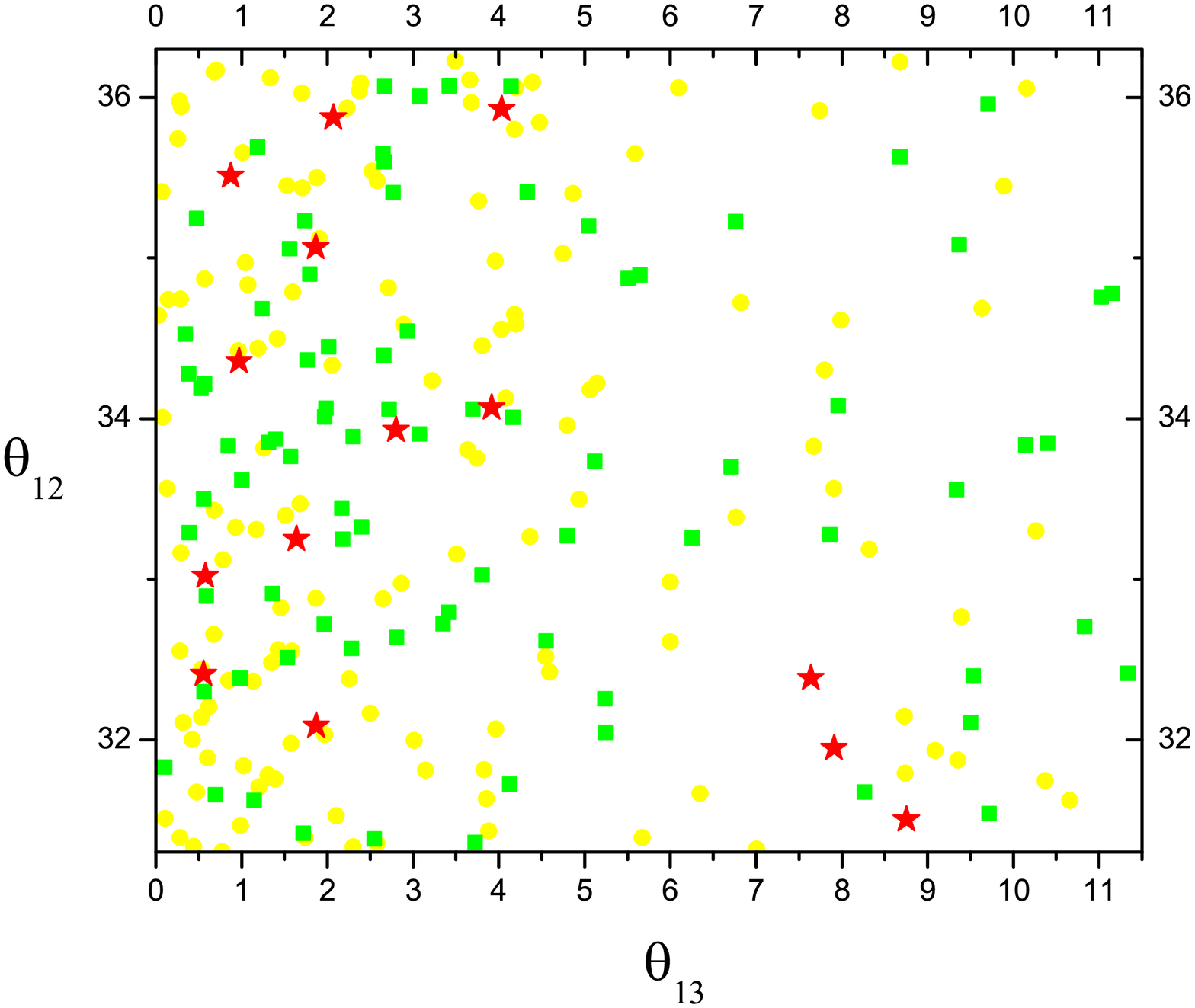,height=48mm,width=54mm}
\hspace{-4mm}
\psfig{file=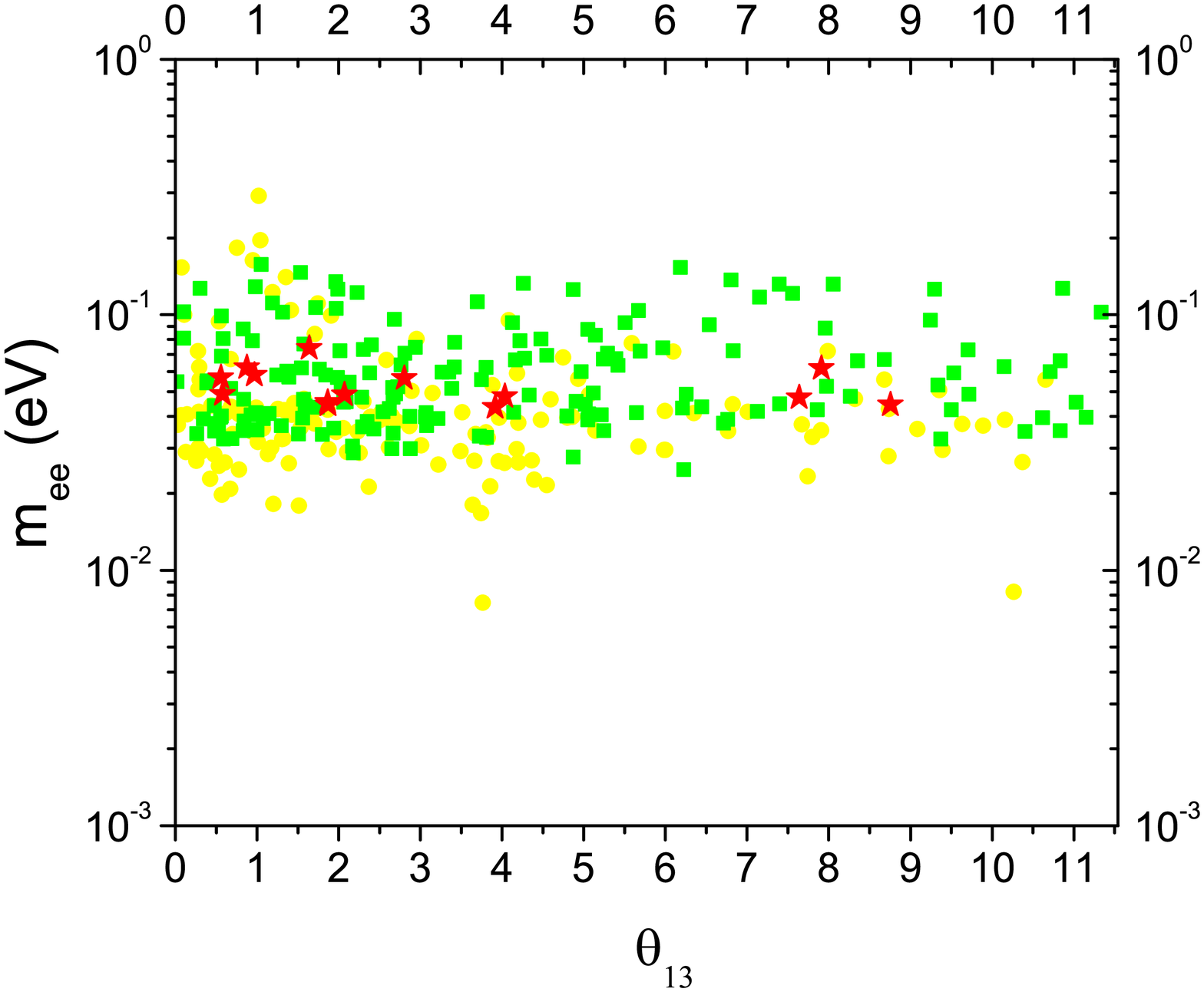,height=48mm,width=54mm}
\end{center}
\vspace{-8mm}
\caption{Global scan, NO, $m_1>0.01 \,{\rm eV}$.
Scatter plot of points in the parameter space that
satisfy successful leptogenesis ($\eta_B>5.9\times 10^{-9}$), for $\a_2=5$ (yellow circles),
$\a_2=4$ (green squares) and $\a_2=1$ (red stars).}
\end{figure}
constraints that will allow to test this scenario during next years
in a quite unambiguous way. First of all from the Fig.~10, thanks to the
very precise values of the Majorana phases, one can notice that there is
a very clear relation between $m_1$ and $m_{ee}$. Second, one can see from the
left panel of Fig.~12 how there is an upper bound $\theta_{23}\lesssim 46^{\circ}$
for $\a_2\leq 4$. For values of $\theta_{13}\simeq (5-6)^{\circ}$, one has even
$\theta_{23}\lesssim 41^{\circ}$. It should be said however that
at these large $m_1$ values, one typically obtains a final asymmetry that
depends on the initial conditions. Since we are assuming vanishing
initial $N_2$ abundance and vanishing initial asymmetry, these constraints
should be regarded as the most stringent ones, but likely also
the best motivated ones.

\subsection{Inverted ordering}

Finally, we repeated the global scan for IO as well
and the results are shown in Fig.~13.
 \begin{figure}
\begin{center}
\psfig{file=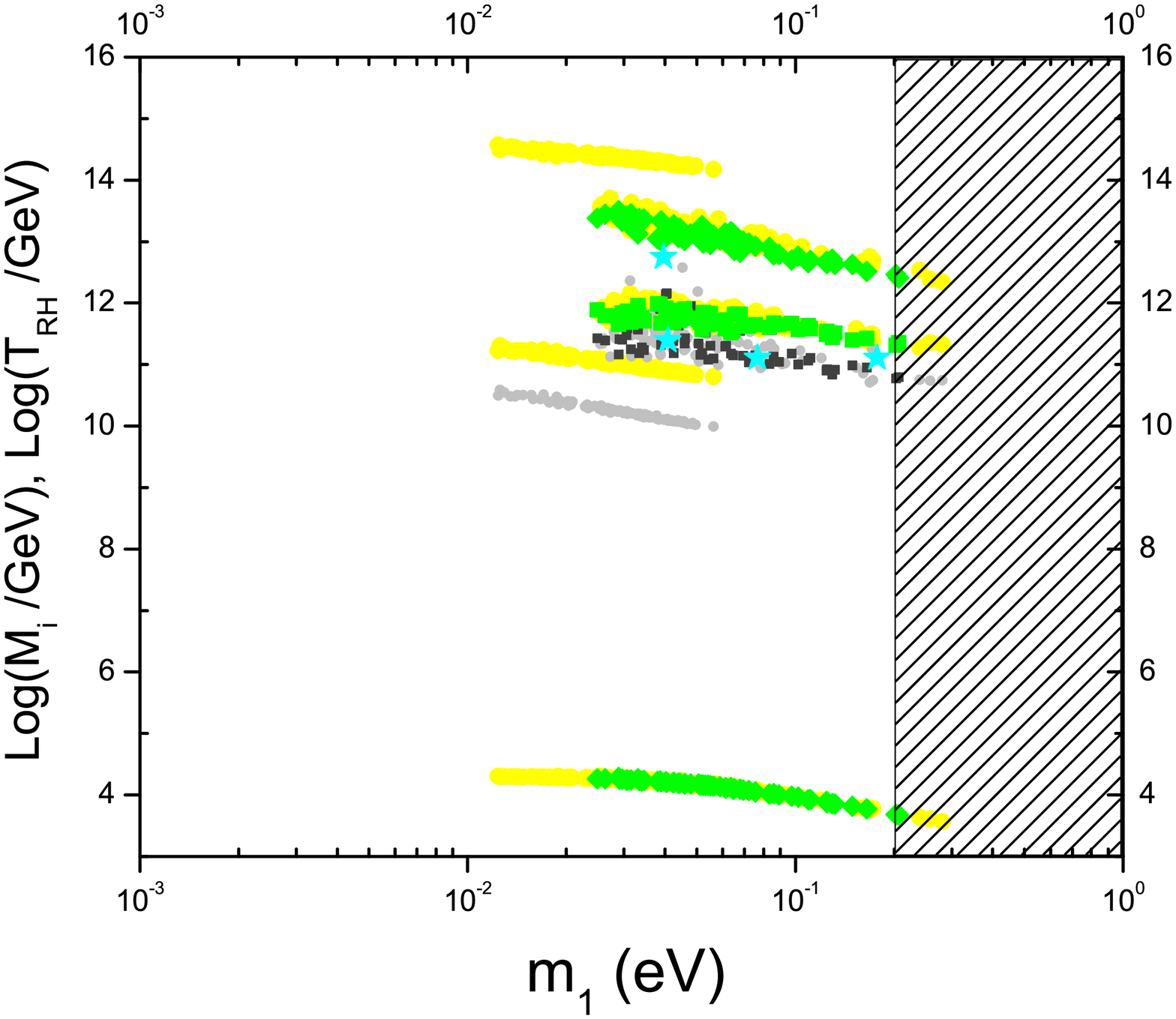,height=48mm,width=54mm}
\hspace{-4mm}
\psfig{file=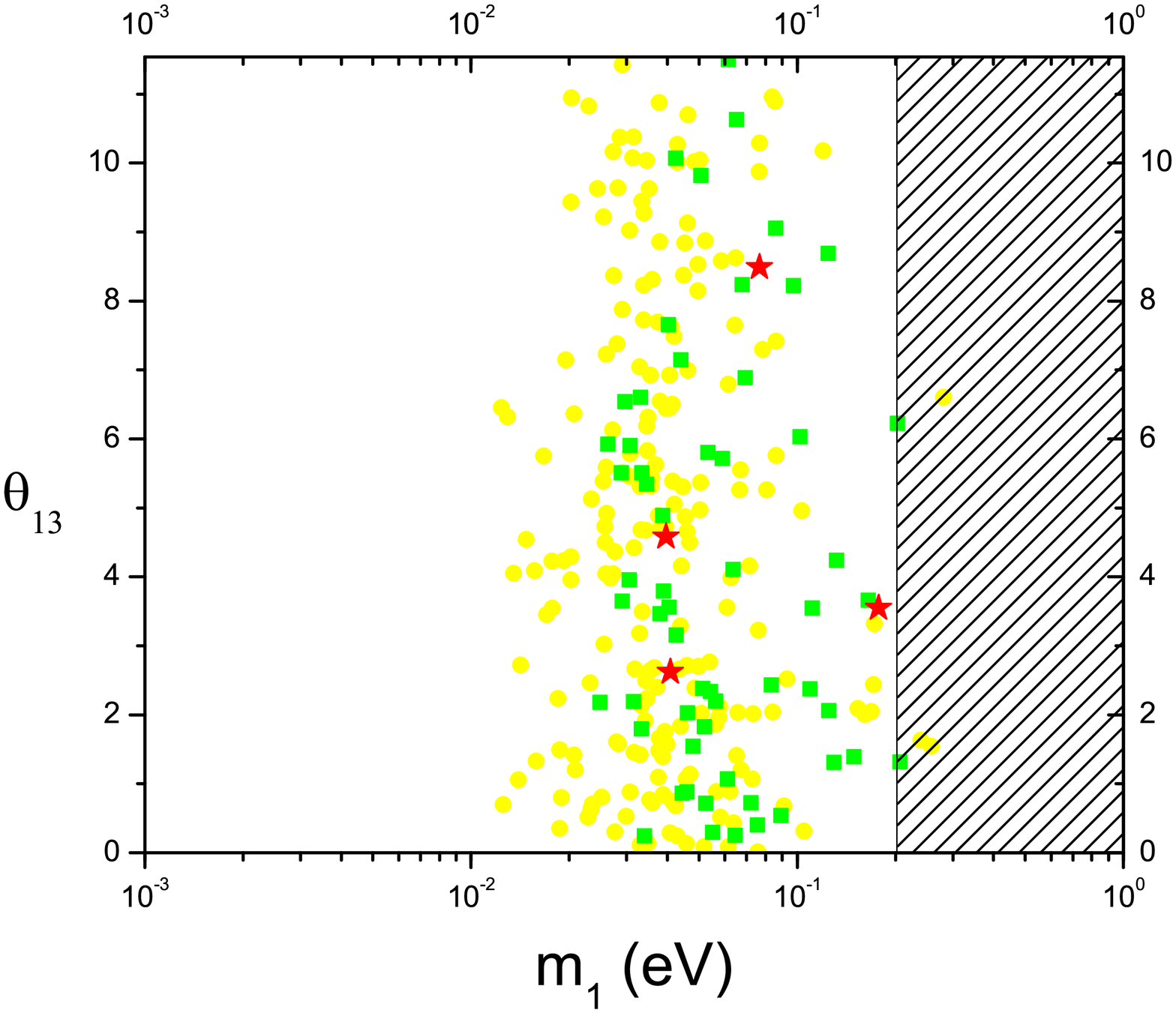,height=48mm,width=54mm}
\hspace{-4mm}
\psfig{file=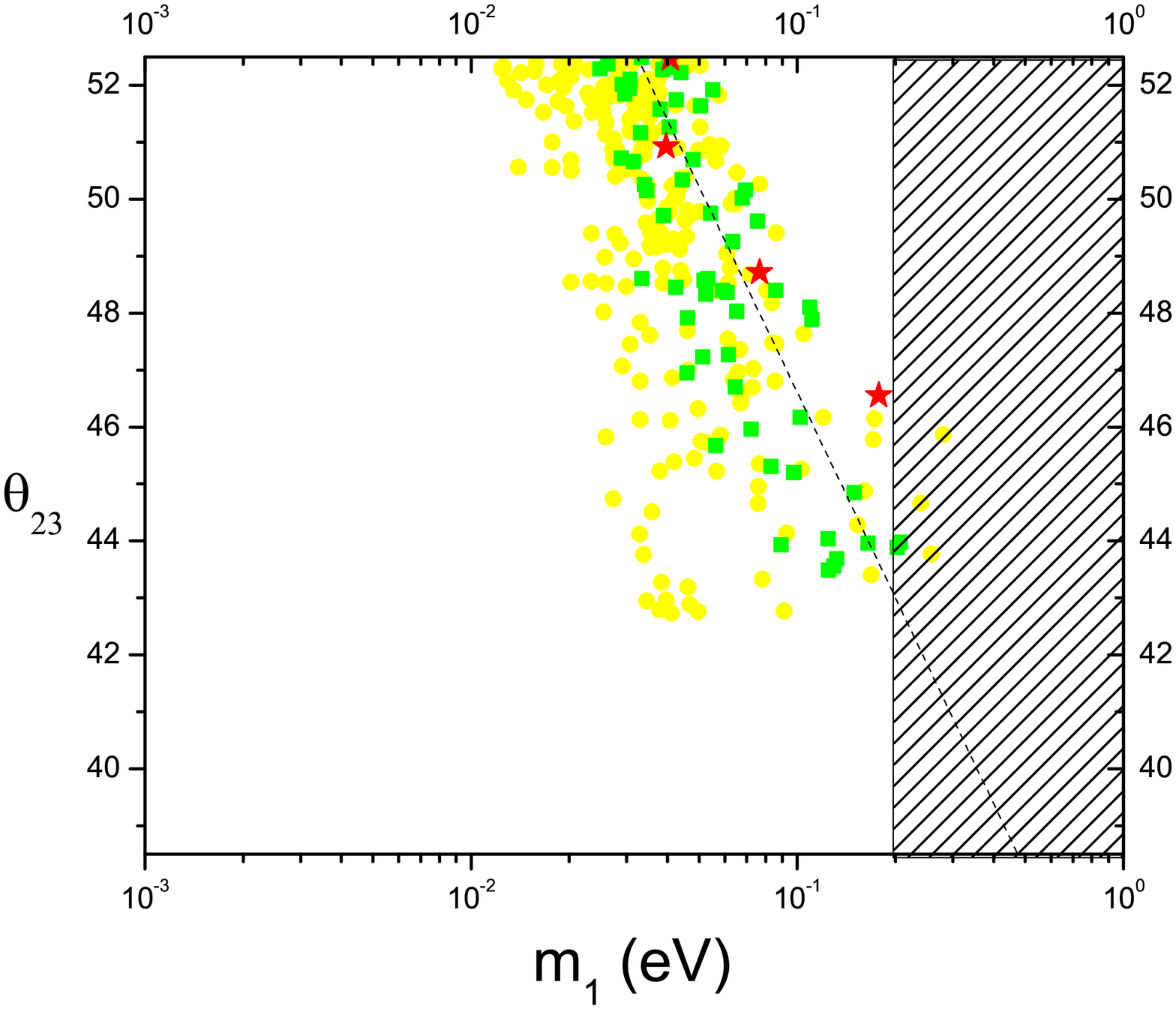,height=48mm,width=54mm} \\
\psfig{file=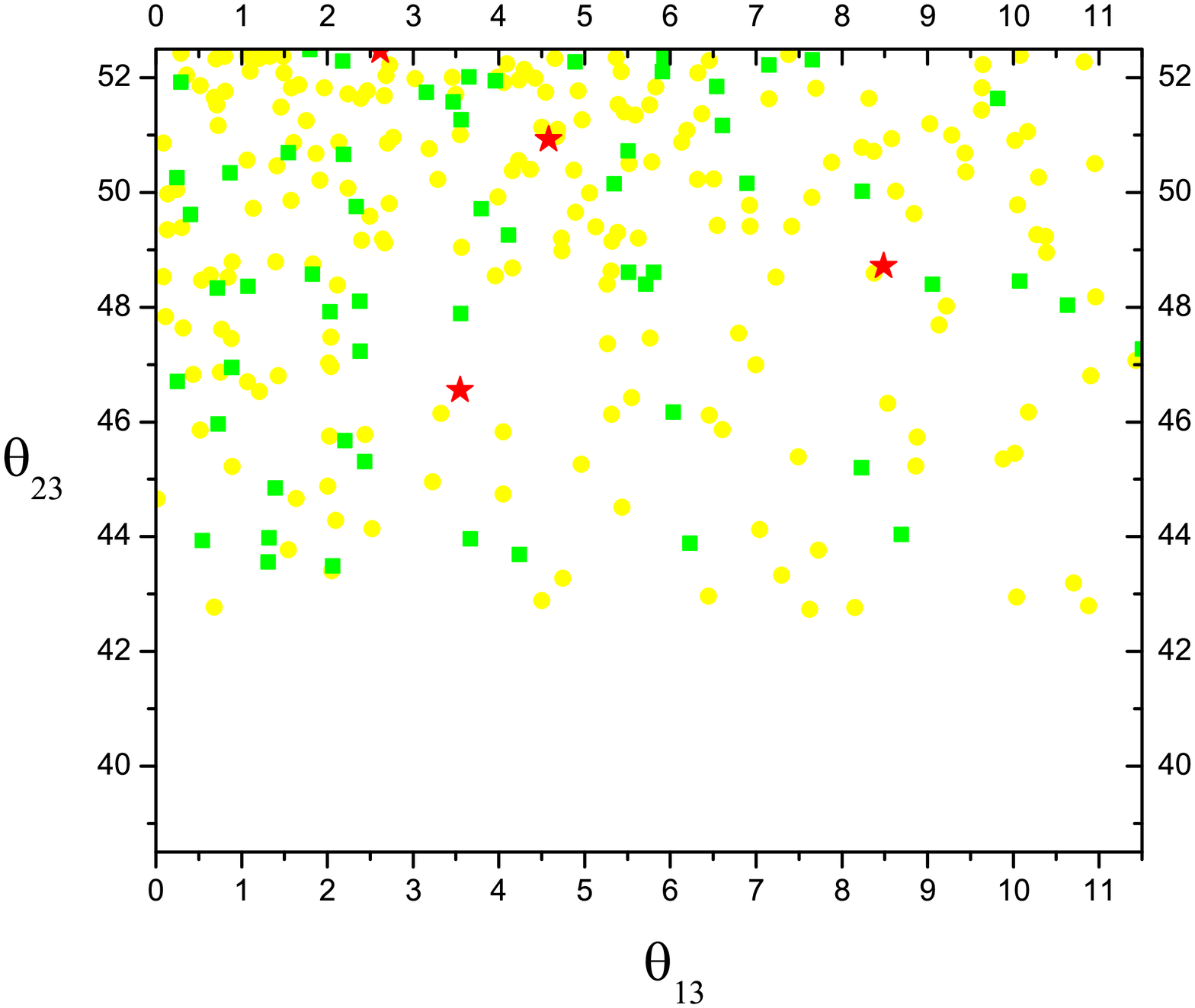,height=48mm,width=54mm}
\hspace{-4mm}
\psfig{file=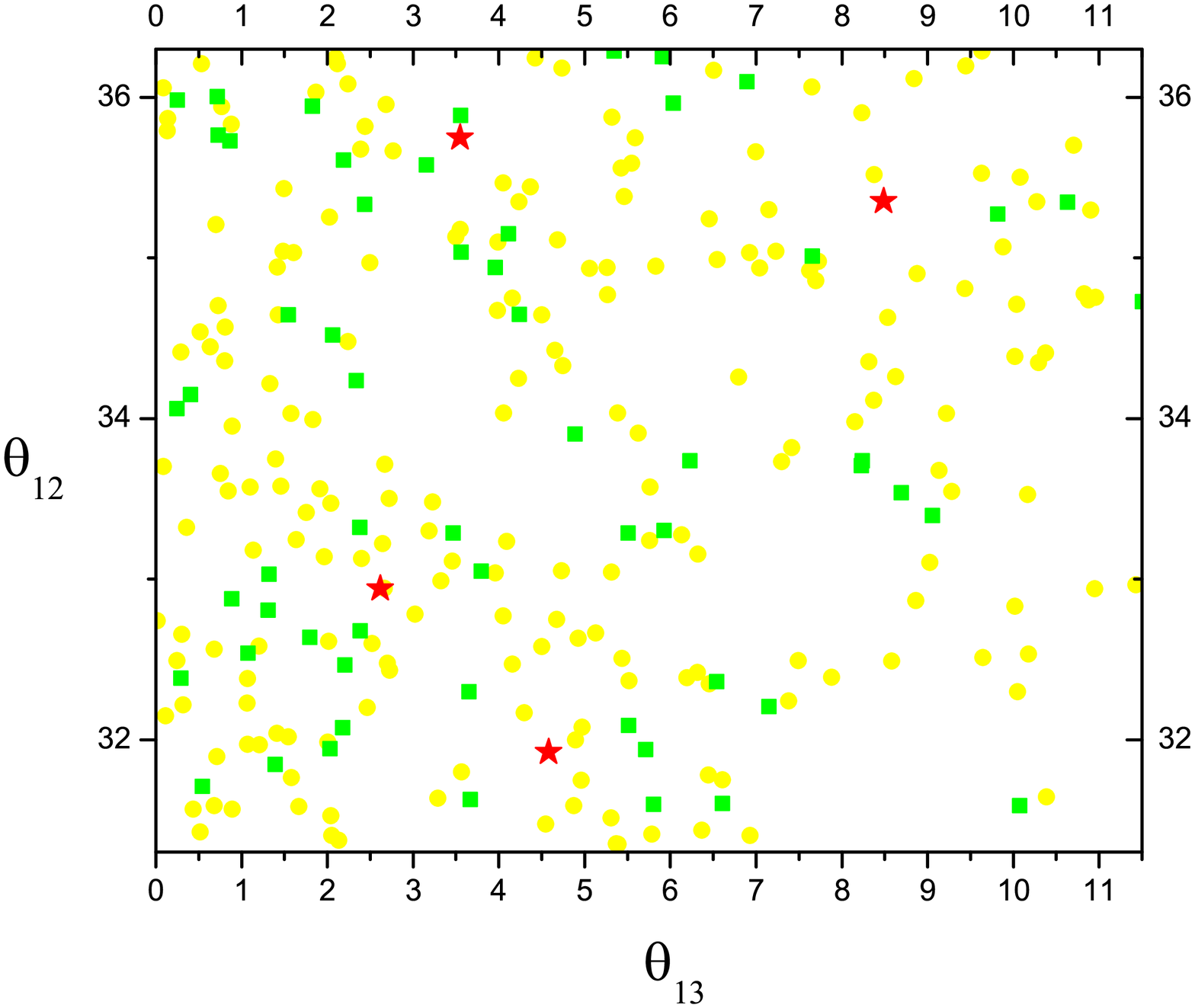,height=48mm,width=54mm}
\hspace{-4mm}
\psfig{file=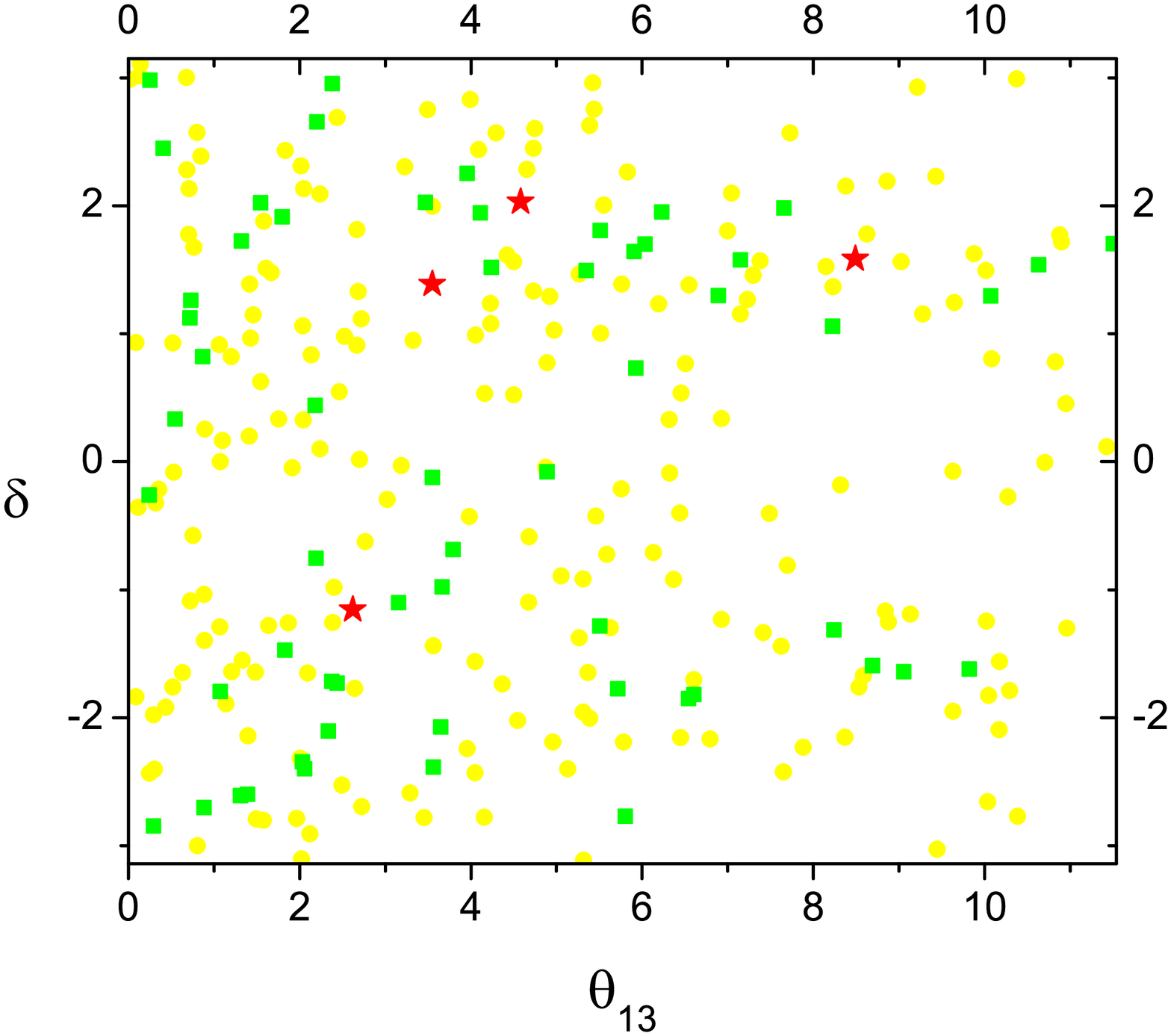,height=48mm,width=54mm} \\
\psfig{file=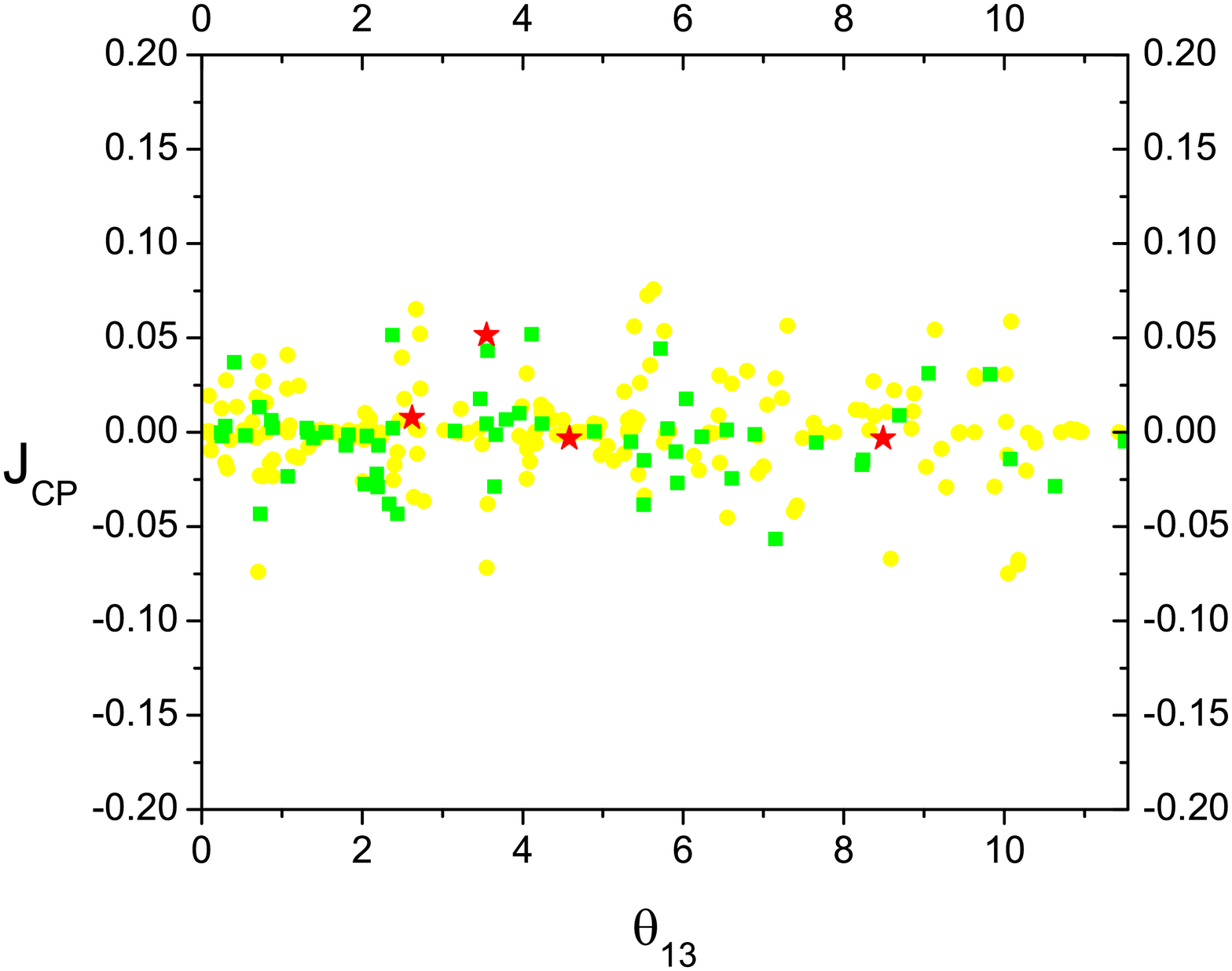,height=48mm,width=54mm}
\hspace{-4mm}
\psfig{file=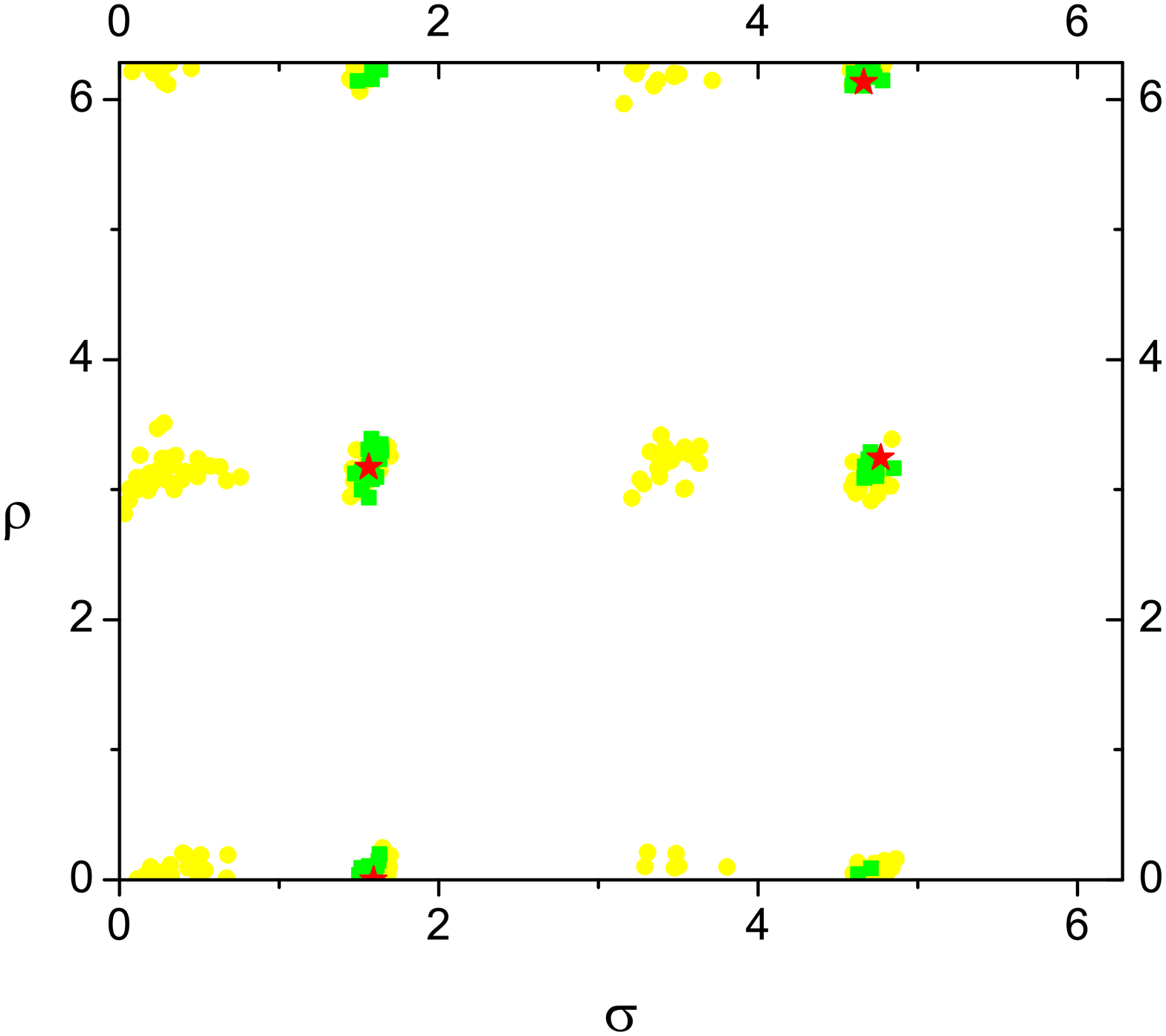,height=48mm,width=54mm}
\hspace{-4mm}
\psfig{file=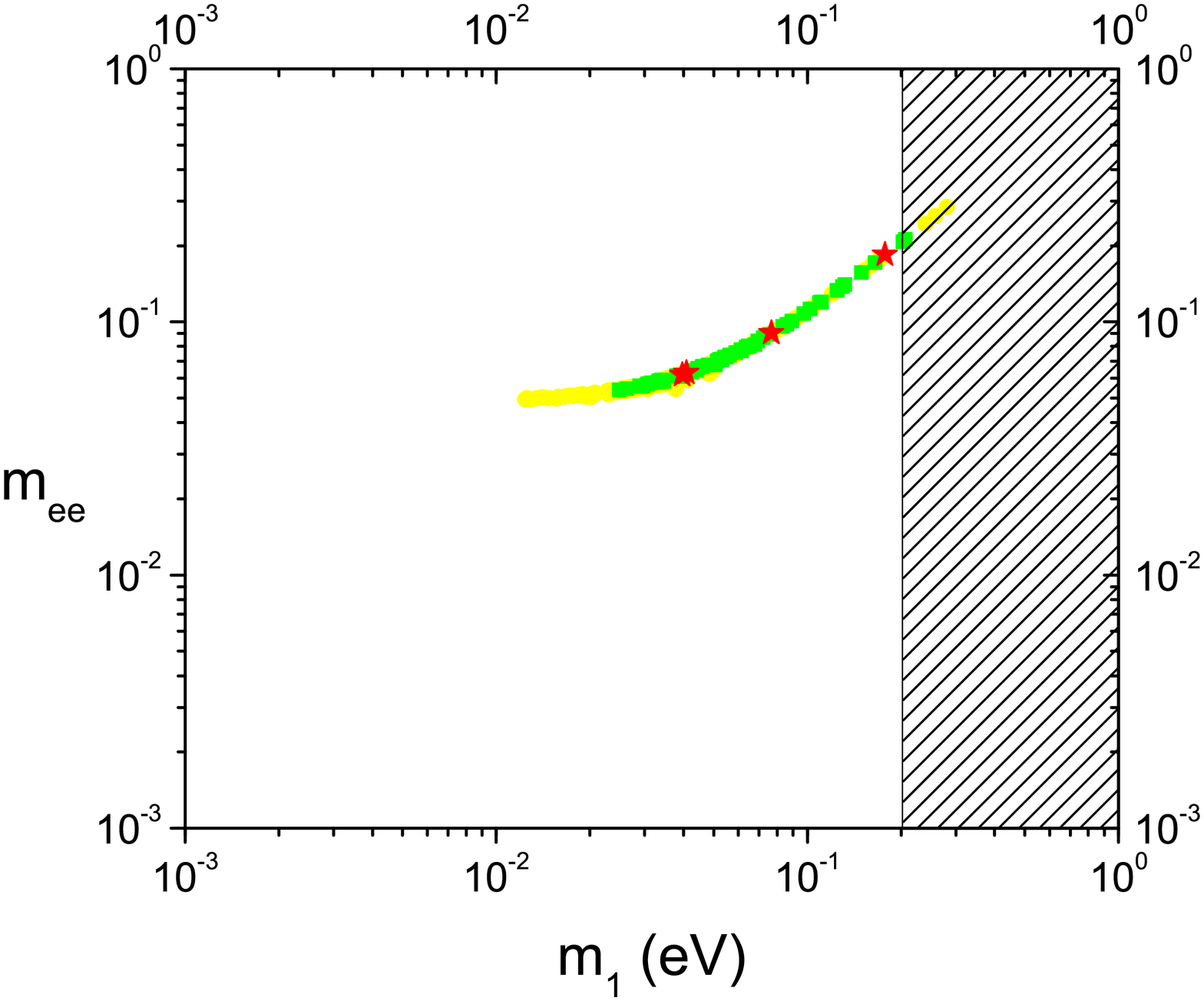,height=48mm,width=54mm} \\
\psfig{file=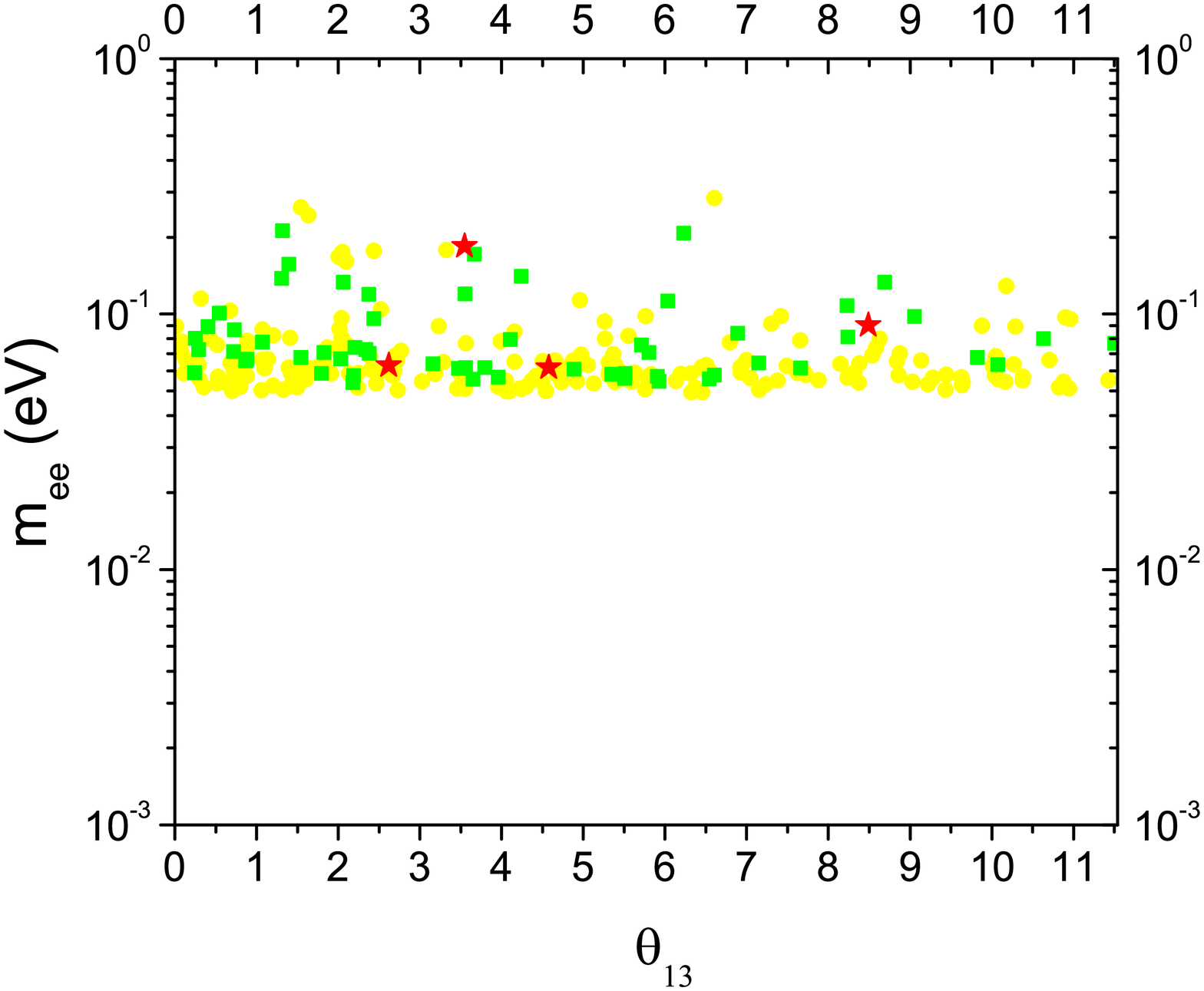,height=48mm,width=54mm}
\end{center}
\caption{Global scan, IO. Scatter plot of points in the parameter space that
satisfy successful leptogenesis ($\eta_B>5.9\times 10^{-9}$), for $\a_2=5$ (yellow circles),
$\a_2=4$ (green squares) and $\a_2=1.5$ (red stars). The dashed line
is the eq.~(\ref{m1th23IO}).}
\end{figure}
One can see how the allowed regions somehow merge those found for the two
extreme cases $V_L=I$ and $V_L=V_{CKM}$. There is therefore nothing really
new. IO is quite strongly constrained and it will be fully tested in next years.
In particular we can notice again how there is a clear lower bound on $\theta_{23}$
rather than an upper bound as in NO. More particularly, one can notice
that the allowed region in the plane $m_1-\theta_{23}$ is approximately described by
\be\label{m1th23IO}
\theta_{23}\simeq 43^{\circ}+12^{\circ}\,\log(0.2\,{\rm eV}/m_1)
\ee
(the dashed line in the upper right panel).
It is then quite interesting that
$SO(10)$ inspired leptogenesis is able to distinguish NO and IO
even at $m_1 \gtrsim 0.01\,{\rm eV}$, when the same values of $m_{ee}$
and of $\sum_i m_i$ (the quantity tested by cosmological observations)
are found both for IO and for NO.
From this point of view $SO(10)$ inspired leptogenesis
provides a way to solve this ambiguity.

%%%%%%%%%%%%%%%%%%%%%%%
\section{Final remarks}
%%%%%%%%%%%%%%%%%%%%%%%

We have derived constraints on the low energy neutrino parameters
from $SO(10)$-inspired leptogenesis.
Our investigation shows  that even minimal leptogenesis, based on a type I seesaw
mechanism and assuming a thermal production of the RH neutrinos and with a traditional
high mass scale RH neutrino spectrum,  can be testable  within a well motivated framework,
where the see-saw parameter space is restricted by the $SO(10)$-inspired conditions.
The role played by the $N_2$ decays is crucial in this respect,
not only in re-opening the viability of these models. The presence in the $N_2$-dominated regime
of a double stage, a production stage and a lightest RH neutrino wash-out stage,
seems to introduce, as shown simultaneously both by the numerical and by the analytical results,
a strong direct dependence on neutrino mixing angles as well, in addition to the
dependence on the absolute neutrino mass scale, already found in usual
$N_1$-dominated leptogenesis \cite{review}.

Interesting  predictions, that can be tested in future years,
with intriguing  correlations involving the absolute neutrino mass scale
and the neutrino mixing angles  emerge.

In the significant case of NO with low $m_1$ values, the neutrinoless
double beta decay effective mass seems to be too small to be measured but not
arbitrary small and in any case
future experimental results can be anyway useful to restrict the allowed regions for the
other parameters and sharpening the predictions.

The results for  $V_L\neq I$ seems also to be sensitive to $V_L$ itself
and they therefore suggest that there is an opportunity
to  gain  information on it,
an interesting point within studies of specific $SO(10)$ models.
It is quite interesting that there is an allowed region in the parameter space that
allows large values of $\theta_{13}$ testable with on-going reactor neutrino
experiments and that for these large values the models favours either large or small
$\theta_{23}$ values depending whether $m_1\lesssim 0.01\,{\rm eV}$ or
$m_1\gtrsim 0.01\,{\rm eV}$.

In the small $m_1$ range it is also interesting that
the constraints are completely independent of any assumption
on the initial conditions, a point that maybe makes
this option more attractive. It is actually quite interesting that
this conclusion is also supported by completely independent and general
considerations based on the possibility to reproduce, without a particularly fine tuned $U_R$ matrix,
the observed atmospheric to solar neutrino mass ratio, $m_{\rm atm}/m_{\rm sol}\simeq 6$,
starting from hierarchical neutrino Yukawa couplings. It is found \cite{casas2}
that this experimental observation is far more natural if the lightest neutrino presents
a much stronger hierarchy than the the two heavy ones, as it occurs in the region
that we have found at small $m_1$. It should be also stressed again, that since
our results are independent of $\alpha_3$ and $\alpha_1$, as far as $M_3\gtrsim 10^{12}\,{\rm GeV}$
and $M_1\lesssim 10^9{\rm GeV}$, they hold even for a Yukawa couplings hierarchy milder than
in the case of up quark masses. This can help to make even more natural to
reproduce the result $m_{\rm atm}/m_{\rm sol}\simeq 6$ without a fine tuned $U_R$.

A more precise measurement of $\theta_{12}$ could also play
a relevant role in testing these models, a point that should be
addressed by a more involved multi-parameter analysis.
A future accurate determination of the neutrino mixing angles
will be therefore crucial to test $SO(10)$-inspired leptogenesis
and could even yield some interesting information on the matrix $V_L$.
In conclusion, it seems  that $SO(10)$-inspired leptogenesis provides
an interesting well justified example that gives some hopes
about the possibility of testing minimal
leptogenesis even only with low energy neutrino experiments.
It will be then quite interesting in next years to compare the
experimental results with the constraints and the predictions
from $SO(10)$-inspired models that we discussed.

\subsection*{Acknowledgments}

We wish to thank S.~Blanchet, M.~Chun Chen, F.~Feruglio, Carlo Giunti, S.~King,
R.~Mohapatra, M.~Schmidt for useful discussions.
P.D.B. acknowledges financial support from the NExT Institute and SEPnet.

\end{document}